\documentclass[%
 reprint,
 amsmath,amssymb,
 aps,
]{revtex4-2}

\usepackage{graphicx}
\usepackage{dcolumn}
\usepackage{bm}

\begin{document}

\preprint{APS/123-QED}

\title{Statistical Properties of three-dimensional Hall Magnetohydrodynamics Turbulence}

\author{Sharad K Yadav$^1$}
\email{sharadyadav@iisc.ac.in; Current address:Department of Physics, Sardar Vallabhbhai National Institute of Technology (SVNIT), Surat 395007, India.}
\author{Hideaki Miura$^2$}
\email{miura.hideaki@nifs.ac.jp}
\author{Rahul Pandit$^1$}
\email{rahul@iisc.ac.in; Also at Jawaharlal Nehru Centre for Advanced Scientific Research (JNCASR), Jakkur, Bangalore, India.}
\affiliation{$^1$ Centre for Condensed Matter Theory, Department of Physics, Indian Institute of Science (IISc.), Bangalore 560012, India}
\affiliation{$^2$National Instittue for Fusion Science (NIFS), Toki, Gifu 509-5292, Japan}




\date{\today}

\begin{abstract}

The three-dimensional (3D) Hall magnetohydrodynamics (HMHD) equations are often
used to study turbulence in the solar wind. Some earlier studies have
investigated the statistical properties of 3D HMHD turbulence by using
simple shell models or pseudospectral direct numerical simulations (DNSs)
of the 3D HMHD equations; these DNSs have been restricted to modest spatial
resolutions and have covered a limited parameter range. To explore the
dependence of 3D HMHD turbulence on the Reynolds number $Re$ and
the ion-inertial scale $d_{i}$, we have carried out detailed
pseudospectral DNSs of the 3D HMHD equations
and their counterparts for 3D MHD ($d_{i} = 0$). We present several
statistical properties of 3D HMHD turbulence, which we compare with 3D
MHD turbulence by calculating (a) the temporal evolution of the
energy-dissipation rates and the energy, (b) the wave-number dependence
of fluid and magnetic spectra, (c) the probability distribution
functions (PDFs) of the cosines of the angles between various pairs of
vectors, such as the velocity and the magnetic field, and (d) various
measures of the intermittency in 3D HMHD and 3D MHD turbulence.


\end{abstract}

\maketitle


\section{\label{sec:int} Introduction}

Electrically conducting fluids can exhibit turbulence that is characterised not
only by fluctuations in the fluid velocity and the vorticity, but also in the
magnetic field and the current. At the simplest level, such fluids can be
modelled by using the equations of magnetohydrodynamics
(MHD)~\cite{choudhuri1998physics,krishan1999astrophysical,rudiger2006magnetic,goedbloed2004principles,biskamp2003magnetohydrodynamic,verma2004statistical,
davidson2002introduction,sahoo2011systematics,basu2014structure}. Examples of such
conducting-fluid flows can be found in liquid metals, in the interiors of
planets or in
laboratories~\cite{moffatt2019self,roberts2000geodynamo,shew2005liquid,monchaux2007generation},
in solar or stellar 
settings~\cite{choudhuri1998physics,krishan1999astrophysical,rudiger2006magnetic,goedbloed2004principles},
in the solar
wind~\cite{salem2009solar,podesta2007spectral,podesta2009scale,weygand2007taylor,
pouquet2020interplay,pouquet2020coupling},
and in the interstellar
medium~\cite{choudhuri1998physics,krishan1999astrophysical,rudiger2006magnetic,goedbloed2004principles,falceta2014turbulence}.
The MHD description of a plasma is based on a single-fluid approximation.
However, in plasmas like the solar wind, this single-fluid assumption is not
valid, especially at small scales comparable to or smaller than the
ion-inertial length scale $d_{i}=c/{\omega_{pi}}$, with $c$ the velocity of
light and the ion plasma frequency $\omega_{pi} \equiv \sqrt{4\pi n_{i}Z^{2}
e^{2}/m_{i}}$, where $Z$ is the charge state, $n_{i}$ the ion density, and
$m_{i}$ the mass of the ion. Hall magnetohydrodynamics (HMHD) is a simplified
fluid description of a plasma that accounts for two-fluid effects, to some
extent; HMHD includes the Hall term in Ohm's
law
~\cite{lighthill1960studies,
gomez2010hall,hnat2005compressibility,horbury2008anisotropic,martin2012energy,
mininni2002dynamo,mininni2003dynamo,mininni2005direct,
mininni2007energy,boffetta1999intermittency,chapman2008solar,
shaikh20093d,hori2008spectrum,miura2014structure,miura2016hall,
miura2012coarse,miura2019hall,banerjee2013multiscaling,
galtier2007multiscale,galtier2008karman,meyrand2012spontaneous,
meyrand2018coexistence,krishan2004magnetic,gomez2005mhd};
and it reduces to MHD if $d_{i}=0$. Shell models have also been developed to
study the statistical properties of HMHD
turblence~\cite{hori2005shell,banerjee2013multiscaling}. The HMHD partial
differential equations (PDEs) pose several challenges for mathematicians, who
study the regularity properties of solutions of these
PDEs~\cite{chae2014well,alghamdi2018regularity}, for fluid dynamicists and
statistical mechanicians, who seek to characterise the statistical properties
of HMHD turbulence, and for astrophysicists, who use these PDEs to model
turbulence in astrophysical systems such as the solar wind.
The solar
wind~\cite{goldstein1995magnetohydrodynamic,verma1996nonclassical,
krishan1999astrophysical,salem2009solar,podesta2007spectral,podesta2009scale,
weygand2007taylor,bruno2013,bruno2016turbulence,kiyani2009global,
matthaeus1982measurement,matthaeus2015intermittency,alexandrova2013solar,
sahraoui2020magnetohydrodynamic}
has been described as a turbulence laboratory~\cite{bruno2013}, for it is in a
highly turbulent state: the magnetic Reynolds number $Re_M$ lies in the range
$10^{5} \lesssim Re_M \lesssim 10^{9}$ and the magnetic Prandtl number $Pr_M
\simeq 1$; the kinetic- and magnetic-energy spectra, $E_{u}(k)$ and $E_{b}(k)$,
respectively, extend over many decades of the wave number $k$; for time-domain
measurements, $k$ is replaced by the frequency $f$. Satellite observations of
solar-wind-plasma turbulence
\cite{kiyani2009global,matthaeus1982measurement,matthaeus2015intermittency,podesta2007spectral,alexandrova2013solar,alexandrova2008small,alexandrova2007solar,zimbardo2010magnetic,sahraoui2020magnetohydrodynamic}
have shown that, in the inertial range, $E_{u}(f) \sim f^{-\alpha}$, with
$\alpha \simeq 5/3$, the scaling exponent that follows from the Kolmogorov
hypotheses~\cite{frish1995} of $1941$ (henceforth K41). In contrast, the
magnetic-energy spectrum has two different scaling or inertial ranges
(henceforth, we refer to them as the \textit{inertial} and
\textit{intermediate-dissipation} ranges): (i)
for $f_I \ll f \ll f_{ci}$, $E_{b}(f) \sim f^{-\alpha}$, where the frequency $f_I$
is related inversely to the integral length scale of the turbulence,
$f_{ci}$, the ion-cyclotron frequency, is related inversely to $d_{i}$ and
$\alpha \simeq 5/3$ is consistent with the K41
value; (ii) for $f_{ci} \ll f \ll f_d < $, where  the frequency $f_d$ is related
inversely to the dissipation length scale at which viscous losses become
significant, $E_{b}(f) \sim f^{-\alpha_1}$, with $\alpha_1$ in the range of $1
\lesssim \alpha_1 \lesssim 4$. The solar wind is unlike a well-controlled
experiments in a laboratory, so it is not surprising that spectral exponents
show a range of values (see, e.g., Ref.~\cite{vasquez2007numerous}, Fig. $5$ in
Ref.~\cite{matthaeus2012review}, and Fig. $3$ of
Ref.~\cite{sahraoui2020magnetohydrodynamic}); this variability of exponents has
been attributed to transients, unsteady conditions, anisotropies, and effects
that lie beyond an incompressible-MHD description~\cite{matthaeus2012review}.
Solar-wind-turbulence data have also been analysed to uncover (a) intermittency
and multiscaling of velocity and magnetic-field structure
functions~\cite{wan2011investigation,kiyani2009global} and (b) the alignment of
velocity and magnetic-field fluctuations~\cite{podesta2009scale}. In
particular, the study of Ref.~\cite{kiyani2009global} has found
structure-function multiscaling (simple scaling), in the first (second)
frequency range mentioned above. The magnetosheath is another near-Earth space
plasma; for a comparison of plasma turbulence in the solar wind and in the
magnetosheath we refer the reader to
Refs.~\cite{zimbardo2010magnetic,sahraoui2020magnetohydrodynamic}.

There has been a steady stream of theoretical studies and direct numerical
simulations (DNSs) of
HMHD~\cite{lighthill1960studies,krishan2004magnetic,gomez2005mhd,gomez2010hall,
hnat2005compressibility,horbury2008anisotropic,martin2012energy,mininni2002dynamo,mininni2003dynamo,
mininni2005direct,mininni2007energy,boffetta1999intermittency,chapman2008solar,
shaikh20093d,hori2008spectrum,miura2014structure,miura2016hall,miura2012coarse,
miura2019hall,banerjee2013multiscaling,galtier2007multiscale,galtier2008karman,
meyrand2012spontaneous,meyrand2018coexistence}; most of these concentrate on
three-dimensional (3D) HMHD or related shell-model or large-eddy-simulation
investigations. These studies yield a spectral exponent $\alpha \simeq 5/3$,
which is consistent with K41; however, the values of the spectral exponent
$\alpha_1$, suggested in different theoretical and DNS studies, lie in a
large range: $1 \lesssim \alpha_1 \lesssim 5.5$; clearly, it is more
challenging to develop an understanding of 3D HMHD turbulence than of its MHD
and fluid counterparts. Some of these works (see, e.g.,
Refs.~\cite{krishan2004magnetic,hori2008spectrum,miura2019hall,galtier2007multiscale})
have provided phenomenological arguments for the values of $\alpha_1$ that have been
obtained in different DNSs. It has been suggested that, in addition to $d_{i}$ and
the Reynolds and  Prandtl numbers, the statistical properties of 3D HMHD might
well depend on other parameters like the Alfv\'en number, which is the ratio
of kinetic and magnetic energies. Furthermore, DNSs have explored (a) the
intermittency and multiscaling of velocity and magnetic-field structure
functions~\cite{boffetta1999intermittency,wan2011investigation,banerjee2013multiscaling,rodriguez2013intermittency,matthaeus2015intermittency}
and (b) the alignment of velocity and magnetic-field
fluctuations~\cite{podesta2009scale}.

Given the uncertainties in spectral exponents and the statistical properties of
3D HMHD turbulence, it behooves us to initiate systematic investigations of
these properties of the type that have been carried out for 3D MHD
turbulence~\cite{sahoo2011systematics,dallas2013structures}. We present such a
study. In particular, we use extensive pseudospectral DNSs, with two different
types of initial conditions [henceforth, the initial conditions A and B (see
below)], to obtain the statistical properties of turbulence in the unforced 3D
HMHD equations; a comparison of such properties provides valuable insights into
the initial-condition dependence of the exponent $\alpha_1$ and multiscaling
exponents. Before we present the details of our work, we provide a qualitative
overview of the principal results from our DNSs:

\begin{itemize}

\item {\bf Spectra:} In our 3D MHD and 3D HMHD DNSs, in the inertial
range mentioned above, both
\begin{eqnarray}
\label{eq:alpha}
E_{u}(k) &\sim& k^{-\alpha} ; \nonumber \\
E_{b}(k) &\sim& k^{-\alpha} ;
\end{eqnarray}
the value of the spectral exponent $\alpha$
consistent with the K41 result $5/3$. In the intermediate-dissipation
range, with lengths $l$ in the range $d_{i} \ll l \ll
\eta_{d}^{b}$, where $\eta_{d}^{b}$ is the magnetic-dissipation
length scale,
\begin{equation}
\label{eq:alpha1}
E_{b}(k)\sim k^{-\alpha_1},
\end{equation}
with the value of $\alpha_1$ is consistent with  (A) $11/3$, for the initial condition
A, and (B) $7/3$, for the initial condition B (see
Sec.~\ref{sec:model} for precise definitions of these spectra
and the initial conditions). We also explore the $k$-dependence of
other spectra and of the wave-number-dependent Alfv\'en number
$E_{b}(k)/E_{u}(k)$ for these two initial conditions.
\item {\bf Probability distribution functions (PDFs):}
\begin{itemize}
\item We compute the PDFs of the cosines of the angles between various fields,
such as the velocity ${\bf u}$, vorticity $\bf{\omega}
=\nabla\times\bf{u}$, magnetic field ${\bf b}$, and current
density ${\bf j=\nabla \times b}$, to highlight the importance
of the Hall term in suppressing the tendency of alignment (or
antialignment) of these fields for both the initial conditions (A)
and (B).
\item We also explore intermittency in 3D HMHD turbulence
(and compare it with its 3D MHD counterpart) by calculating the PDFs of the velocity and
magnetic-field increments as a function of the separation
length scale $l$. We find evidence of small-scale intermittency
in our 3D HMHD plasma turbulence DNSs (as in 3D MHD plasma
turbulence DNSs); our DNSs, especially those for the initial
condition (A), show clearly that intermittency is suppressed
significanty in the second scaling range of 3D HMHD plasma
turbulence, in agreement with the results of solar-wind
measurements~\cite{kiyani2009global}.
\end{itemize}
\item {\bf Structure functions}: We compute the $l$-dependence of velocity and
magnetic-field structure functions and, therefrom, their order-$p$
multiscaling exponents $\zeta_{p}^{u}$ and $\zeta_{p}^{b}$,
respectively. In the inertial range, $\zeta_{p}^{u}$ and
$\zeta_{p}^{b}$ are nonlinear, monotone increasing functions of
$p$; this is a clear signature of multiscaling. By contrast, in
the intermediate-dissipation range,  $\zeta_{p}^{u}$ and $\zeta_{p}^{b}$
increase linearly with $p$, a hallmark of simple scaling; this linear
dependence is in consonance with solar-wind results~\cite{kiyani2009global}.
\label{eq:evofu}
\end{itemize}
The remainder of this paper is organized as follows. In Sec.~\ref{sec:model} we
present the 3D HMHD PDEs (Subsection~\ref{subsec:basiceq}), the pseudospectral
DNSs we employ to solve these PDEs, and the definitions
of various statistical measures (Subsection~\ref{subsec:method}) that we use
to characterise 3D HMHD turbulence.
In Sec.~\ref{sec:results} we provide results from our DNSs in three subsections: In Subsection~\ref{subsec:temevo}
we discuss the temporal evolution of the energy-dissipation rates and the
energy and the wave-number dependence of spectra, such as $E_{u}(k)$ and
$E_{b}(k)$. In Subsection~\ref{subsec:pdf} we compute PDFs of the cosines of the 
angles between the following pairs of vectors:
\{$\bf{u},\bf{b}$\}, \{$\bf{u},\bf{j}$\}, \{$\bf{u},\bf{\omega}$\},
\{$\bf{b},\bf{j}$\}, \{$\bf{b},\bf{\omega}$\}, and \{$\bf{\omega},\bf{j}$\};
by using these PDFs we quantify the degree of alignment between these pairs
of vectors.  In Subsection~\ref{subsec:intermitt} we characterise intermittency in
3D HMHD turbulence by examining the $l$ dependence PDFs of velocity- and
magnetic-field increments and of the order-$p$ structure functions of these increments.
Finally, we discuss the implications of our study in Sec.~\ref{sec:conclusion}.
In the Appendix we present some joint PDFs.

\section{Model and methods} 
\label{sec:model}

We begin with the 3D HMHD PDEs (Subsection~\ref{subsec:basiceq}); then we
present an outline of our pseudospectral DNS method and the definitions of
statistical measures (Subsection~\ref{subsec:method}) for 3D HMHD turbulence.

\subsection{Basic Equations}
\label{subsec:basiceq}

Three-dimensional (3D) HMHD is described by the following set of coupled PDEs
for ${\bf u}$ and ${\bf b}$: 

\begin{equation}
\label{eq:evofu}
 \frac{\partial{\bf u}}{\partial t}+ \left({\bf u}\cdot {\nabla}\right){\bf u} =
-{\nabla}\bar{p}+\left({\bf b}\cdot{\nabla}\right){\bf b}+\nu\nabla^{2}{\bf u}
+{\bf f_{u}} ;
\end{equation}
\begin{equation}
\label{eq:evofb}
\frac{\partial{\bf {b}}}{\partial t}= {\nabla}\times\left({\bf u}\times {\bf b}-d_{i}{\bf j}\times{\bf b}\right) + \eta\nabla^{2}{\bf b}+{\bf f_{b}};
\end{equation}
\begin{equation}
\label{eq:incomp}
{\bf \nabla \cdot u}=0 ; {\bf \nabla \cdot b}=0 .
\end{equation}
Equation~(\ref{eq:evofu}) is the momentum equation; this includes a
contribution from the Lorentz force ${\bf j} \times {\bf b}$, with the current
density ${\bf j=\nabla \times b}$, which can be separated into the
magnetic-tension term $\left({\bf b}\cdot{\nabla}\right){\bf b}$ and the
magnetic pressure, which we combine with the pressure $p$ to obtain the total
pressure $\bar{p} \equiv p+|{\bf b}|^{2}/4\pi$.  The induction
equation~(\ref{eq:evofb}) uses the generalized form of Ohm's law, which
includes the Hall term $\sim {\bf j} \times {\bf b}$, whose coefficient $d_{i}$
is the ion-inertial length; $\eta$ and $ \nu$ are, respectively, the kinematic
viscosity and the magnetic resistivity (or diffusivity); ${\bf f_{u}}$ and
${\bf f_{b}}$ are the forcing terms.  The Poisson equation for $\bar{p}$
follows from the divergence of Eq.~(\ref{eq:evofu}) and the incompressibility
condition $\nabla \cdot \bf u=0$:
\begin{equation}
\label{eq:4}
\nabla^{2} \bar{p}=\nabla\cdot\left[({\bf b}\cdot \nabla){\bf b}-({\bf u}\cdot\nabla){\bf u}\right] + \nabla \cdot {\bf f_{u}}
\end{equation}
We study decaying HMHD turbulence, so we set ${\bf f_{u}} = 0$ and ${\bf f_{b}} = 0$. 
Thus, the final form for the pressure-Poisson equation is
\begin{equation}
\label{eq:5}
\nabla^{2} \bar{p}=\nabla\cdot\left[\left({\bf b}\cdot \nabla \right){\bf b}-\left({\bf u}\cdot\nabla\right){\bf u}\right].
\end{equation}

\subsection{Direct Numerical Simulations and Statistical Measures}
\label{subsec:method}

We solve Eqs.~(\ref{eq:evofu}) and (\ref{eq:evofb}) by using the pseudospectral
method in a cubical domain of side $L=2\pi$, with periodic boundary conditions
(see, e.g., Ref.~\cite{sahoo2011systematics} for 3D MHD). We remove the
aliasing error, because of the nonlinear terms, by using the $2/3$ dealiasing
method. For time integration we employ the second-order slaved, Adams-Bashforth
scheme.

\begin{table*}
\hspace*{-1.8cm}
\begin{center}
\resizebox{!}{0.09\textheight}{
\begin{tabular} {c c c c c c c c c c c c c c}
\hline
\hline
{\bf $Run$} &
{\bf $N$} &
{\bf $\nu$} &
{\bf $P_{rm}$} &
{\bf $d_{i}$} &
{\bf $\delta t$} &
{\bf $u_{rms}$} &
{\bf $l_{I}$} &
{\bf $\lambda$} &
{\bf $Re_{\lambda}$} &
{\bf $t_{c}$} &
{\bf $k_{max}\eta_{d}^{u}$} &
{\bf $k_{max}\eta_{d}^{b}$} \\
\hline
$Run1$ &
 $256$ &
 $10^{-3}$ &
 $1$ &
 $0.0$ &
 $5.0\times 10^{-4}$ &
 $0.29$ &
 $1.193$ &
 $0.13$ &
 $39$ &
 $6.76$ &
 $2.09$ &
 $1.89$ \\
$Run2$ &
 $256$ &
 $10^{-3}$ &
 $1$ &
 $0.05$ &
 $5.0\times 10^{-4}$ &
 $0.13$ &
 $0.12$ &
 $0.128$ &
 $17$ &
 $7.4$ &
 $2.1$ &
 $1.8$ \\
 $Run3$ &
 $512$ &
 $5.0\times10^{-4}$ &
 $1$ &
 $0.0$ &
 $10^{-4}$ &
 $0.28$ &
 $0.29$ &
 $0.25$ &
 $142$ &
 $9.13$ &
 $1.82$ &
 $1.65$ \\
 $Run4$ &
 $512$ &
 $5.0\times10^{-4}$ &
 $1$ &
 $0.05$ &
 $10^{-4}$ &
 $0.29$ &
 $0.32$ &
 $0.28$ &
 $165$ &
 $9.65$ &
 $1.8$ &
 $1.6$ \\  
 $Run5a$ &
 $1024$ &
 $5.0\times10^{-4}$ &
 $1$ &
 $0.0$ &
 $10^{-4}$ &
 $1.0$ &
 $0.35$ &
 $0.07$ &
 $143$ &
 $0.5$ &
 $1.53$ &
 $1.37$ \\ 
 $Run5b$ &
 $1024$ &
 $5.0\times10^{-4}$ &
 $1$ &
 $0.025$ &
 $10^{-4}$ &
 $1.0$ &
 $0.37$ &
 $0.08$ &
 $153$ &
 $0.5$ &
 $1.60$ &
 $1.30$ \\
 $Run5c$ &
 $1024$ &
 $5.0\times10^{-4}$ &
 $1$ &
 $0.05$ &
 $10^{-4}$ &
 $1.0$ &
 $0.36$ &
 $0.09$ &
 $176$ &
 $0.5$ &
 $1.71$ &
 $1.25$ \\
\hline
\hline
\end{tabular}%
}
\end{center}
\caption{{\it Parameters in our DNSs $Run1$, $Run2$, $Run3$,
$Run4$, $Run5a$, $Run5b$, and $Run5c$:} $N^{3}$ is the number of collocation
points; $\nu$ is the kinematic viscosity; $Pr_{M}$ is the magnetic Prandtl
number; $d_{i}$ is the ion-inertial length (it is $0$ for the MHD runs); 
$\delta t$ is the time step;
$u_{rms}$, $l_{I}$, $\lambda$, and $Re_{\lambda}$ are the root-mean-square
velocity, the integral scale, the Taylor microscale, and the Taylor-microscale
Reynolds number, respectively. Most of these quantitites are obtained at the
cascade-completion time $t_{c}$ (see text); $\eta_{d}^{u}$ and $\eta_{d}^{b}$
are, respectively, the Kolmogorov dissipation length scales for the velocity
and magnetic fields. $k_{max}$ is the magnitude of the largest wave numbers in
our DNSs ($k_{max} \simeq 85.33, \, 170.67,$ and $343.33$ for $N=256$, $512$,
and $1024$, respectively). 
}
\label{table:1}
\end{table*}

We perform four sets of simulations, $Run1$, $Run2$, $Run3$, and 
$Run4$, in which the initial energy spectra (initial condition A) for the velocity and magnetic 
fields are as follows:
\begin{equation}
\label{eq:initA}
E_{u}^{0}(k)=E_{b}^{0}(k)=E^{0}k^{4}\exp(-2k^{2}),
\end{equation}
where $E^{0} \simeq 10$ is the initial amplitude in our DNSs.
For the DNSs $Run5a$, $Run5b$, and $Run5c$ these initial spectra (initial condition B) are
\begin{equation}
\label{eq:initB}
E_{u}^{0}(k)=E_{b}^{0}(k)=E^{0}k^{2}\exp(-2k^{2}). 
\end{equation}
The phases of the Fourier modes of the velocity and magnetic fields in
Eqs.~\ref{eq:initA} and \ref{eq:initB} are distributed randomly and uniformly
on the interval $[0, 2\pi)$.  Some data from the DNSs $Run5a$, $Run5b$, and
$Run5c$ have been published~\cite{miura2014structure}, in a different context;
we carry out a detailed comparison of results from $Run1$-$Run4$ and
$Run5a$-$Run5c$. The values of various parameters from our DNSs are listed
in Table~\ref{table:1}.

To characterize the statistical properties of 3D MHD and 3D HMHD turbulence, we
compute the following (cf. Ref.~\cite{sahoo2011systematics} for 3D MHD turbulence): 
\begin{itemize}
\item At time $t$ we obtain the kinetic-energy, magnetic-energy,  
kinetic-energy-dissipation-rate, magnetic-energy-dissipation-rate, and 
pressure spectra, which are, respectively:
\begin{eqnarray}
\label{eq:spectra}
E_{u}(k,t) &\equiv& \sum \limits_{k-\frac{1}{2}\leq k^{\prime}\leq k+\frac{1}{2}}|\tilde{\bf u}({\bf k}^{\prime},t)|^{2}; \nonumber \\ 
E_{b}(k,t) &\equiv& \sum \limits_{k-\frac{1}{2}\leq k^{\prime}\leq k+\frac{1}{2}}|\tilde{\bf b}({\bf k}^{\prime},t)|^{2}; \nonumber \\
\epsilon_{u}(k,t)&=&\nu k^{2}E_{u}(k,t); \nonumber \\ 
\epsilon_{b}(k,t)&=&\eta k^{2}E_{b}(k,t); \nonumber \\ 
P(k,t) &\equiv& \sum \limits_{k-\frac{1}{2}\leq k^{\prime}\leq k+\frac{1}{2}}|\tilde{\bar{p}}({\bf k}^{\prime},t)|^{2};
\end{eqnarray}
here, tildes denote spatial Fourier transforms, and 
$k^{\prime} \equiv \mid {\bf k}^{\prime} \mid$; we compute the time evolution of 
different energies and dissipation rates by summing the corresponding spectrum over
the wave number $k$; e.g., $E_{u}(t) = \sum_k E_{u}(k,t)$. In the inertial and 
intermediate-dissipation ranges mentioned above, these spectra show power-law dependences 
on $k$, which we  elucidate below.
\item We compute PDFs of the cosine of the 
angles between the following pairs of vectors, to characterise the degree of 
alignment between them: \{$\bf{u},\bf{b}$\}, \{$\bf{u},\bf{j}$\}, 
\{$\bf{u},\bf{\omega}$\}, \{$\bf{b},\bf{j}$\}, \{$\bf{b},\bf{\omega}$\}, and 
\{$\bf{\omega},\bf{j}$\}.
\item To characterise multifractality and the intermittency  
we calculate the order$-p$, equal-time longitudinal structure functions 
at time $t$ :
\begin{eqnarray}
\label{eq:strucfun}
S_{p}^{a}(l) &=& \langle |\delta a_{||}({\bf x},l)|^{p} \rangle ; \nonumber \\ 
\delta a_{||}({\bf a},l) &=& [{\bf a(x+l,t) -a(x,t)}\cdot \frac{\bf l}{l}];
\end{eqnarray}
here, $\bf a_{||}$ is the longitudinal component of $\bf a$, which is 
$\bf u$ or $\bf b$ for velocity and magnetic-field structure functions
and $\delta a_{||}({\bf a},l)$ the longitudinal component of its increment; 
we suppress $t$ in the arguments of structure functions for notational convenience. 
In the inertial range
\begin{eqnarray}
\label{eq:zetap}
S_{p}^{u}(l)\sim l^{\zeta_{p}^{u}};  \nonumber \\ 
S_{p}^{b}(l)\sim l^{\zeta_{p}^{b}}; 
\end{eqnarray}
$\zeta_{p}^{u}$ and $\zeta_{p}^{b}$ are the velocity and magnetic-field 
multiscaling exponents of order $p$. As we show below, in 3D HMHD turbulence, 
the magnetic-field spectra and structure functions exhibit two different scaling 
regions called \textit{inertial} and \textit{intermediate-dissipation} regions; e.g., 
\begin{eqnarray}
\label{eq:zetap12}
d_{i} \ll l \ll L ; S_{p}^{b}(l)\sim l^{\zeta_{p}^{b,1}};  \nonumber \\
\eta_{d}^{b} \ll l \ll d_{i}; S_{p}^{b}(l)\sim l^{\zeta_{p}^{b,2}};
\end{eqnarray}
here, $\zeta_{p}^{b,1}$ and $\zeta_{p}^{b,2}$ are the multiscaling exponents 
in these two scaling regions. To study intermittency we also compute PDFs of 
velocity and magnetic field increments and the hyperflatnesses
\begin{equation}
\label{eq:f6}
F_{6}^{a}(l)=S_{6}^{a}(l)/[S_{2}^{a}(l)]^{3}. 
\end{equation}
\end{itemize}

\section{Results} 
\label{sec:results} 
		
We present the results of our DNSs in three subsections: In
Subsection~\ref{subsec:temevo} we discuss the temporal evolution of the
energy-dissipation rates and the energy; we then present the spectra that we
have defined above.  In Subsection~\ref{subsec:pdf} we compute probability
distribution functions (PDFs) of the cosine of the angles between the following
pairs of vectors: \{$\bf{u},\bf{b}$\}, \{$\bf{u},\bf{j}$\},
\{$\bf{u},\bf{\omega}$\}, \{$\bf{b},\bf{j}$\}, \{$\bf{b},\bf{\omega}$\}, and
\{$\bf{\omega},\bf{j}$\}; these PDFs help us to quantify the degree of
alignment between these pairs.  In Subsection~\ref{subsec:intermitt} we
characterise intermittency in 3D HMHD turbulence by examining PDFs of velocity
and magnetic-field increments and the $l$ dependences of order-$p$ structure
functions of these increments.

\subsection{Temporal Evolution and Spectra}
\label{subsec:temevo}

\begin{figure*}[t]
\centering 
\begin{tabular}{c c c}
\textbf{(a)} & \textbf{(b)} & \textbf{(c)}  \\
\includegraphics [scale=0.4]{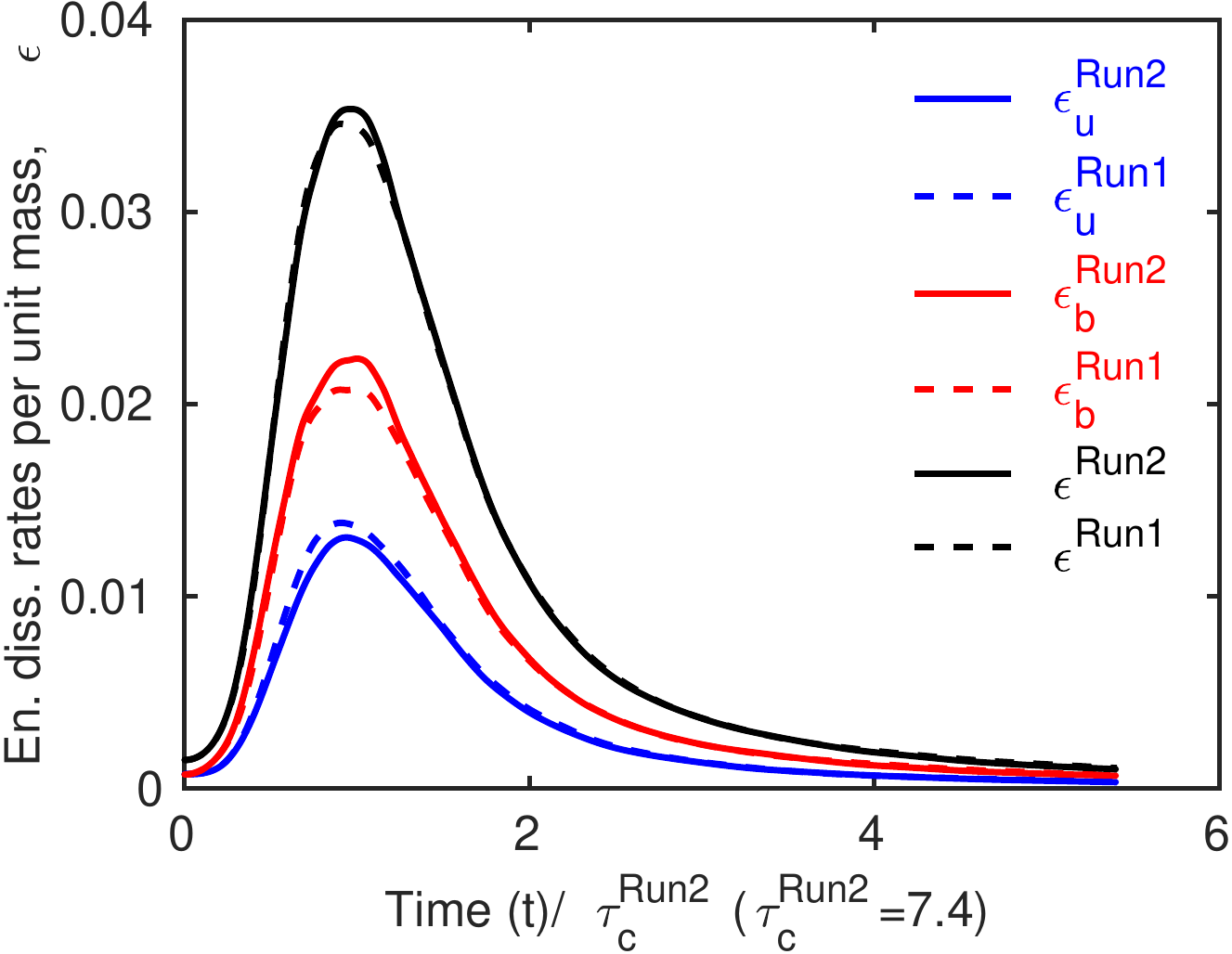} & 
\includegraphics [scale=0.4]{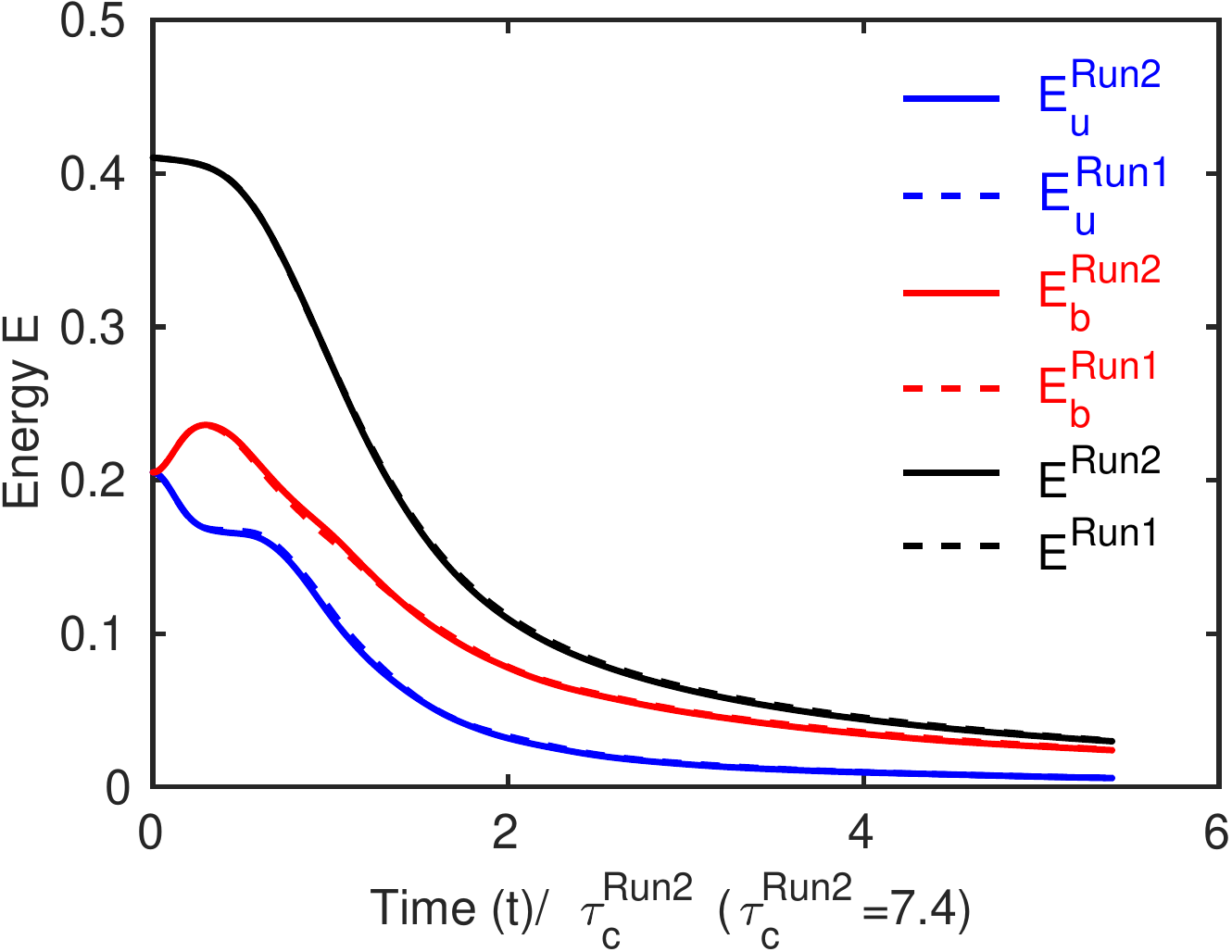} &
\includegraphics [scale=0.4]{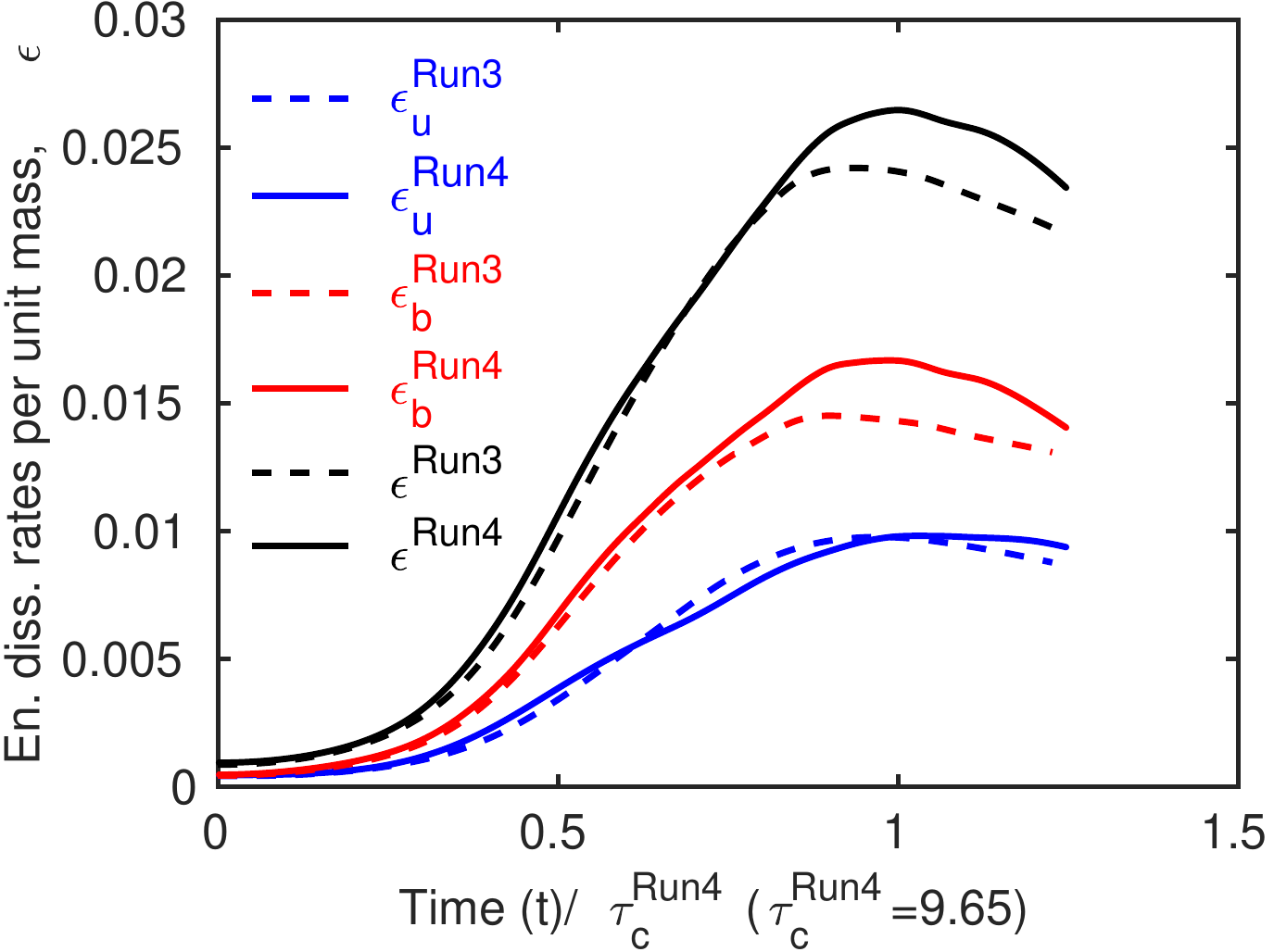}
\end{tabular}
%
\begin{tabular}{c c c}
\textbf{(d)} & \textbf{(e)} & \textbf{(f)}  \\
\includegraphics [scale=0.4]{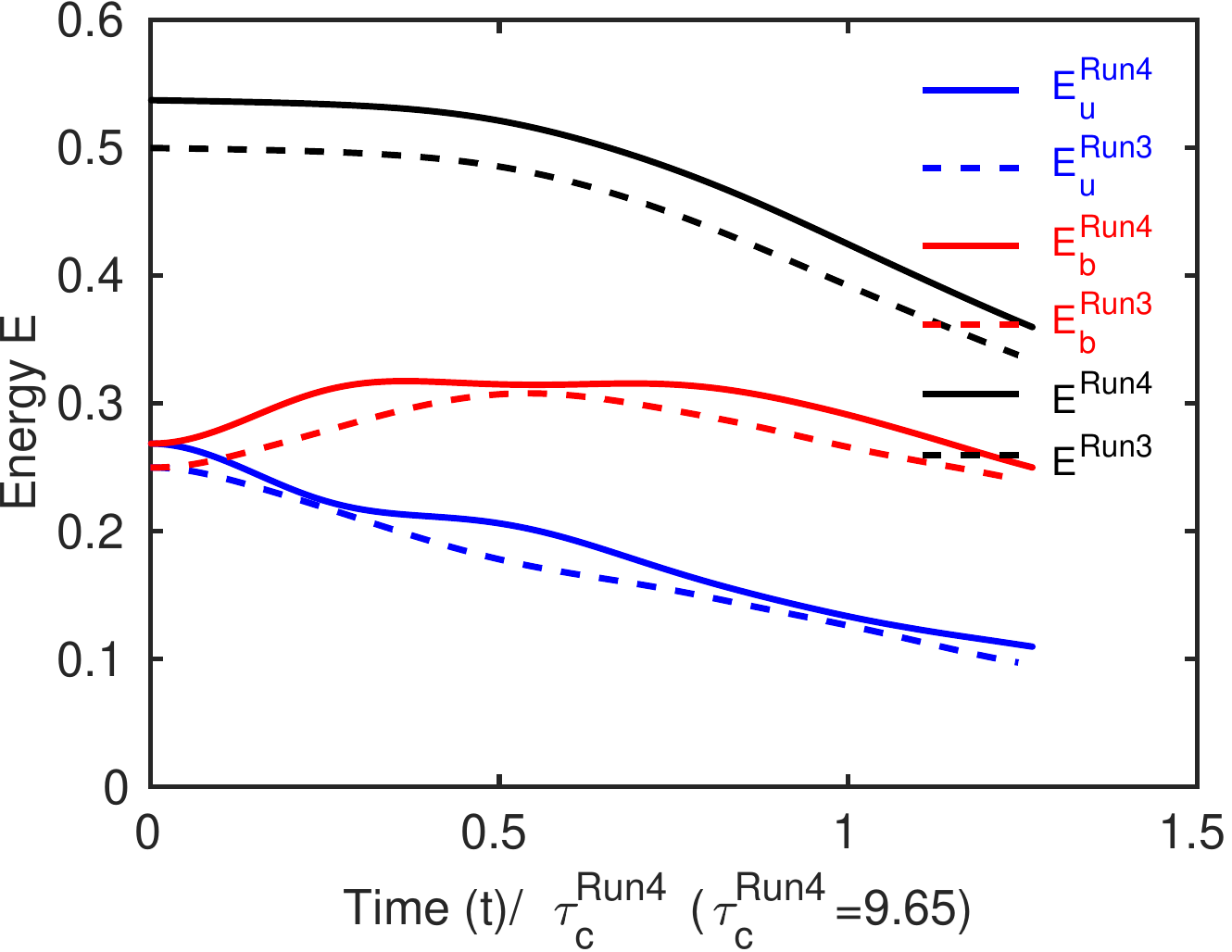} &
\includegraphics [scale=0.4]{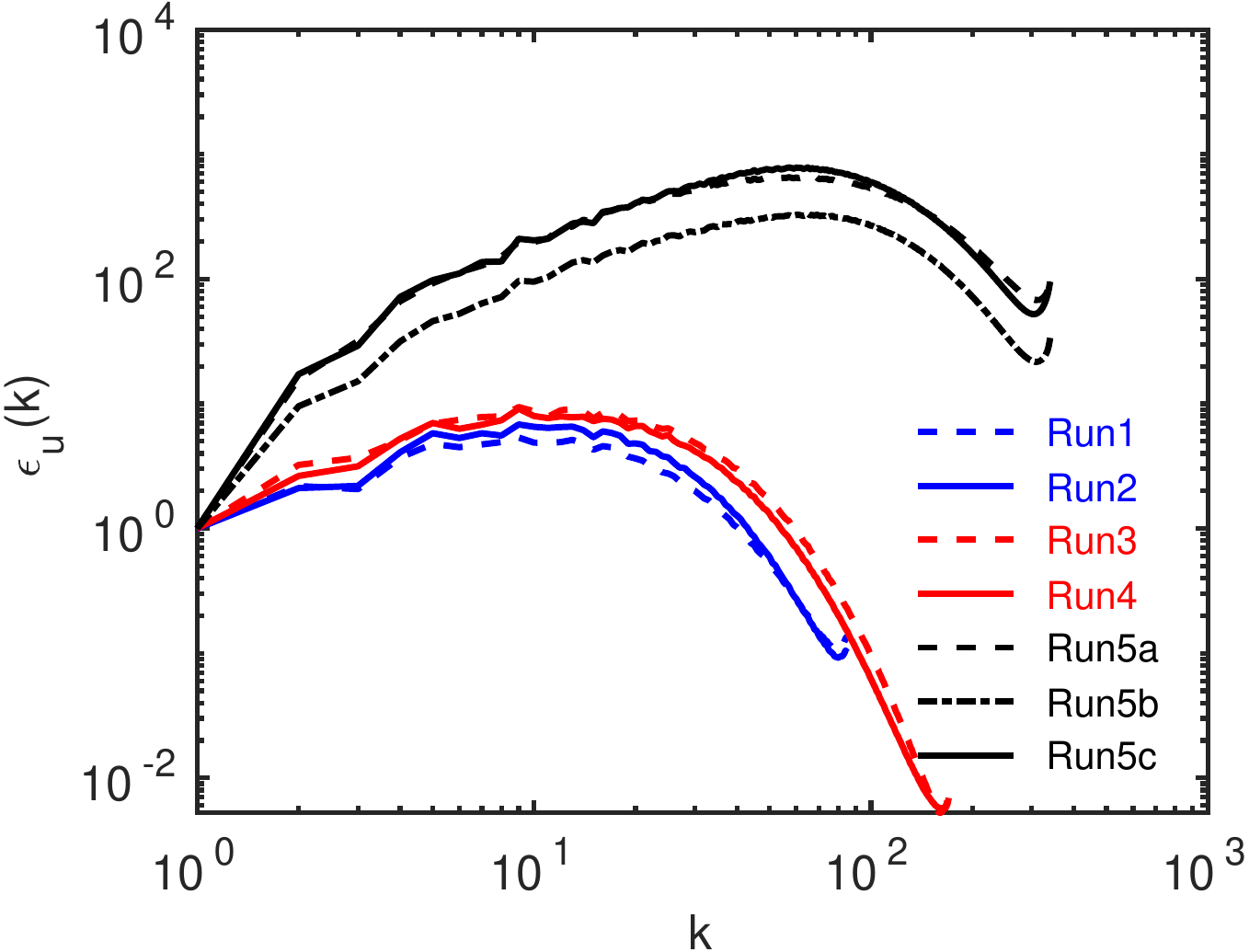} & 
\includegraphics [scale=0.4]{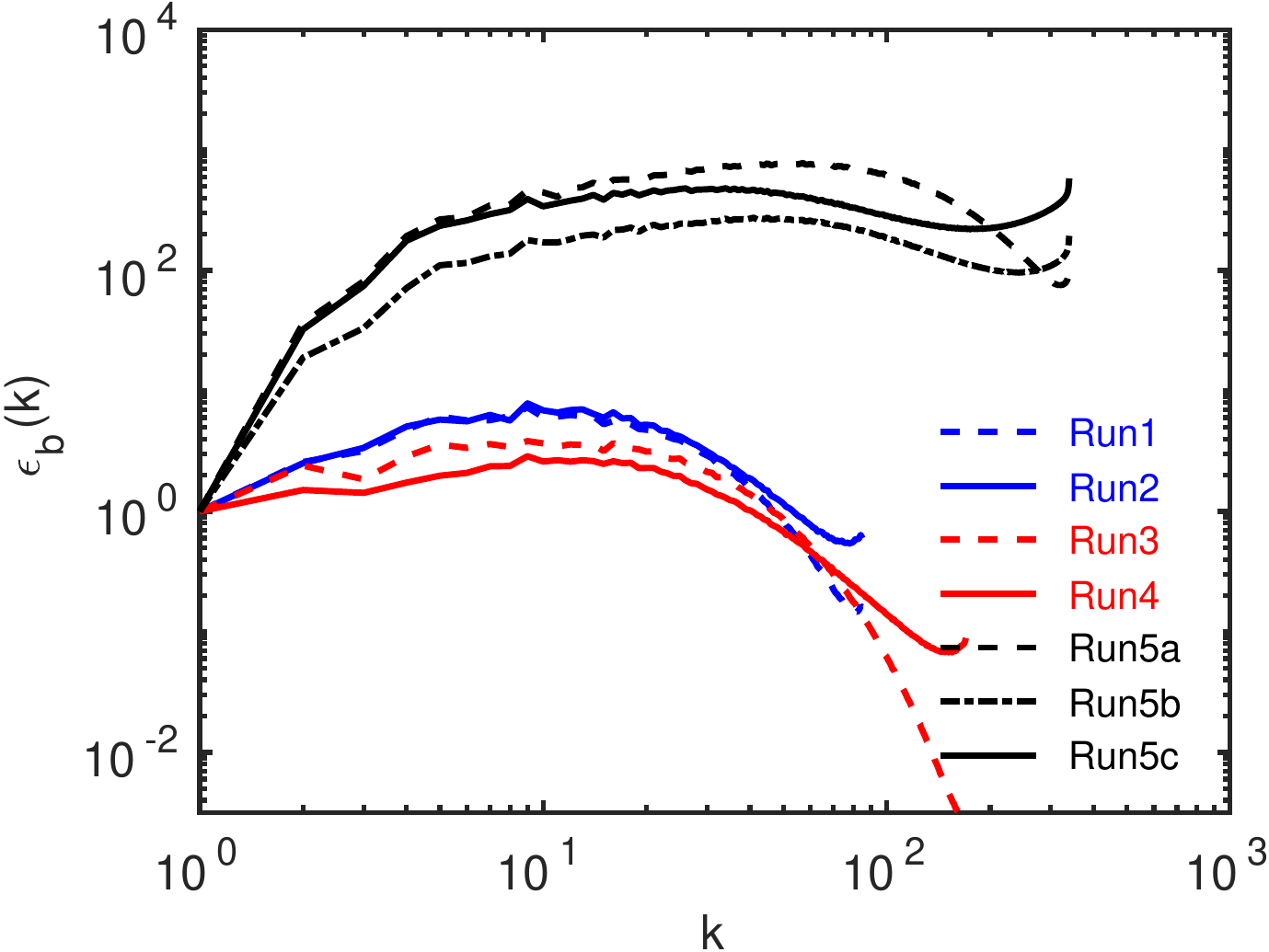}
\end{tabular}
\vskip -.1cm

\caption{(Color online) Plots versus the scaled time $t$ of (a) the
kinetic-energy dissipation rate $\epsilon_{u}$ (blue curves),
magnetic-energy dissipation rate $\epsilon_{b}$ (red curves), and the
total-energy dissipation rate $\epsilon$ (black curves) for $Run1$
(dashed lines) and $Run2$ (solid lines); and (b) the kinetic energy
$E_{u}$ (blue curve), magnetic energy $E_{b}$ (red curve), and the
total energy $E$ (black curve) for $Run1$ (dashed lines) and $Run2$
(solid lines). (c) and (d) are, respectively, the counterparts of (a)
and (b) for $Run3$ and $Run4$. Log-log (base $10$) plots versus the
wave number $k$ of (e) the  kinetic-energy-dissipation-rate spectra
$\epsilon_{u}(k)$ and (f)  the  magnetic-energy-dissipation-rate
spectra $\epsilon_{b}(k)$ from $Run1$ (dashed-blue curve), $Run2$
(solid-blue curve), $Run3 $ (dashed-red curve), $Run4 $ (solid-red
curve), $Run5a$ (dashed-black curve), $Run5b$ (dashed-dot black curve),
and $Run5c$ (solid-black curve).}

\label{fig:1}
\end{figure*}

In Fig.~\ref{fig:1}$\left(a\right)$ we display the time evolution of the
kinetic-energy dissipation rate (blue curves), the magnetic-energy dissipation
rate (red curves), and total-energy dissipation rate (black curves) from $Run1$
(dashed lines) and $Run2$ (solid lines) for 3D MHD and 3D HMHD turbulence,
respectively. We remark that these dissipation rates increase sharply, up until
a cascade-completion time $\tau_{c}$, and then they decay slowly. The peak
positions are nearly the same for all these curves; the peaks for $Run1$ occur
marginally earlier than they do in $Run2$.  In Fig.~\ref{fig:1}$\left(b\right)$
we present the time dependence of the fluid, magnetic, and total energies in
simulations $Run1$ and $Run2$.  Figures~\ref{fig:1}$\left(c\right)$ and
\ref{fig:1}$\left(d\right)$ are the counterparts of
Figs.~\ref{fig:1}$\left(a\right)$ and \ref{fig:1}$\left(b\right)$ for $Run3$
and $Run4$. Figures~\ref{fig:1}$\left(e\right)$ and~\ref{fig:1}$\left(f\right)$
display the spectra of the kinetic-energy-dissipation rate and the magnetic
energy-dissipation rate, respectively, at $\tau_{c}$ for $Run1$, $Run2$,
$Run3$, $Run4$, $Run5a$, $Run5b$, and $Run5c$; the well-developed peaks in
these spectra show that our DNSs are well resolved; we also summarise this
in Table~\ref{table:1}, which shows that $k_{max}\eta_{d}^{u} > 1$ and 
$k_{max}\eta_{d}^{b} >1 $ for all these runs.

\begin{figure*}[t]
\centering 
\begin{tabular}{c c c}
\textbf{(a)} & \textbf{(b)} & \textbf{(c)} \\
\includegraphics [scale=0.4]{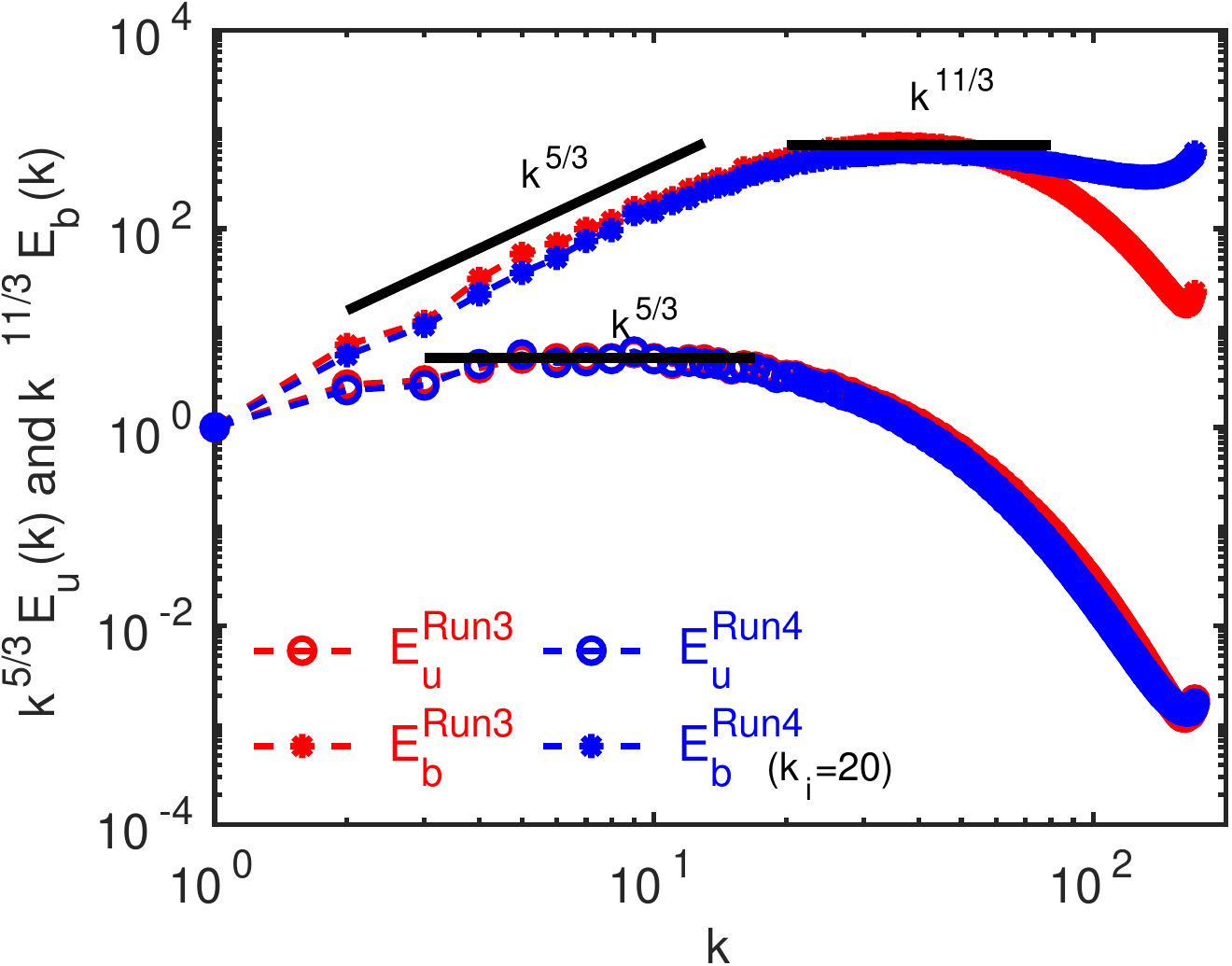} & 
\includegraphics [scale=0.4]{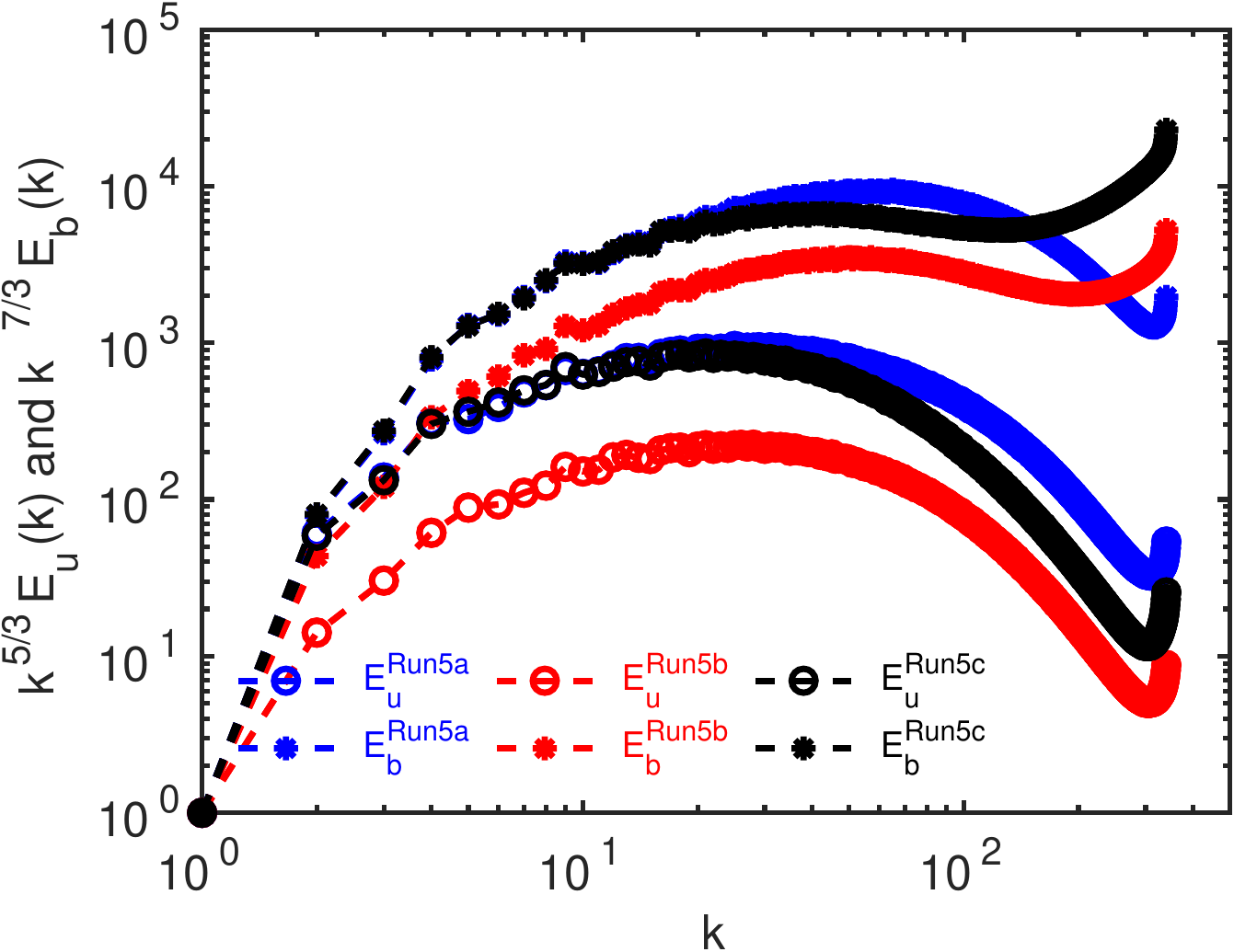} &
\includegraphics [scale=0.4]{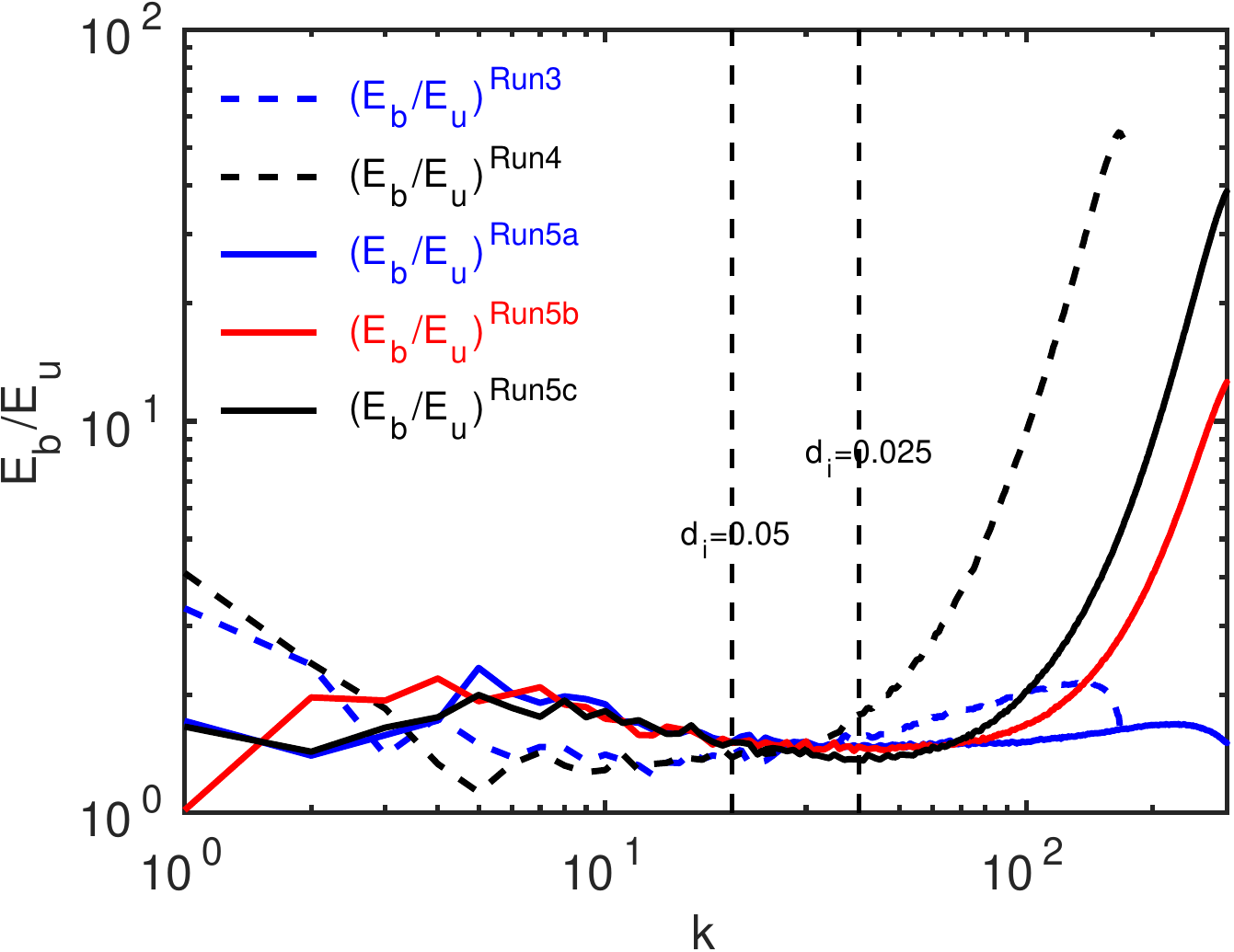}
\end{tabular}
\vskip -.1cm
\caption{Log-log (base $10$) plots versus the wave number $k$
of: (a) the compensated kinetic energy spectra $k^{5/3}E_{u}(k)$ (lower two curves) 
and the compensated magnetic energy spectra $k^{11/3}E_{b}(k)$ (upper two curves) 
from $Run3$ (red curves) and $Run4$ (blue curves); solid black lines indicate
different power-law regions; (b) the compensated kinetic energy spectra 
$k^{5/3}E_{u}(k)$ (lower three curves) and the compensated magnetic energy spectra 
$k^{7/3}E_{b}(k)$ (upper three curves) from $Run5a$ (blue curves), $Run5b$ 
(red curves), and $Run5c$ (black curves); solid black lines indicate
different power-law regions; (c) the wave-number-dependent Alfv\'en  ratios 
$(E_{b}(k)/E_{u}(k))$ for $Run3$, $Run4$, $Run5a$, $Run5b$ and $Run5c$. 
We compute these spectra at the cascade-completion time $\tau_{c}$; 
and the wavenumber $k_{i}\propto 1/d_{i}$.}
\label{fig:2}
\end{figure*}
%

In Fig.~\ref{fig:2} we show the compensated spectra of the
kinetic energy  and the magnetic energy $E_{u}(k)$ (open circles) $E_{b}(k)$
(asterisks), respectively, which we have obtained from our DNSs
at the cascade-completion time $\tau_{c}$. Specifically, we present
log-log (base $10$) plots versus the wave number $k$ of the
following: (a) $k^{5/3}E_{u}(k)$ (lower two curves) 
and $k^{11/3}E_{b}(k)$ (upper two curves) 
from $Run3$ (red curves) and $Run4$ (blue curves), with solid black lines 
indicating different power-law regions (Fig.~\ref{fig:2}(a)); 
(b) $k^{5/3}E_{u}(k)$ (lower three curves) and the compensated magnetic energy spectra 
$k^{11/3}E_{b}(k)$ (upper three curves) from $Run5a$ (blue curves), $Run5b$ 
(red curves), and $Run5c$ (black curves); (c) the wave-number-dependent Alfv\'en  ratios 
$E_{b}(k)/E_{u}(k)$ for $Run3$, $Run4$, $Run5a$, $Run5b$ and $Run5c$ 
(Fig.~\ref{fig:2}(c)); 
the wavenumber $k_{i}\propto 1/d_{i}$. From these plots we conclude that the 
spectral exponents (Eqs.~\ref{eq:alpha} and \ref{eq:alpha1}) are consistent with
\begin{eqnarray}
\label{eq:alphavalues}
\alpha &\simeq& 5/3 , \rm{\hspace{0.5cm} the \hspace{0.1cm} K41 \hspace{0.1cm} value}; \nonumber \\
\alpha_1 &\simeq& 11/3 , \rm{\hspace{0.5cm} for \hspace{0.1cm} initial \hspace{0.1cm} condition \hspace{0.1cm} A}; \nonumber \\
\alpha_1 &\simeq& 7/3 , \rm{\hspace{0.5cm} for  \hspace{0.1cm} initial \hspace{0.1cm} condition \hspace{0.1cm} B}.
\end{eqnarray}
The values of $\alpha_1$ are clearly different for initial condition A
(Eq.~\ref{eq:initA} and $Run2$ $\&$ $ Run4$) and  initial condition B
(Eq.~\ref{eq:initB} and $Run5b - Run5c$). These differences stem, in part, from
the disparities in the wave-number-dependent Alfv\'en  ratios
$E_{b}(k)/E_{u}(k)$ (Fig.~\ref{fig:2}(c)) and, as we show in
Subsection~\ref{subsec:pdf}, in the alignment PDFs of various vector fields. In
all our runs, the ratio $E_{b}(k)/E_{u}(k)$ lies in the range $1-2$ at small
values of $k \lesssim k_i \propto d_{i}$; for $k_i < k < k_{max}$ this ratio is
different for different runs. For example, for $Run5c$, $E_{b}(k)/E_{u}(k)$
remains nearly constant up until $ k \sim 60$, i.e., for almost three decades
beyond $k_i$; by contrast, for $Run4$ this ratio rises rapidaly,
with increasing $k$, at large $k$.  For $Run5c$, this near constancy, with
$E_{b}(k)/E_{u}(k)$ in the range $1-2$, indicates approximate equipartition of
the energies in the velocity and magnetic fields.  It has been
recognised~\cite{galtier2007multiscale} that such equipartition suggests
$\alpha_1 = 7/3$. In contrast, $\alpha_1 = 11/3$ can be obtained by equating
$\tau_{nl}$ and $\tau_{h}$ and by using $E_{b}/E_{u}=d_{i}^{2}k^{2}$, where
$\tau_{nl}$ is the energy-transfer because of the nonlinear term in the
momentum equation and $\tau_{h}$ is the energy-transfer time because of the
Hall term in the induction equation~\cite{galtier2007multiscale}. 

\subsection{Probability Distribution Functions} 
\label{subsec:pdf}

\begin{figure*}[t]
\centering  
\begin{tabular}{c c c}
\textbf{(a)} & \textbf{(b)} & \textbf{(c)} \\
\includegraphics [scale=0.4]{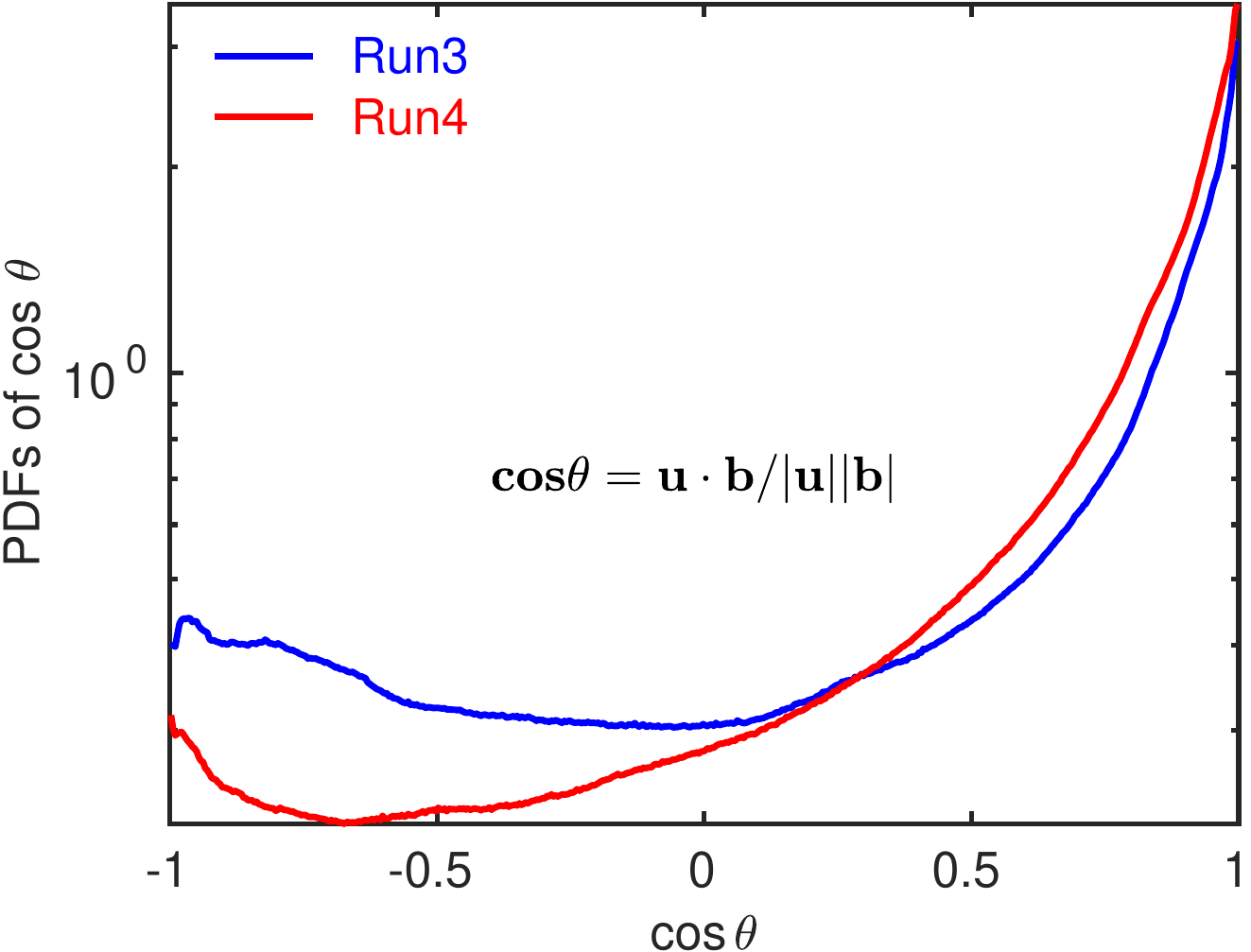} &
\includegraphics [scale=0.4]{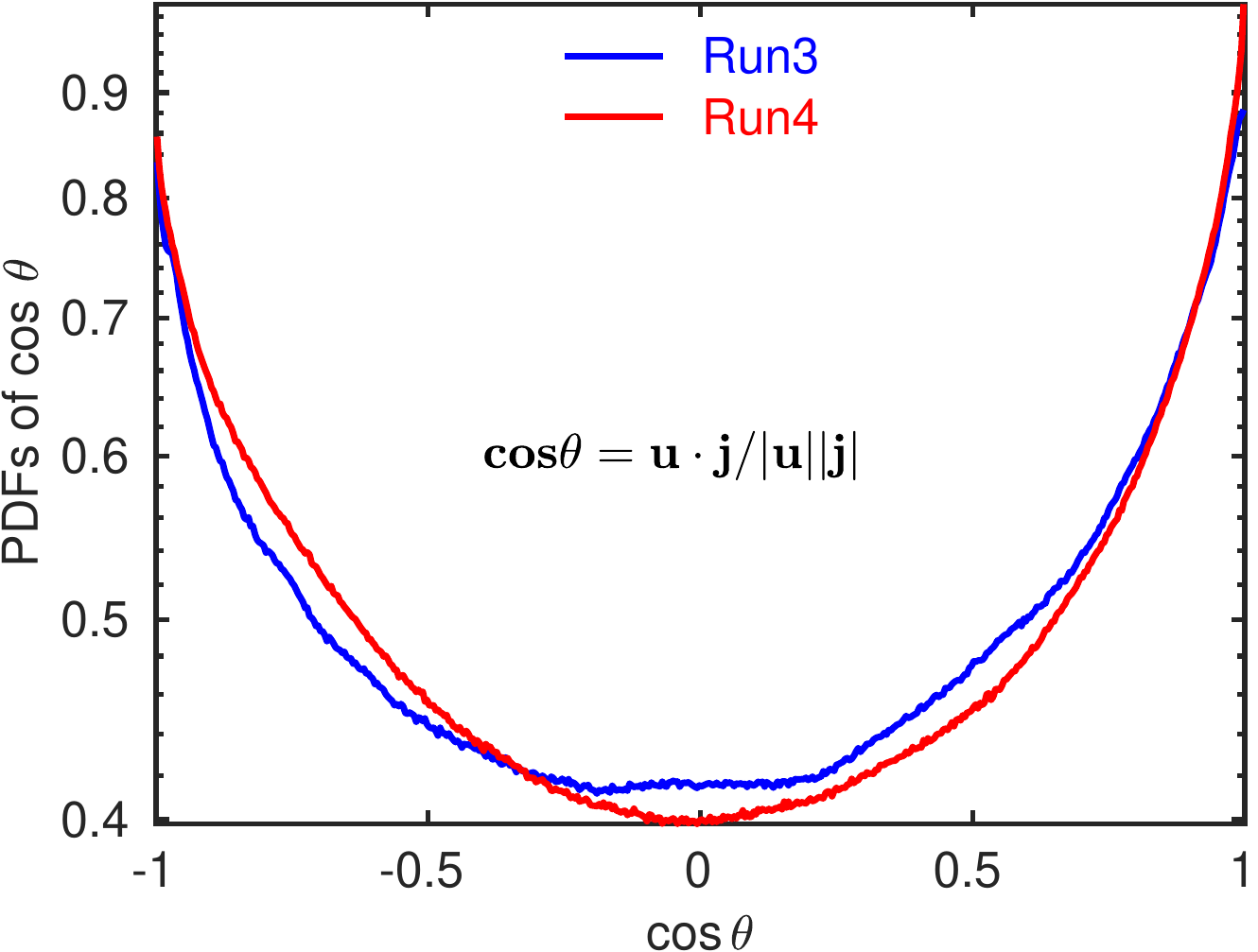} &
\includegraphics [scale=0.4]{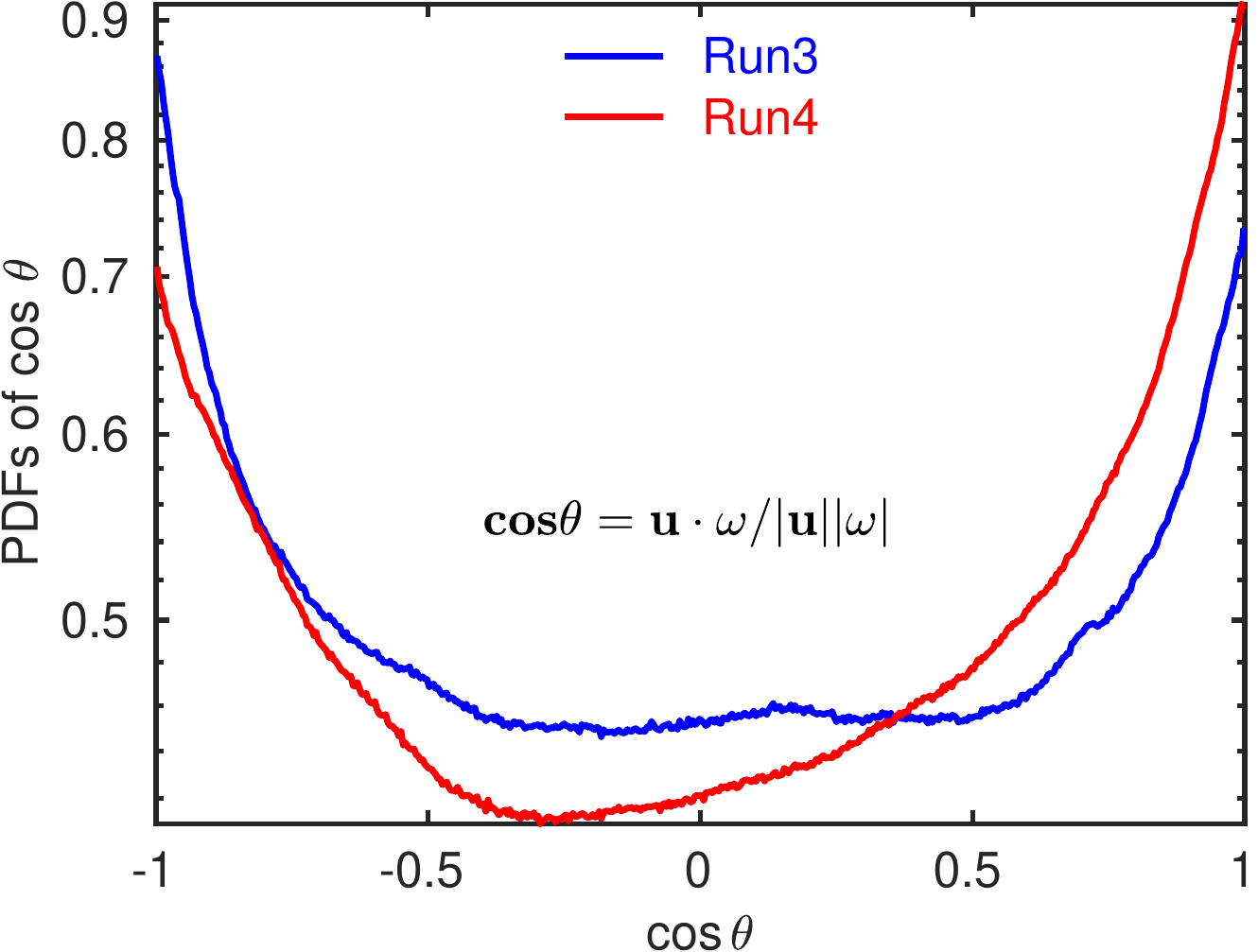}
\end{tabular}
\begin{tabular}{c c c}
\textbf{(d)} & \textbf{(e)} & \textbf{(f)} \\
\includegraphics [scale=0.4]{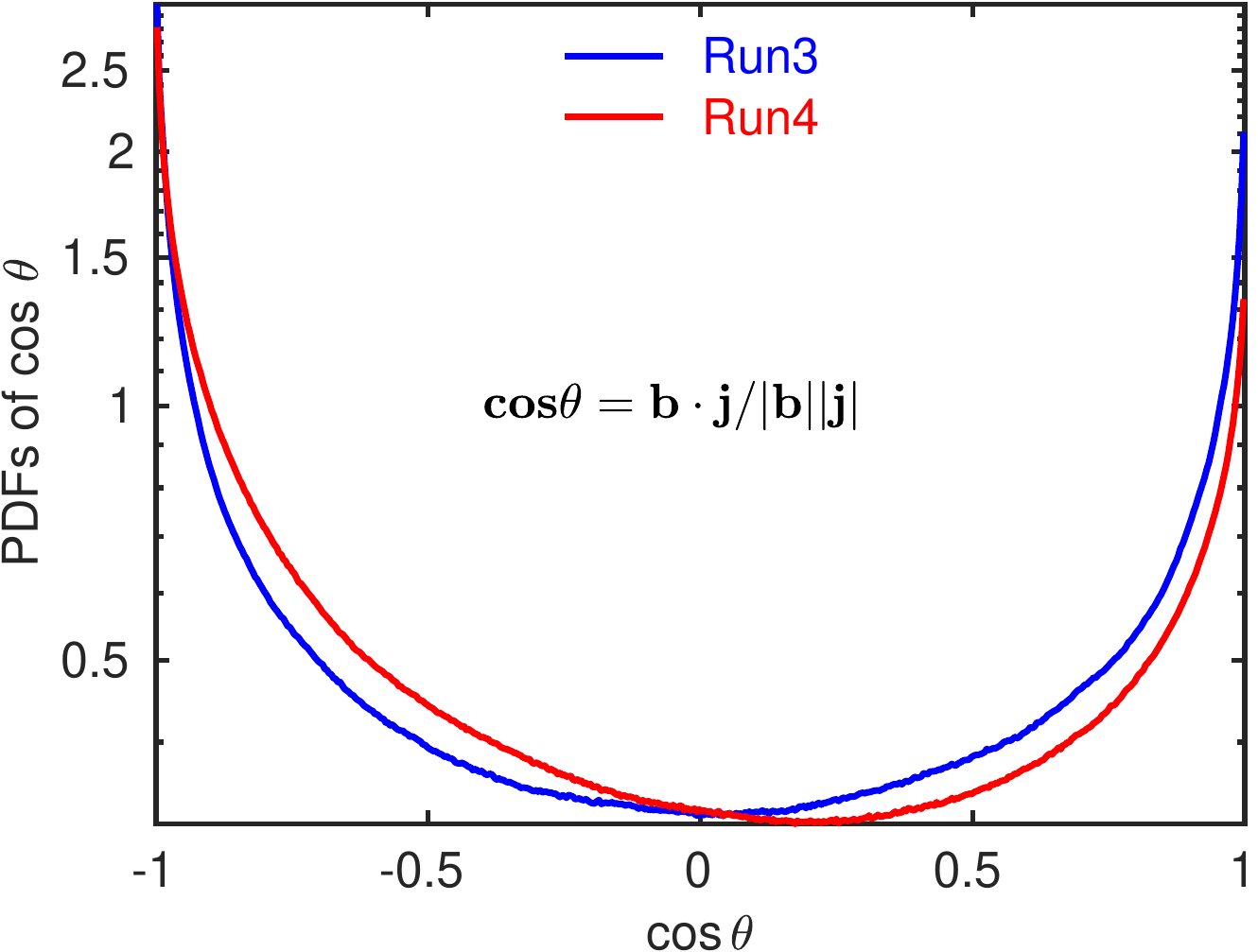} &
\includegraphics [scale=0.4]{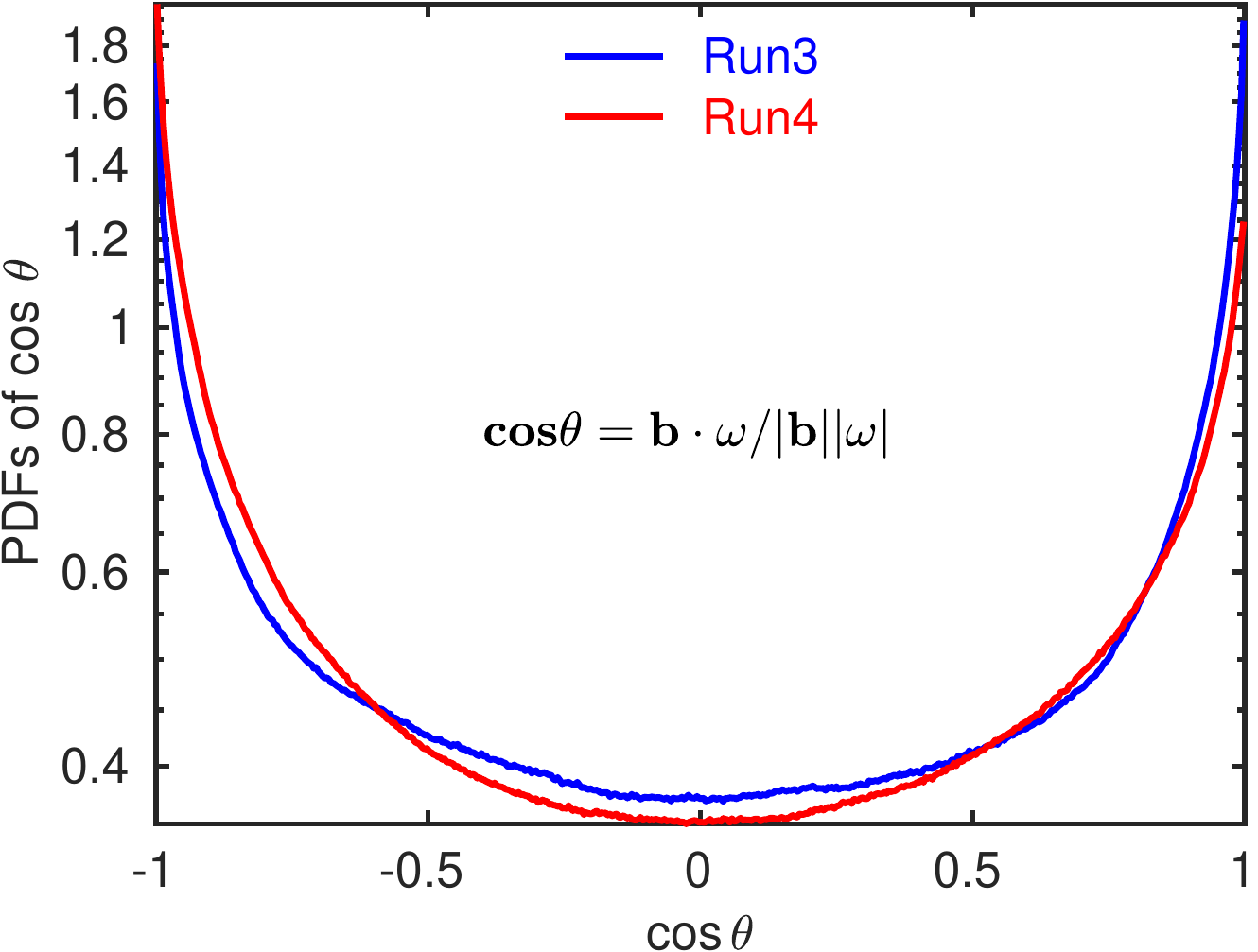} & 
\includegraphics [scale=0.4]{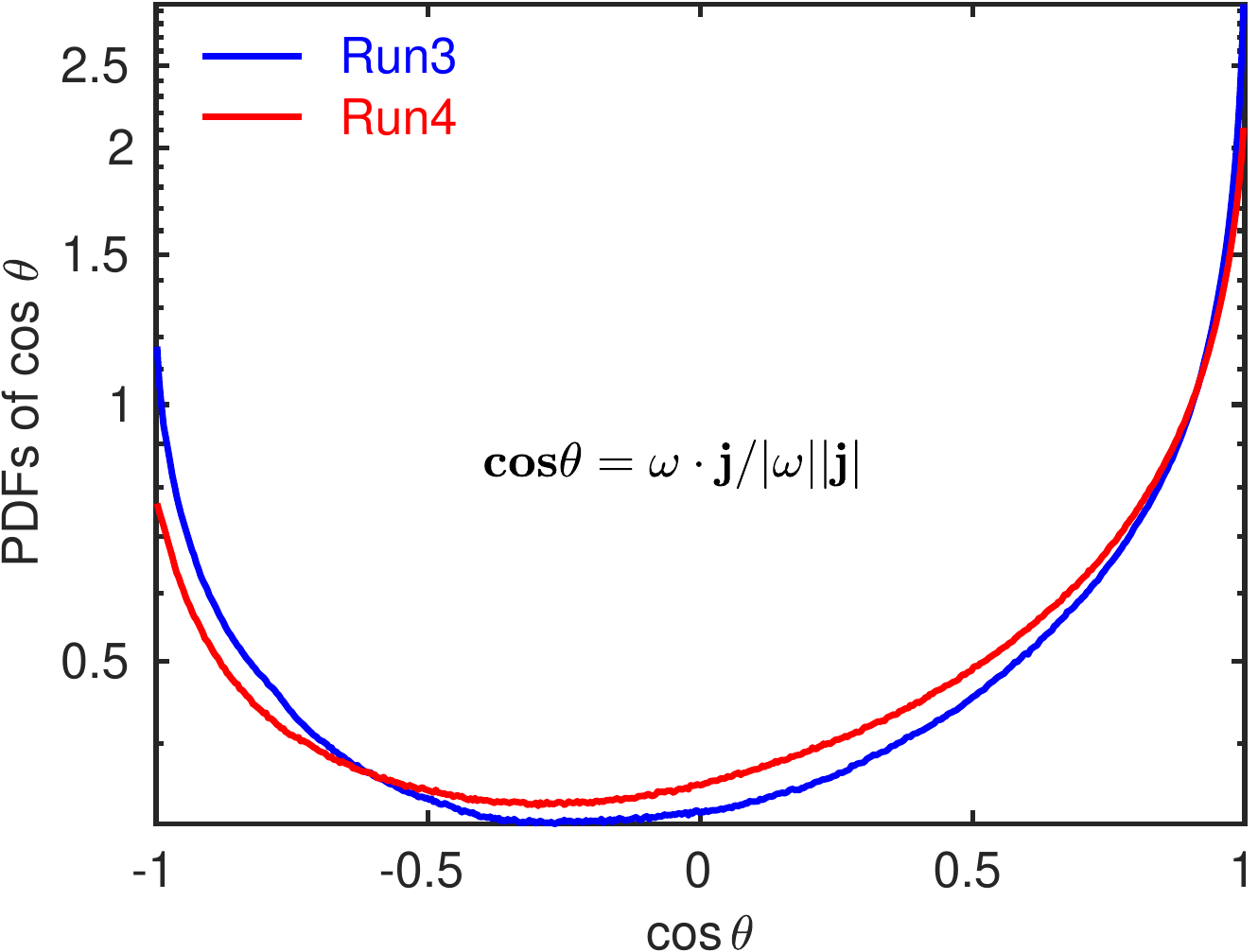}
\end{tabular}
\vskip -.1cm
\caption{Semilog (base 10) plots of PDFs of cosines of angles, denoted by $\theta$, between (a) $\bf u$ and $\bf b$, (b) $\bf u $ and $\bf j$, (c) $\bf u$ and $\bf \omega$, (d) $\bf b$ and $\bf j$,
(e) $\bf b$ and $\bf \omega$, and (f) $\bf \omega$ and $\bf j$ from $Run3$ (blue curves) and $Run4$ (red curves).}
\label{fig:4}
\end{figure*}

\begin{figure*}[t]
\centering 
\begin{tabular}{c c c}
\textbf{(a)} & \textbf{(b)} & \textbf{(c)} \\
\includegraphics [scale=0.4]{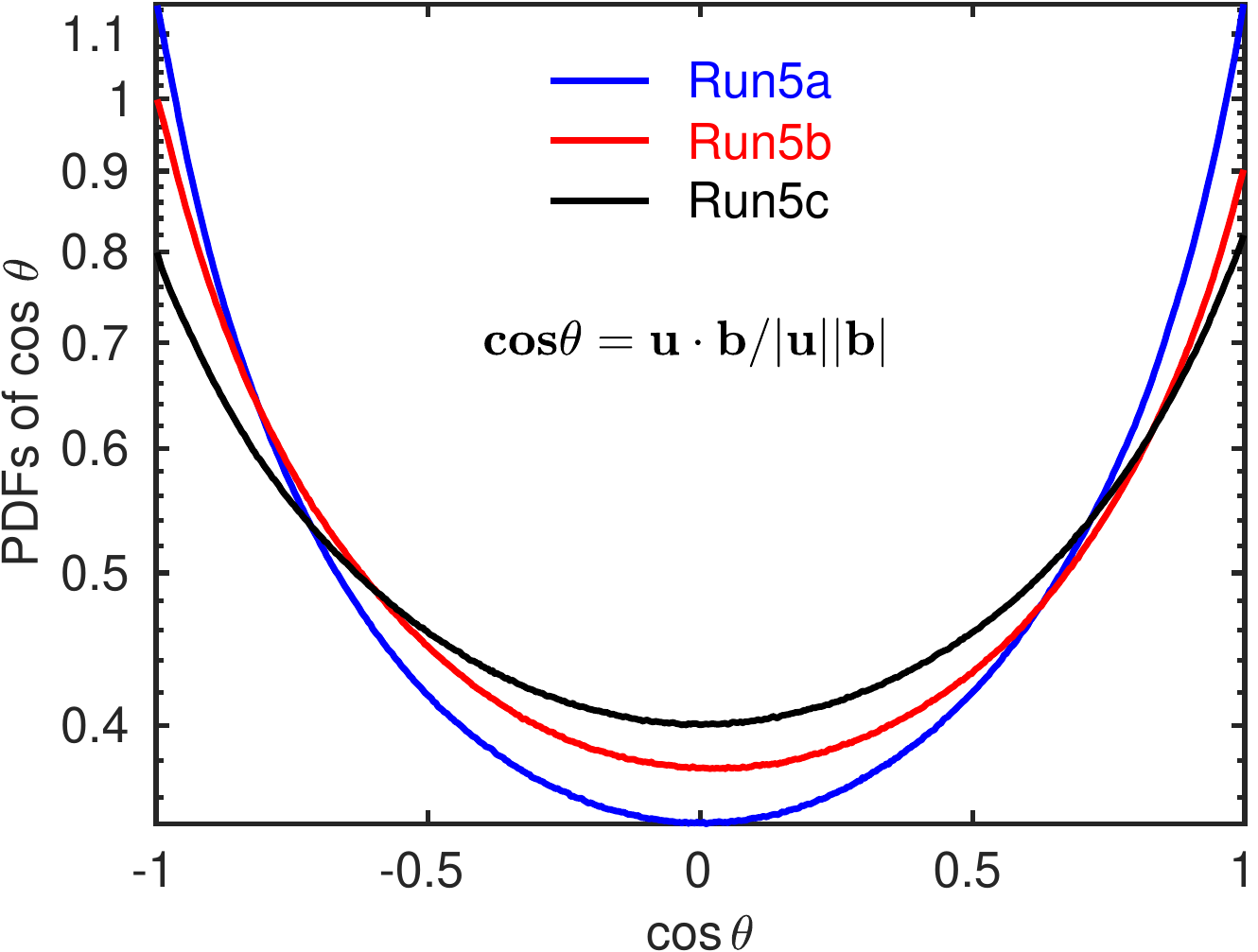} &
\includegraphics [scale=0.4]{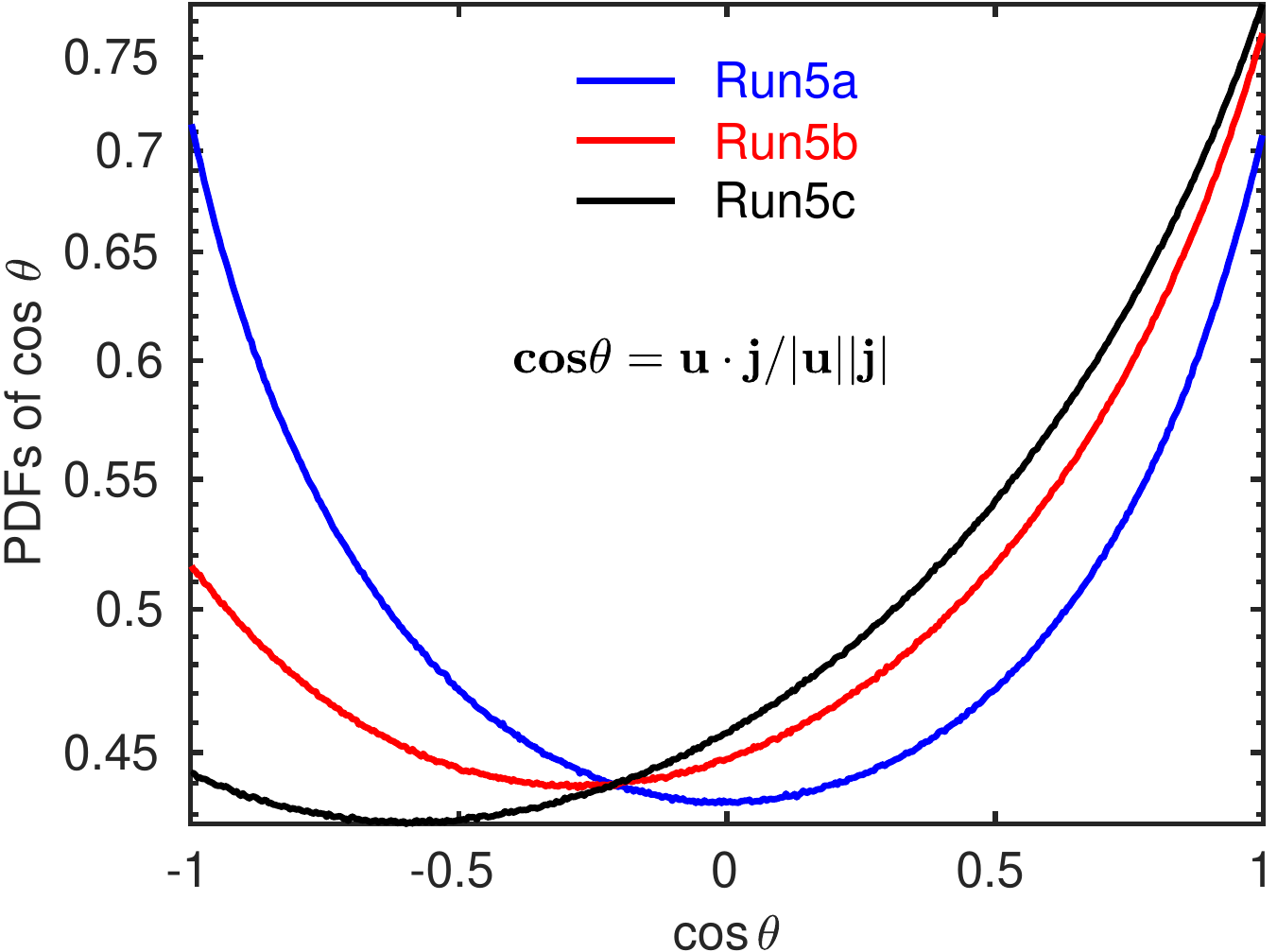} &
\includegraphics [scale=0.4]{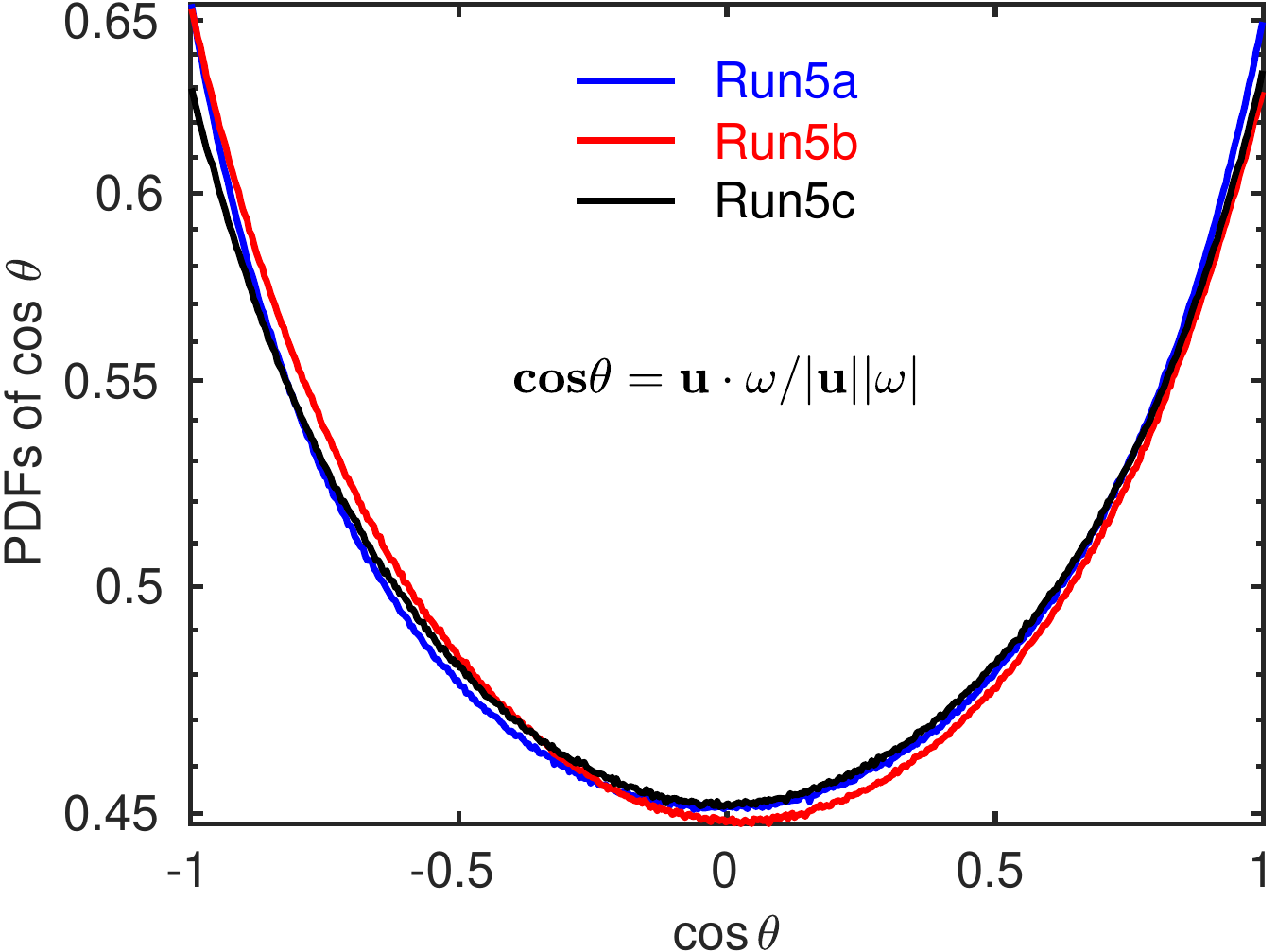}
\end{tabular}
\begin{tabular}{c c c}
\textbf{(d)} & \textbf{(e)} & \textbf{(f)} \\
\includegraphics [scale=0.4]{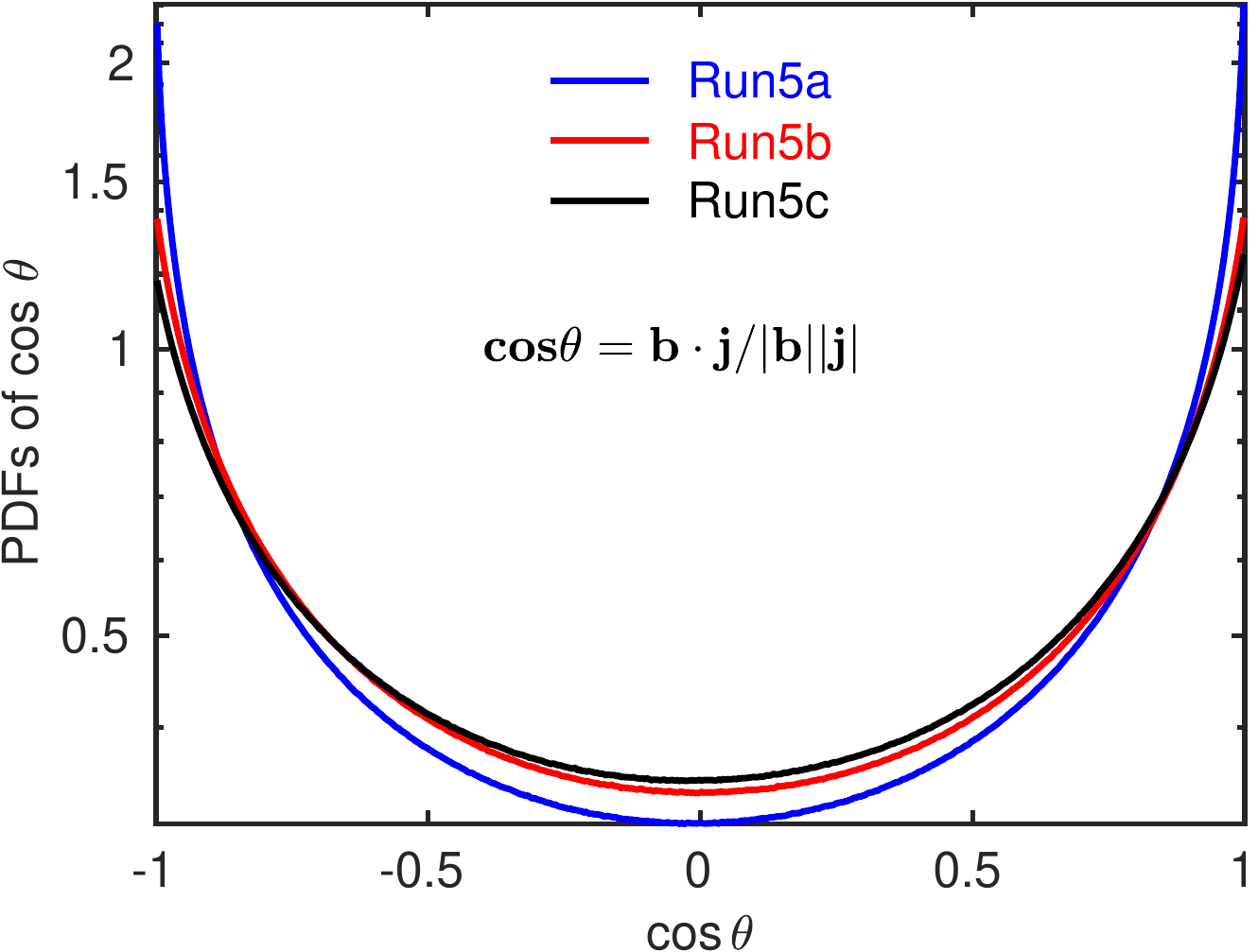} &
\includegraphics [scale=0.4]{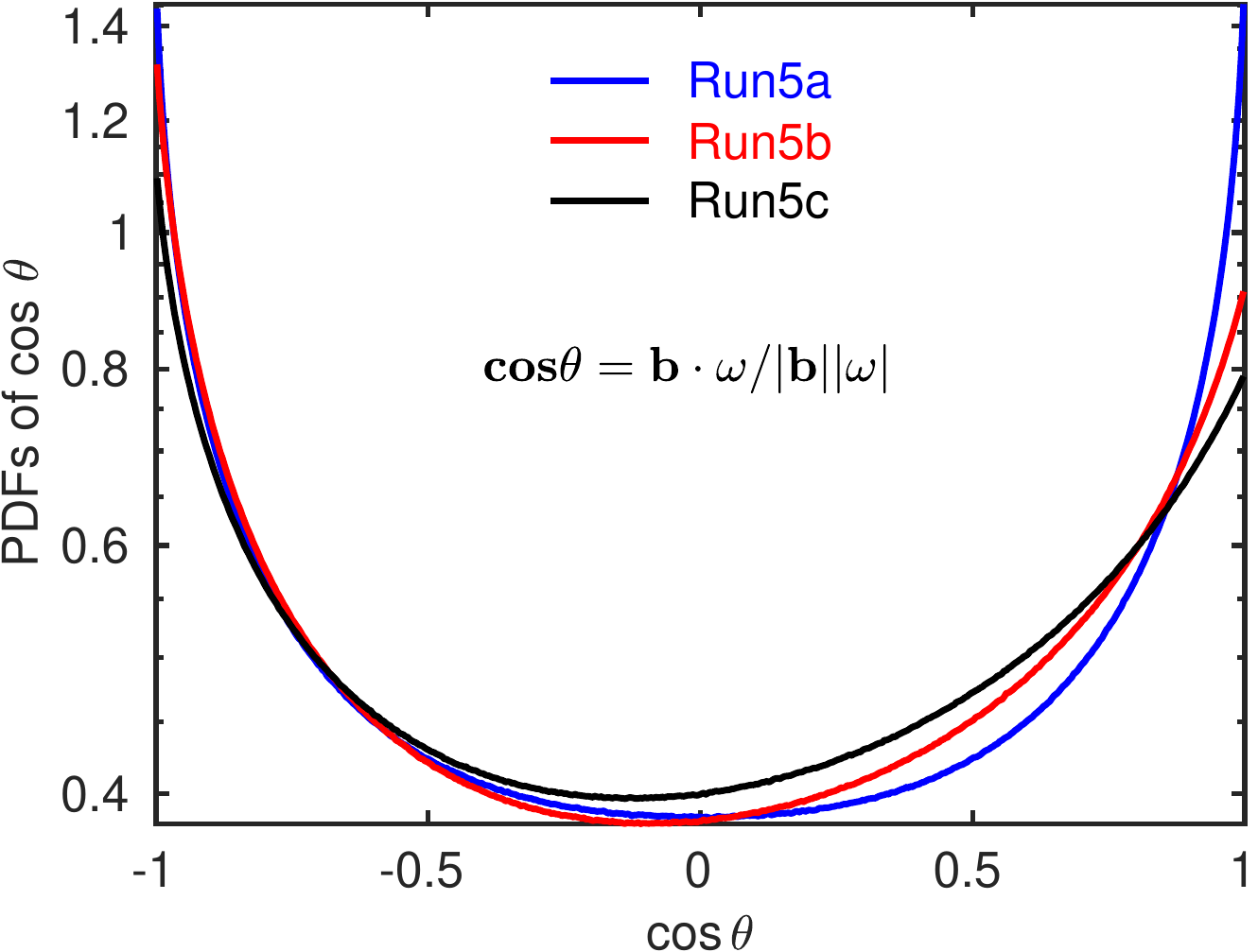} & 
\includegraphics [scale=0.4]{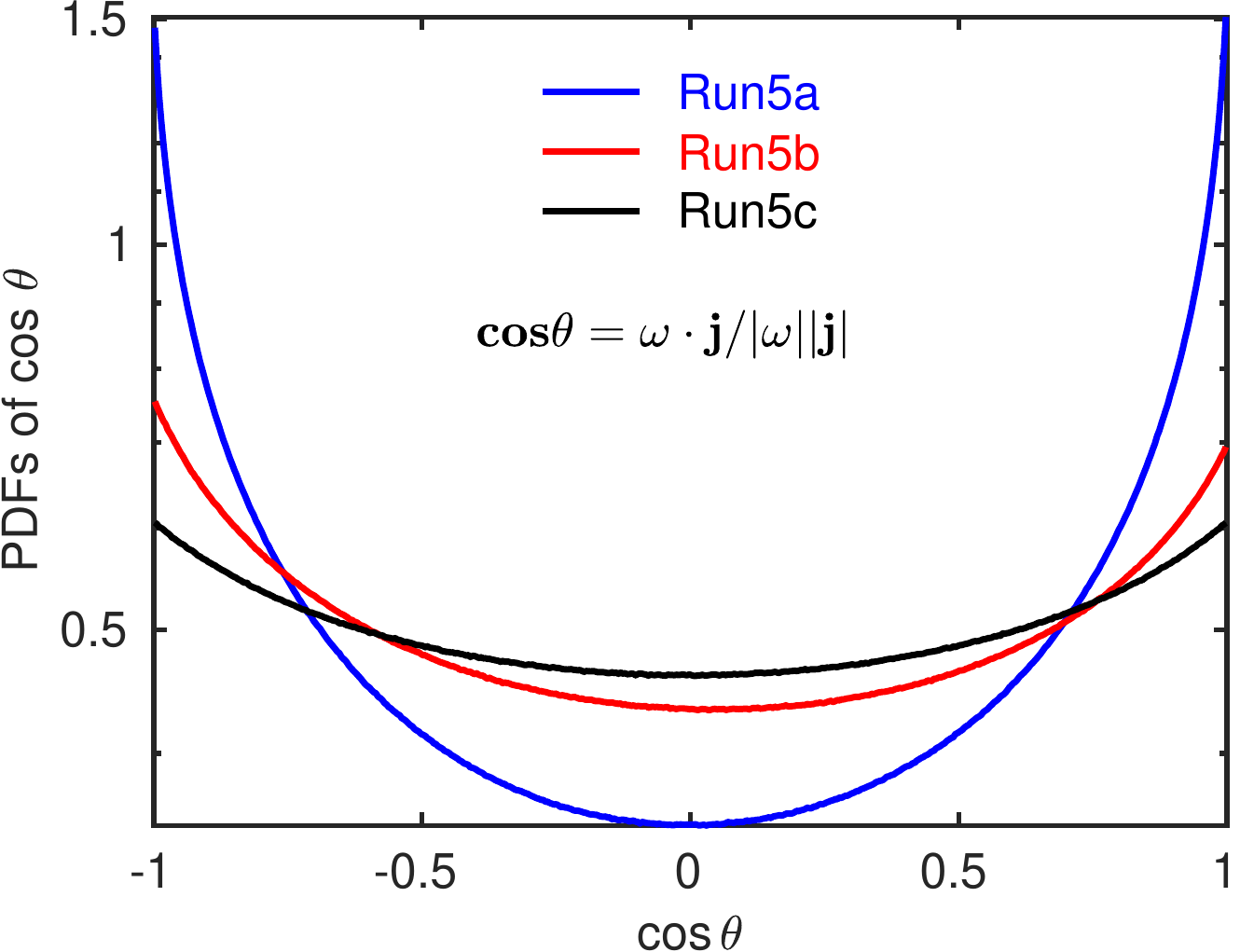}
\end{tabular}
\vskip -.1cm
\caption{Semilog (base 10) plots of PDFs of cosines of angles, denoted by $\theta$, between (a) $\bf u$ and $\bf b$, (b) $\bf u $ and $\bf j$, (c) $\bf u$ and $\bf \omega$, (d) $\bf b$ and $\bf j$,
(e) $\bf b$ and $\bf \omega$, and (f) $\bf \omega$ and $\bf j$ from $Run5a$ (blue curves), $Run5b$ (red curves) and $Run5c$ (black curves).}
\label{fig:5}
\end{figure*}

In Fig.~\ref{fig:4} we display, at the cascade-completion time $\tau_{c}$, the 
PDFs of the cosine of the angles between the  following
pairs of vectors: \{$\bf{u},\bf{b}$\}, \{$\bf{u},\bf{j}$\},
\{$\bf{u},\bf{\omega}$\}, \{$\bf{b},\bf{j}$\}, \{$\bf{b},\bf{\omega}$\} and
\{$\bf{\omega},\bf{j}$\},
for $Run3$ (blue curve) and $Run4$ (red curve). We present their 
counterparts for $Run5a$ (blue curve), $Run5b$ (red curve) and $Run5c$ (black curve)
in Fig.~ \ref{fig:5}. In Figs.~\ref{fig:4} and ~\ref{fig:5} the blue curves are from
the MHD runs $Run3$ and $Run5a$ (see Table~\ref{table:1}).

In all these PDFs, there are, roughly speaking, two peaks at $\cos\theta \simeq
\pm 1$, which quantify the degree of antialignment ($\theta = 180^{o}$)
and of alignment ($\theta = 0^{o}$) between the two vectors. The amplitudes of these peaks
depend upon the parameters in our DNSs. One qualitative trend shows up clearly:
the alignment and antialignment and peaks (Figs.~\ref{fig:4}(a) and ~\ref{fig:5}(a)
for the pair \{$\bf{u},\bf{b}$\}) are more prononunced in the 3D MHD runs 
($Run3$ and $Run5a$) than in their 3D HMHD counterparts; and this peak-suppression
trend appears in most of the PDFs we show in Figs.~\ref{fig:4} and ~\ref{fig:5}.
In 3D MHD, the alignment or antialignment of \{$\bf{u},\bf{b}$\} is associated
with a \textit{depletion of nonlinearity}. This can be seen most simply by writing the 
3D MHD equations in terms of Els\"asser variables~\cite{basu2014structure}. To the extent
that the  alignment or antialignment of peaks in the PDF of the cosine of
the angle between \{$\bf{u},\bf{b}$\} are suppressed in 3D HMHD, relative to 3D MHD,
we conclude that this depletion of nonlinearity is also suppressed.

We remark that the PDFs of the cosines of the angles mentioned above are related to
PDFs of various helicities, which we list below:
\begin{eqnarray}
H_{c} &=& \langle \bf{u} \cdot \bf{b} \rangle ; \nonumber \\
H_{m} &=& \langle \bf{a} \cdot \bf{b} \rangle ; \nonumber \\
H_{u} &=& \langle \bf{u} \cdot \bf{\omega} \rangle ; \nonumber \\
H_{g} &=& H_{m}+2d_{i}H_{c}+d_{i}^{2}H_{u} .
\end{eqnarray}
Here, $H_{c}$ is the cross helicity; $H_{m}$ is the magnetic helicity, with
$\bf {a}$ the vector potential that follows from $\bf{b}=\nabla \times \bf{a}$;
$H_{u}$ is the kinetic helicity; and $H_{g}$ is a generalised helicity that is useful
in 3D HMHD. $H_{u}$ is conserved for an ideal, unforced fluid; in the 
absence of forcing, $H_{m}$ is conserved in both ideal 3D MHD and ideal 3D HMHD;
$H_{c}$ is conserved in ideal, unforced 3D MHD, but not in its 3D HMHD counterpart;
in ideal, unforced 3D HMHD the generalized cross helicity $H_{g}$ is conserved \cite{turner1986conserved}.
It is useful to define the electron velocity ${\bf v_{e}} \equiv {\bf u}-d_{i}{\bf
\nabla\times b}$ \cite{mininni2007energy}. In Figs.~\ref{fig:6}$\left(a\right)$ and
\ref{fig:6}$\left(b\right)$ we present, respectively, the PDFs of the cosines of the 
angles between (a) $\bf v_{e}$ and $\bf j$ and (b) $\bf v_{e}$ and $\bf b$ from runs 
$Run5a$ (blue curve), $Run5b$ (red curve), and $Run5c$ (black curve). 
In Figs.~\ref{fig:6}$\left(c\right)$ and~\ref{fig:6}$\left(d\right)$, we present 
similar plots for $Run3$ (blue curve) and $Run4$ (red curve).  

In the Appendix we present some joint PDFs.

%
%
\begin{figure*}[t]
\centering 
\begin{tabular}{c c}
\textbf{(a)} & \textbf{(b)} \\
\includegraphics [scale=0.5]{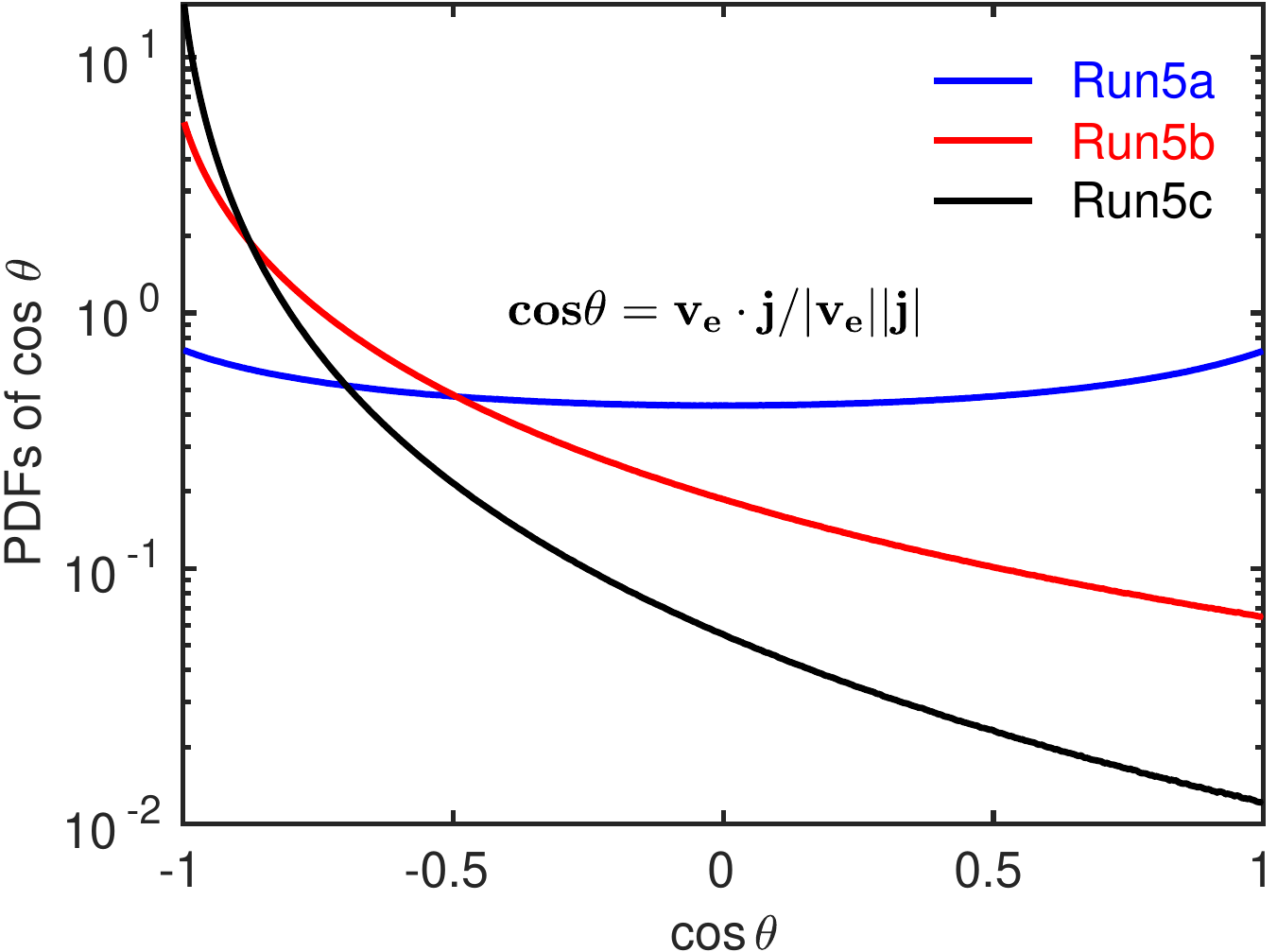} &
\includegraphics [scale=0.5]{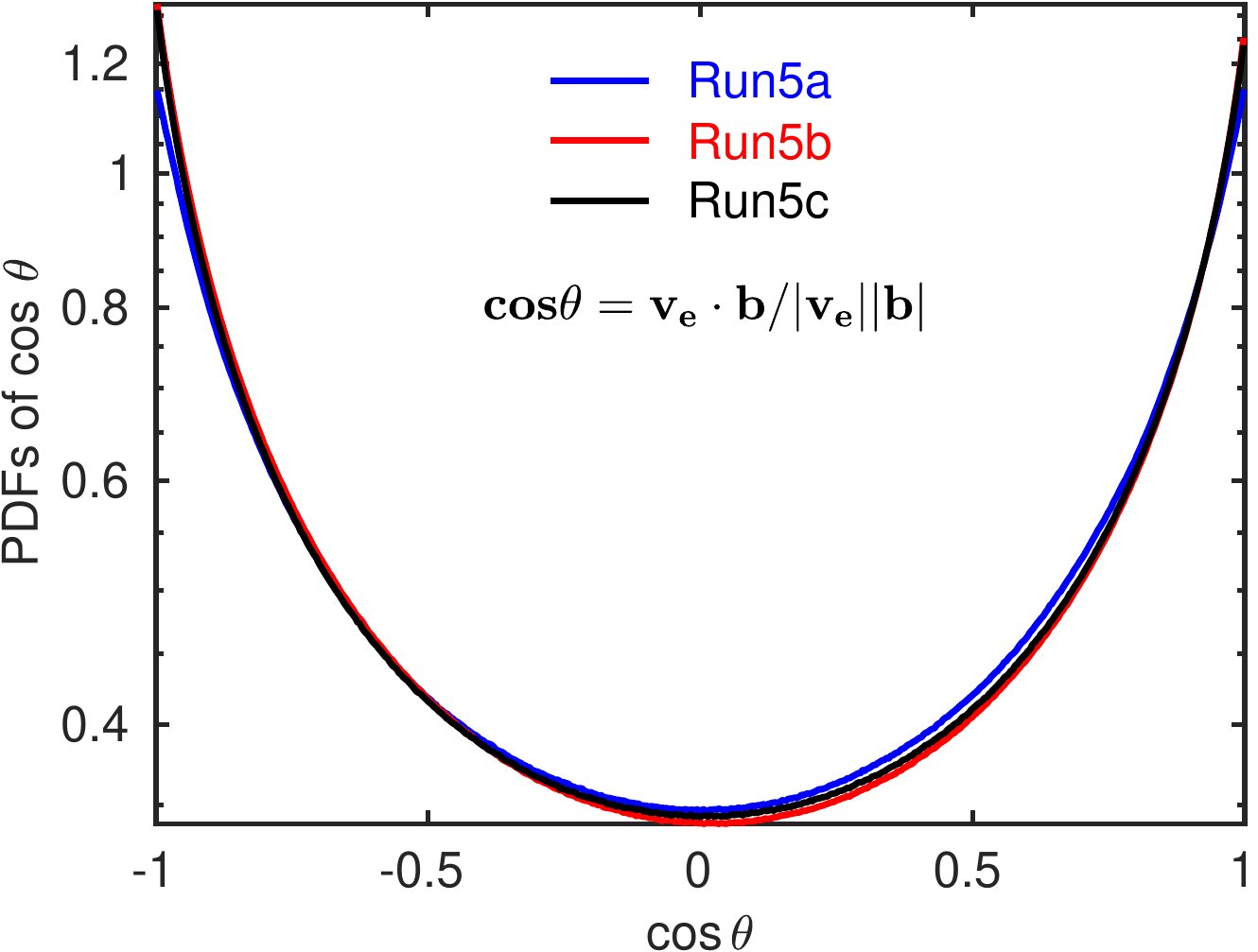}
\end{tabular}
\begin{tabular}{c c}
\textbf{(c)} & \textbf{(d)} \\
\includegraphics [scale=0.5]{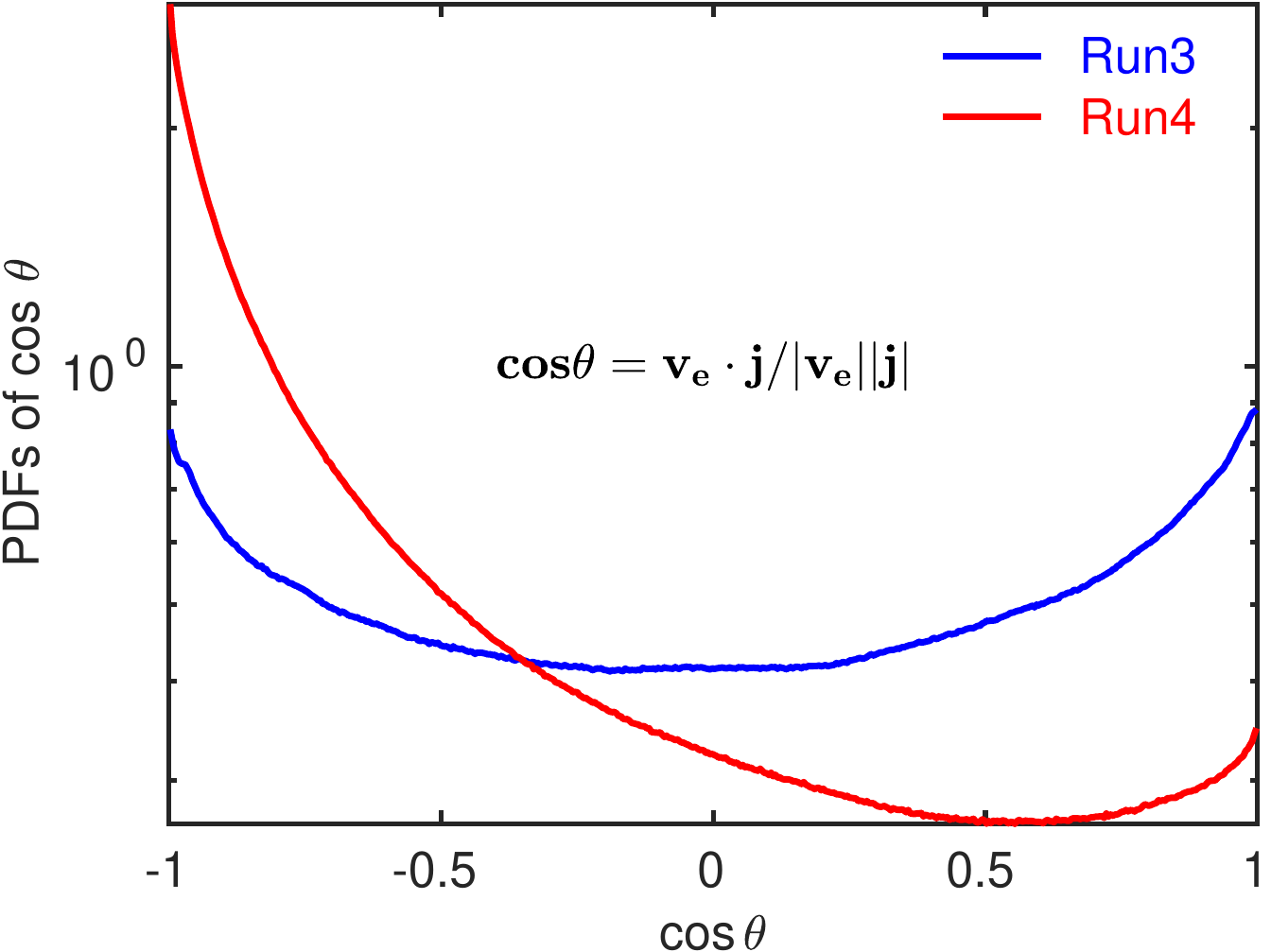} &
\includegraphics [scale=0.5]{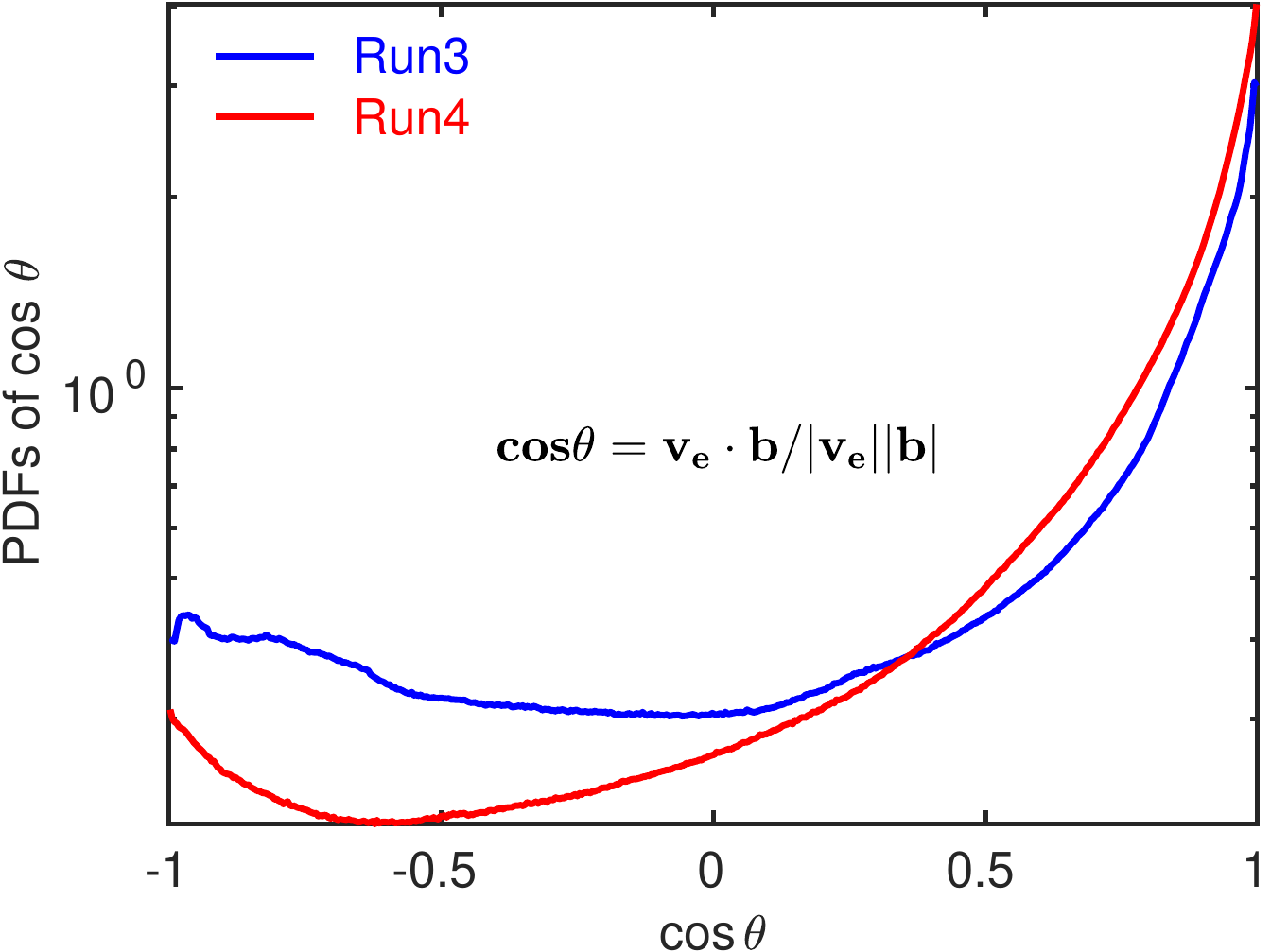} 
\end{tabular}
\vskip -.1cm
\caption{Semilog (base 10) plots of PDFs of cosines of angles, denoted by $\theta$, between (a) $\bf v_{e}$ and $\bf j$, and (b) $\bf v_{e} $ and $\bf b$ from $Run5a$ (blue curves), $Run5b$ (red curves) and $Run5c$ (black curves); $\left(c\right)$ and  $\left(d\right)$ are the similar plots from $Run3$ (blue curves) and $Run4$ (red curves).}
\label{fig:6}
\end{figure*}
%
\subsection{Intermittency}
\label{subsec:intermitt}

At the cascade completion time $\tau_{c}$, we investigate intermittency in 3D
HMHD turbulence, and in its 3D MHD counterpart, by calculating the
length-scale-$l$ dependence of (a) PDFs of the velocity and magnetic-field
increments (Eq.~\ref{eq:strucfun}) and (b) of the velocity and magnetic-field
structure functions (Eq.~\ref{eq:strucfun}) and, therefrom, their order-$p$
multiscaling exponents, Eqs.~\ref{eq:zetap} and \ref{eq:zetap12}, in the
inertial and the intermediate-dissipation ranges. We also examine the
dependence on $l$ of the hyperflatness Eq.~\ref{eq:f6}.

\begin{figure*}[t]
\centering 
\vskip -.1cm 
\begin{tabular}{c c}
\textbf{(a)} & \textbf{(a')} \\
\includegraphics [scale=0.6]{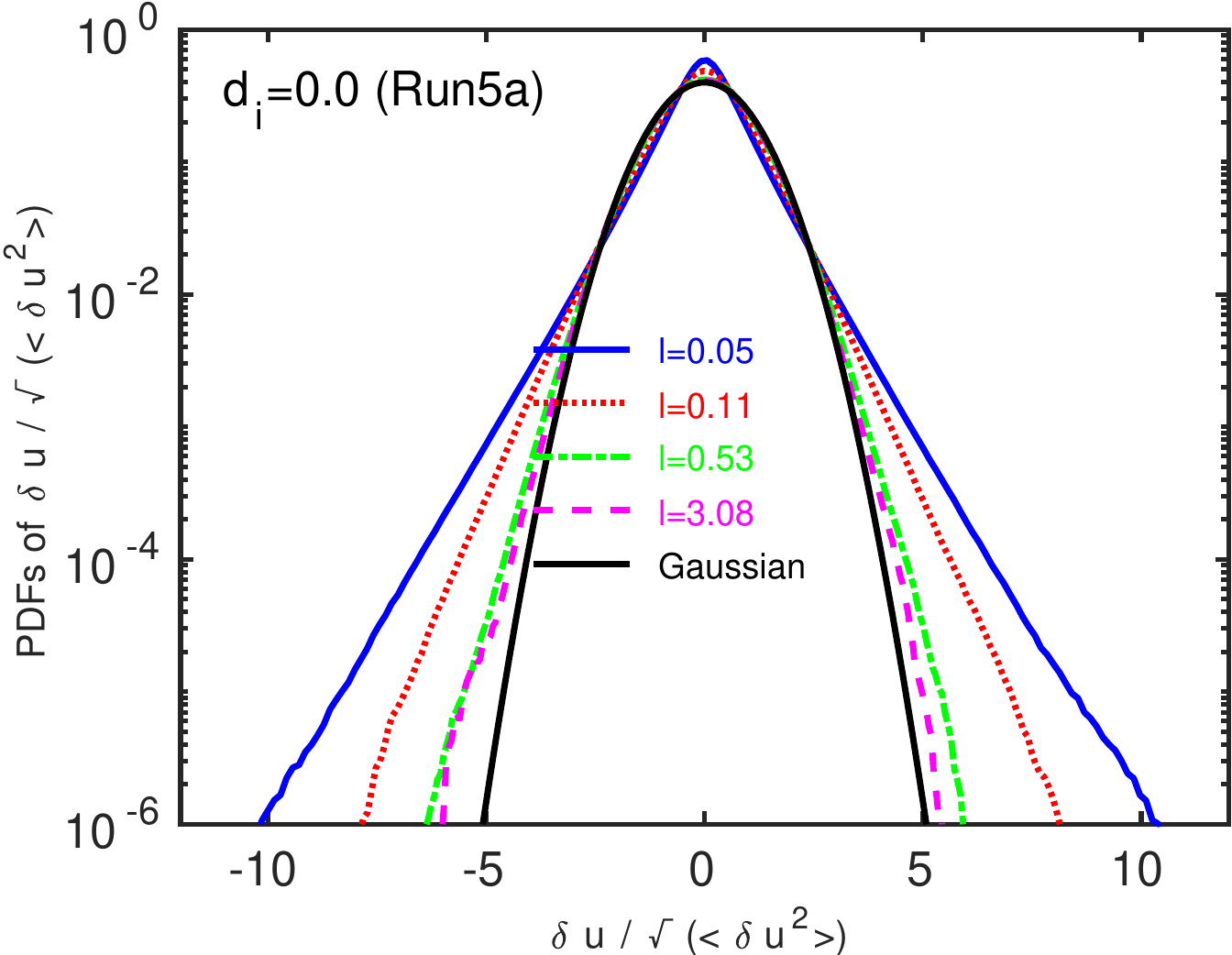} &
\includegraphics [scale=0.6]{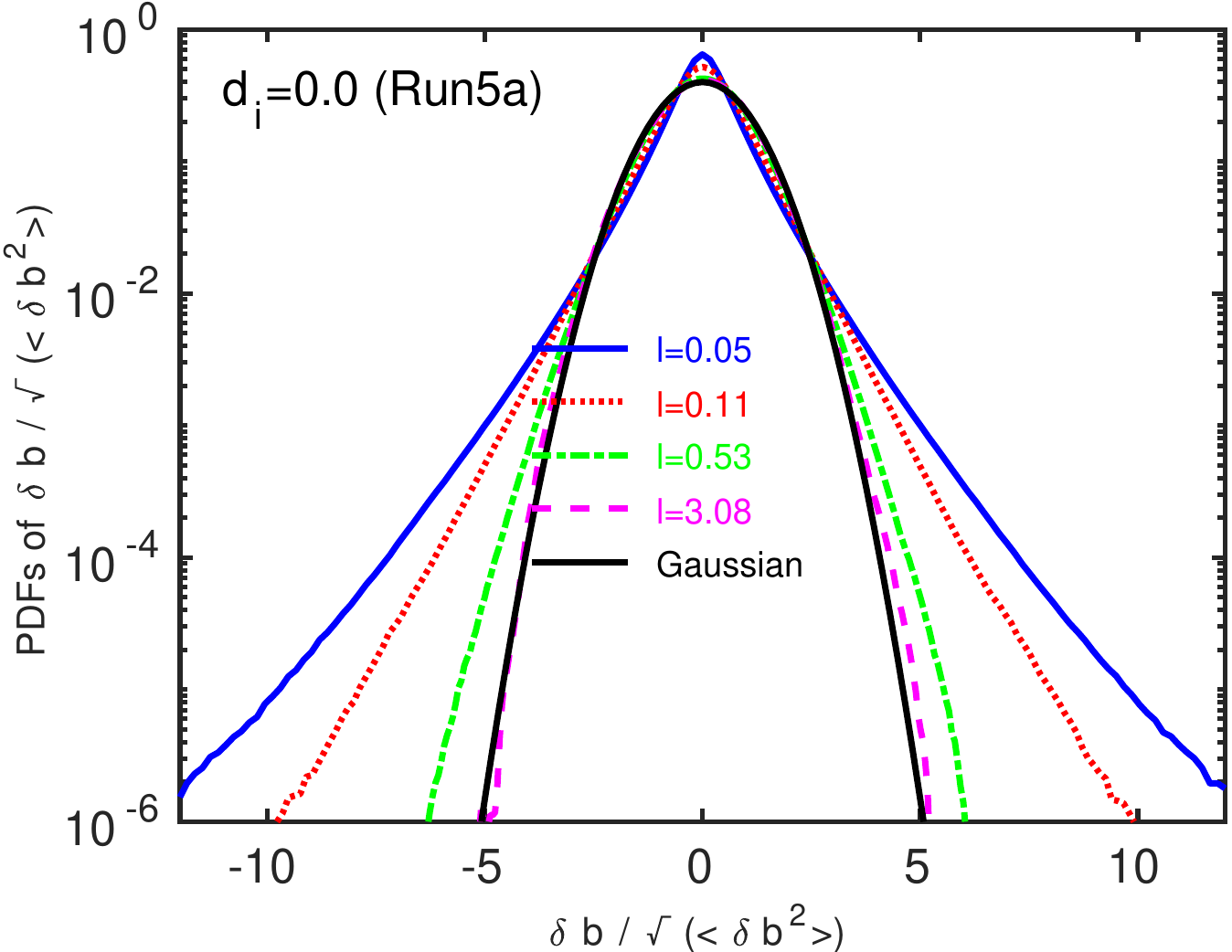}
\end{tabular}
\begin{tabular}{c c}
\textbf{(b)} & \textbf{(b')} \\
\includegraphics [scale=0.6]{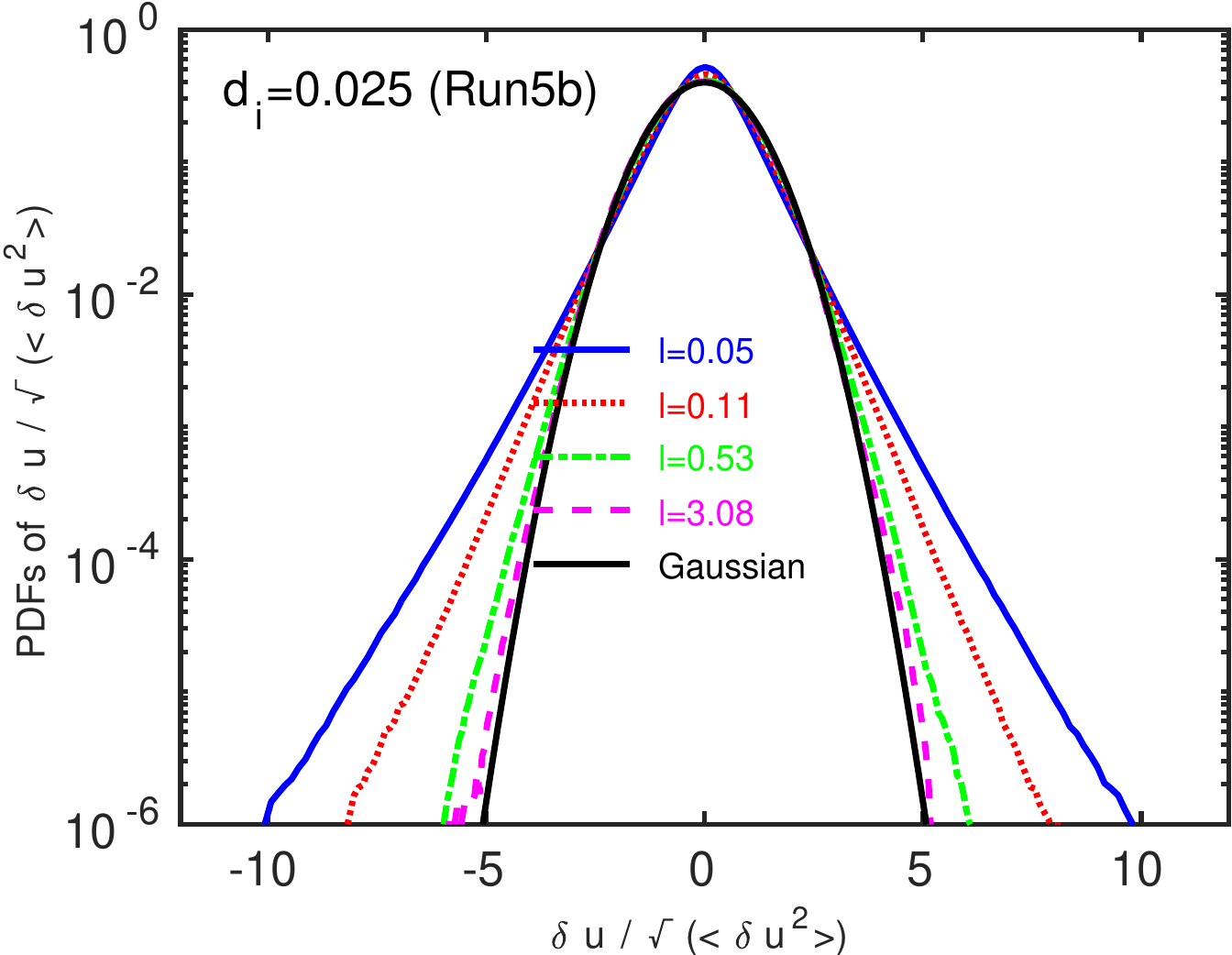} &
\includegraphics [scale=0.6]{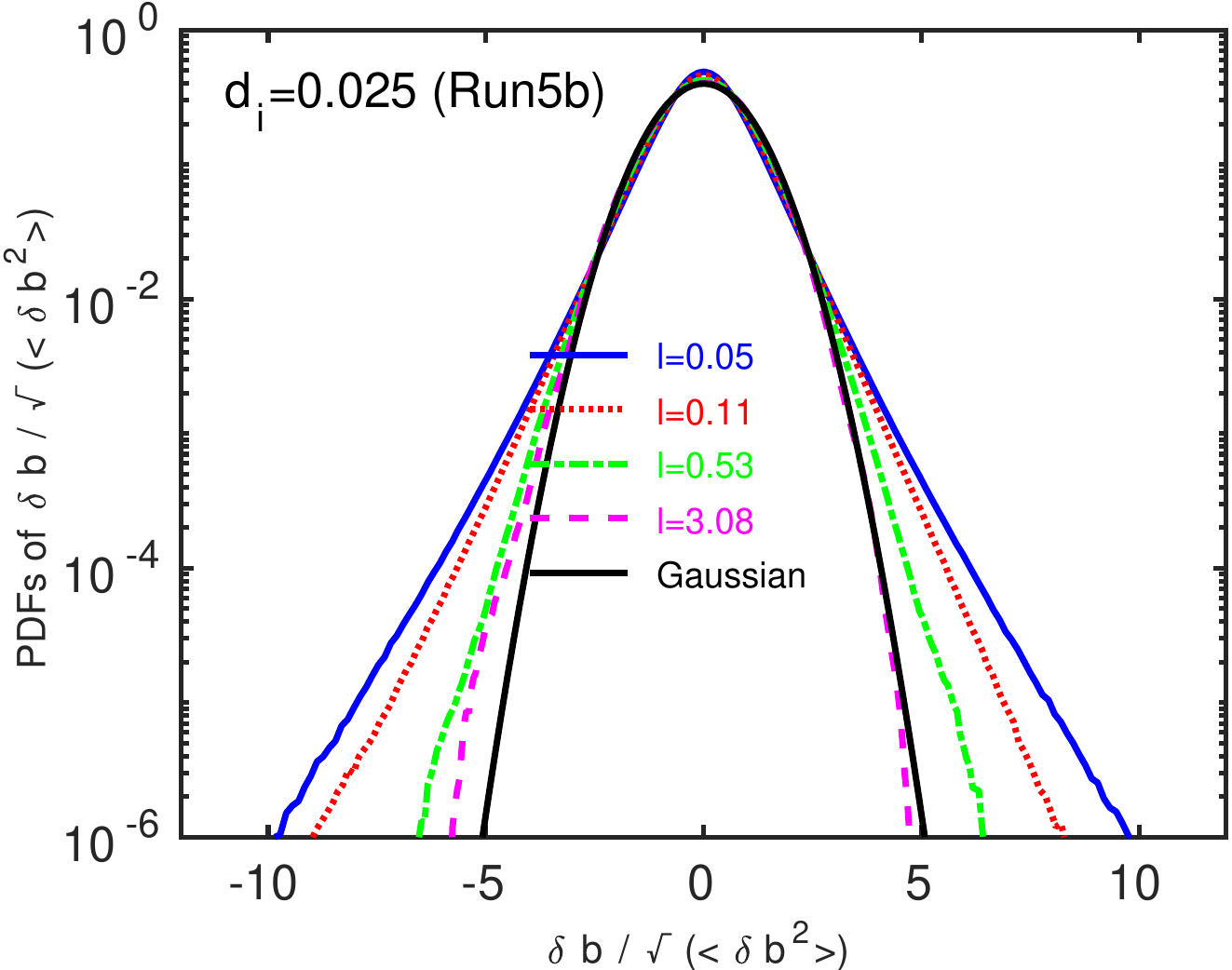}
\end{tabular}
\begin{tabular}{c c}
\textbf{(c)} & \textbf{(c')} \\
\includegraphics [scale=0.6]{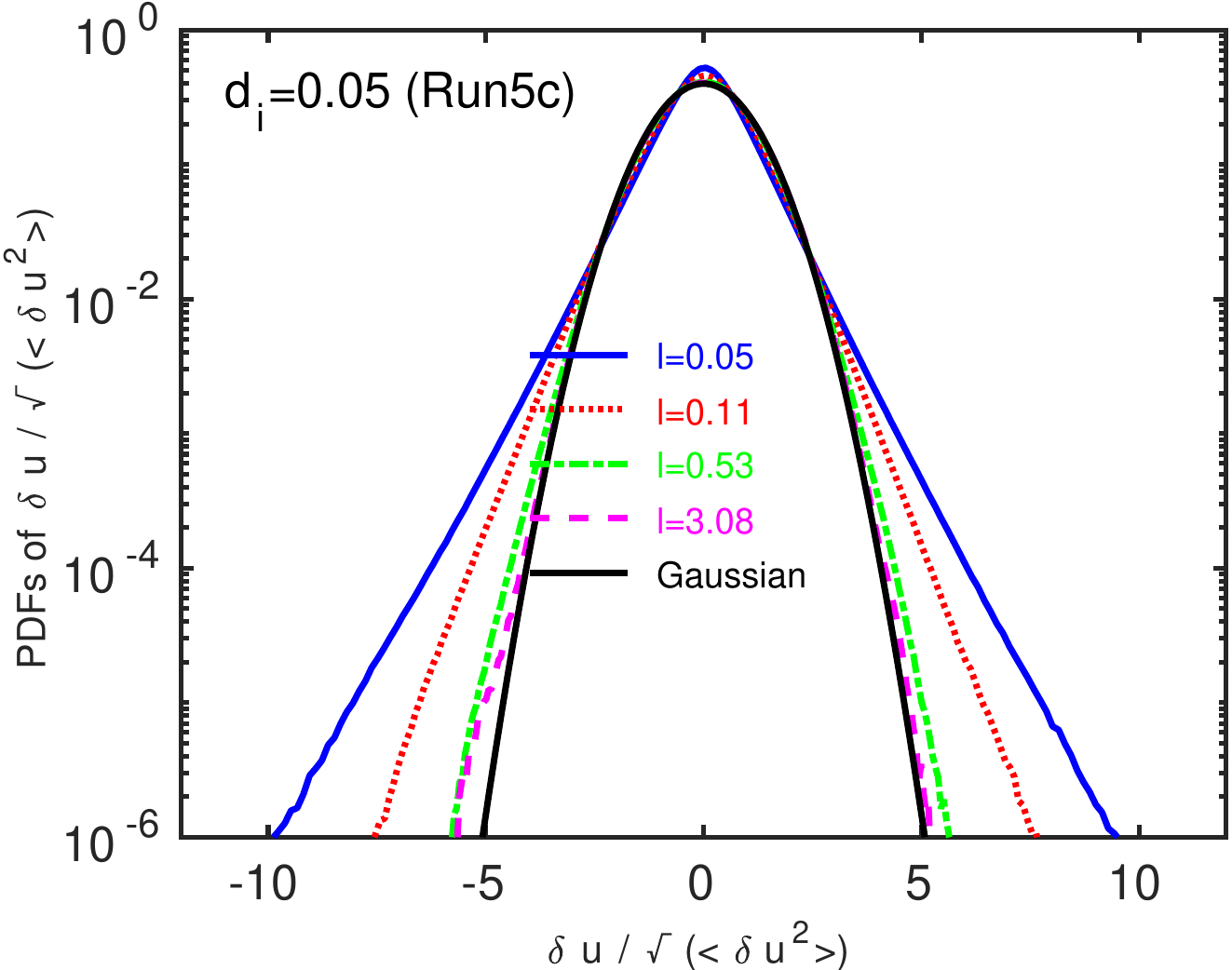} &
\includegraphics [scale=0.6]{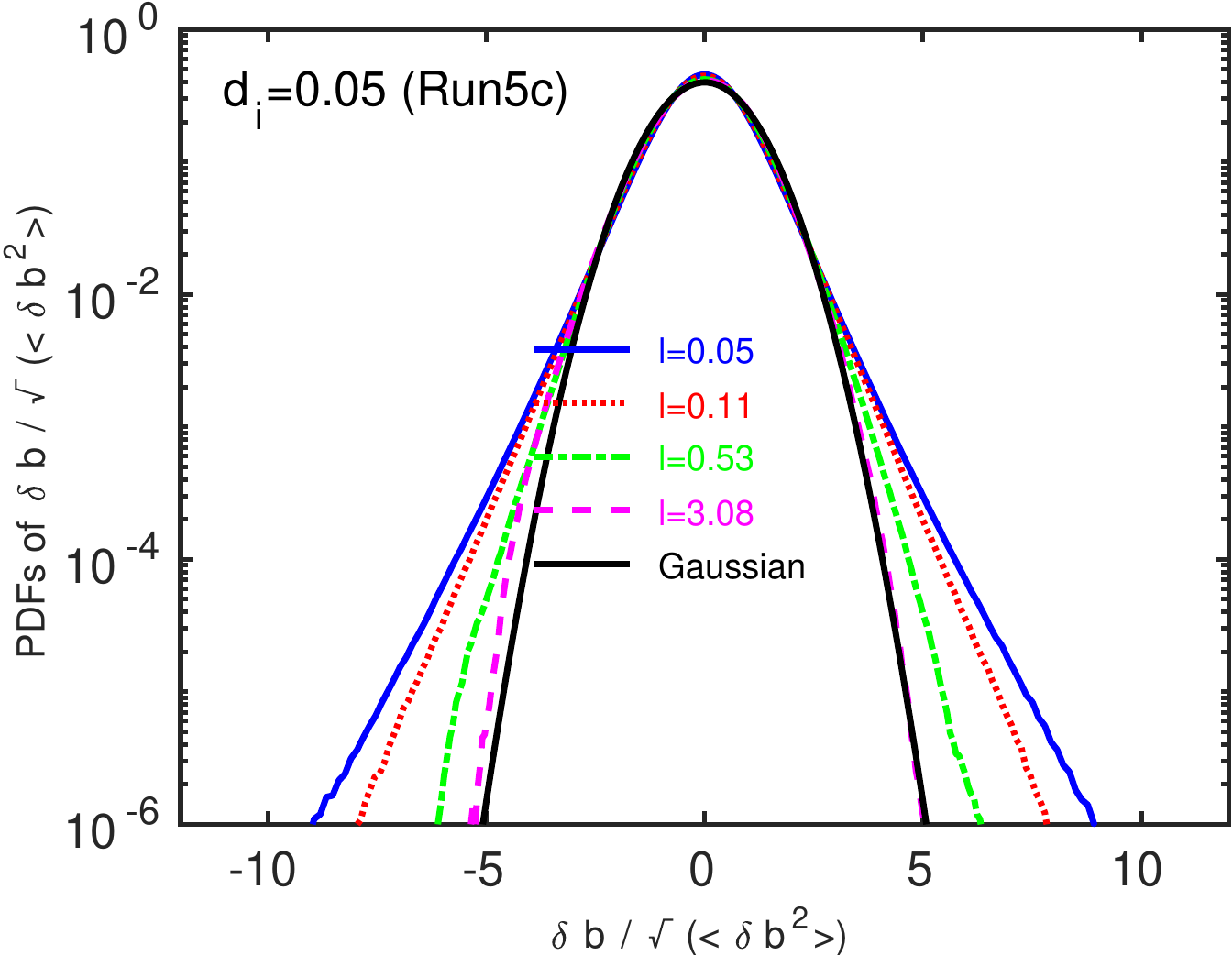}
\end{tabular}
\vskip -.1cm
\caption{Semilog (base 10) plots of PDFs of the velocity- (first column) and 
magnetic-field (second column) increments from $Run5a$ (first row), $Run5b$ (second row),
and $Run5c$ (third row) for $l=0.05$ (blue solid lines), $0.11$ (red dotted lines), 
$0.53$ (green dashed-dot lines) and $3.08$ (magneta dashed lines); for reference,
we also show zero-mean and unit-variance Gaussian PDFs (black lines).}
\label{fig:7}
\end{figure*}

In Figs.~\ref{fig:6} and ~\ref{fig:7} we display the $l$ dependence of the PDFs
of, respectively, the velocity- and magnetic-field increments, from $Run3$,
$Run4$, $Run5a$, $Run5b$, and $Run5c$.  In these PDFs, $l$ goes from the  the
intermediate-dissipation range to the inertial-range: $ l = 0.05$ (blue-solid
curves), $l=0.11$ (red-dotted curves), $l=0.53$ (green dashed-dot curves) and
$l=3.08$ (magenta dashed curves) from $Run3$ and $Run4$; we give similar plots
for  $Run5a$, $Run5b$, and $Run5c$ in Fig.~\ref{fig:7}; we include, for
reference, Gaussian PDFs (black lines) with zero mean and unit variance.  We
note that the field-increment PDFs show tails that deviate significantly from
those of Gaussian PDFs: the smaller the value of $l$, the greater this
deviation, a clear signature of small-scale intermittency (see, e.g.,
Ref.~\cite{pandit2009statistical}).

%
\begin{figure*}[t]
\centering 
\vskip -.1cm 
\begin{tabular}{c c}
\textbf{(a)} & \textbf{(a')}\\
\includegraphics [scale=0.6]{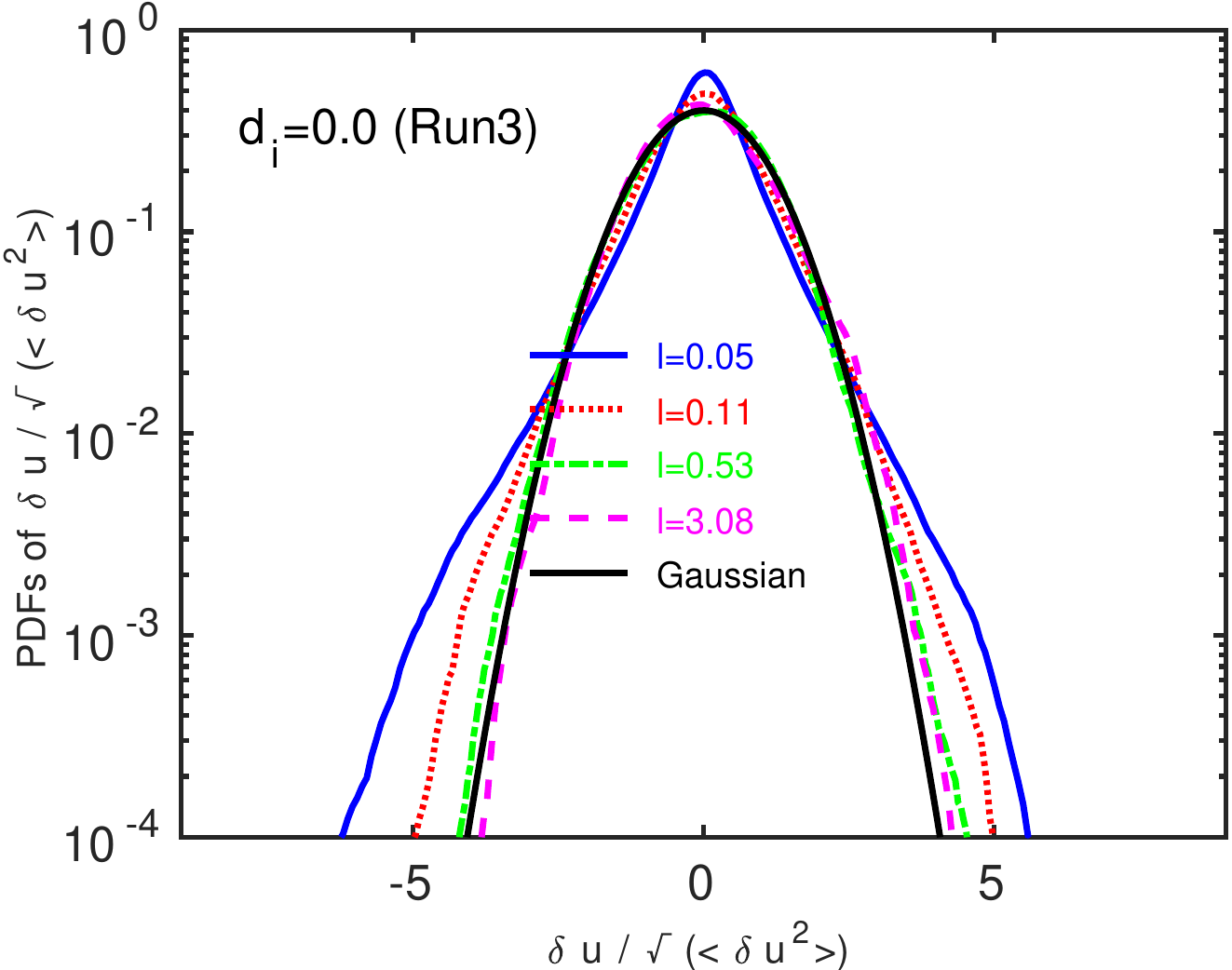} &
\includegraphics [scale=0.6]{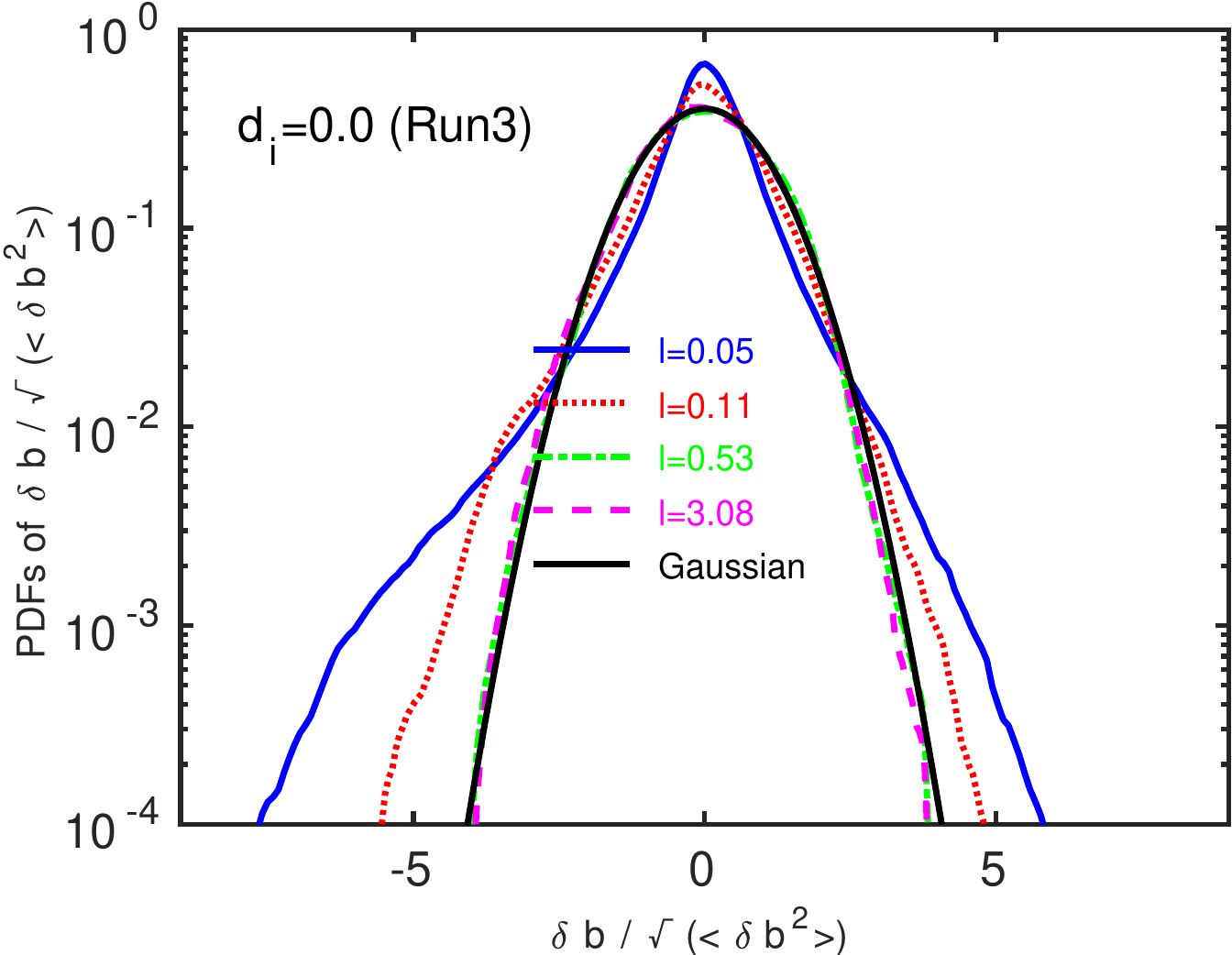}
\end{tabular}
\begin{tabular}{c c}
\textbf{(b)} & \textbf{(b')}\\
\includegraphics [scale=0.6]{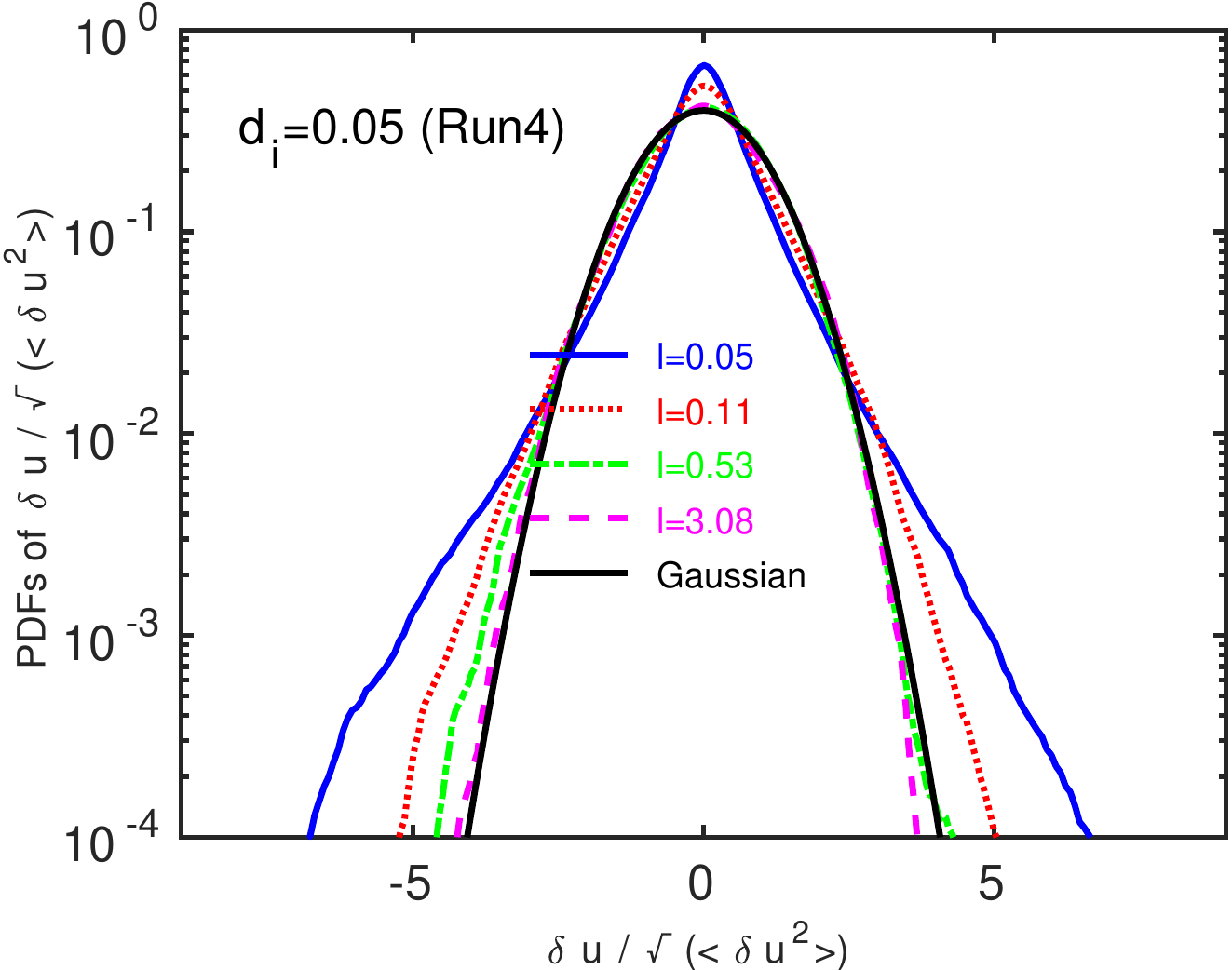} & 
\includegraphics [scale=0.6]{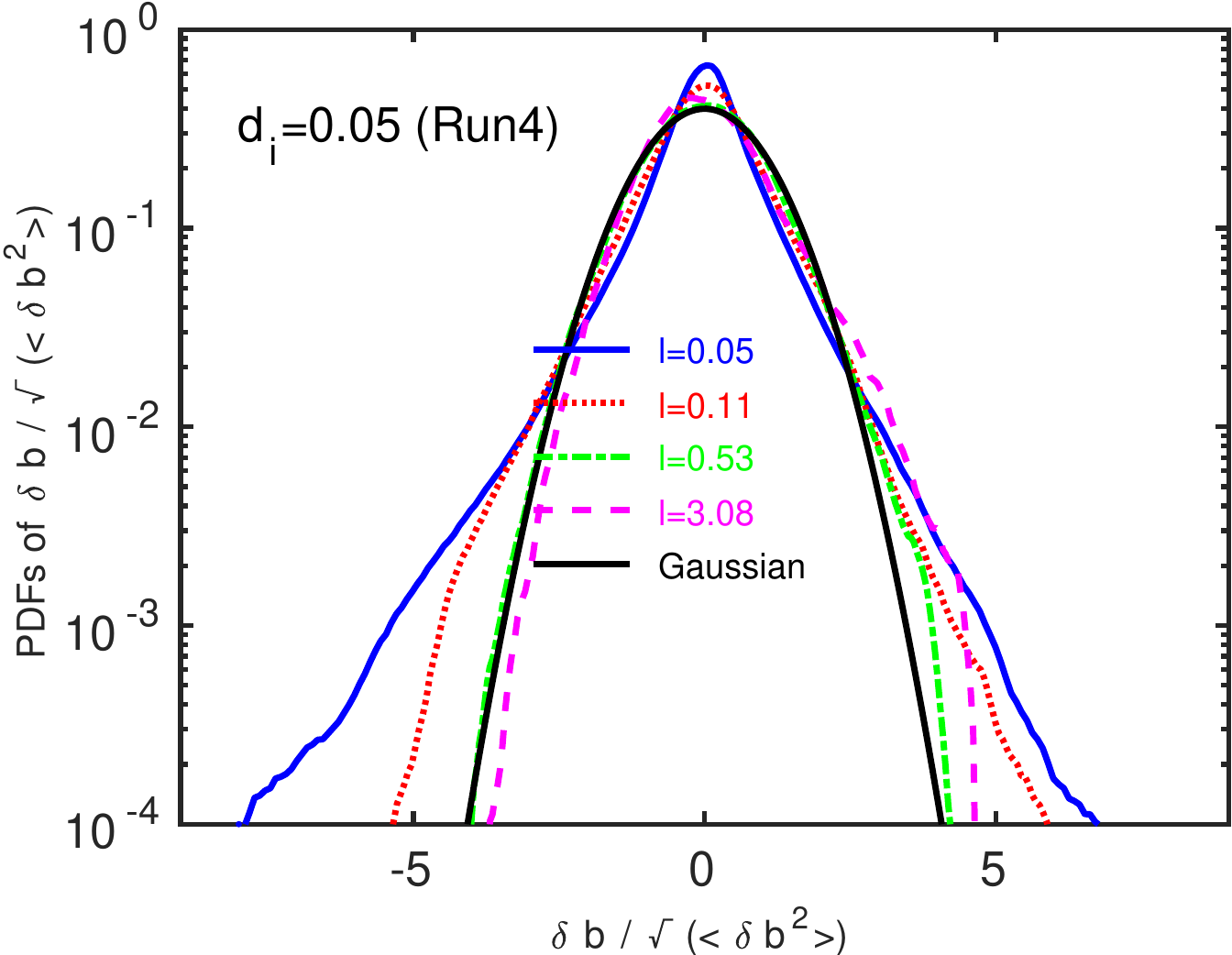}
\end{tabular}
\vskip -.1cm
\caption{Semilog (base 10) plots of PDFs of the velocity- (first column) and 
magnetic-field (second column) increments from $Run3$ (first row) and $Run4$ (second row)
for $l=0.05$ (blue solid lines), $0.11$ (red dotted lines), $0.53$ 
(green dashed-dot lines) and $3.08$ (magneta dashed lines); for reference,
we also show zero-mean and unit-variance Gaussian PDFs (black lines).}
\label{fig:6}
\end{figure*}

%

\begin{figure*}[t]
\begin{tabular}{c c}
\textbf{(a)} & \textbf{(b)} \\
\includegraphics [scale=0.5]{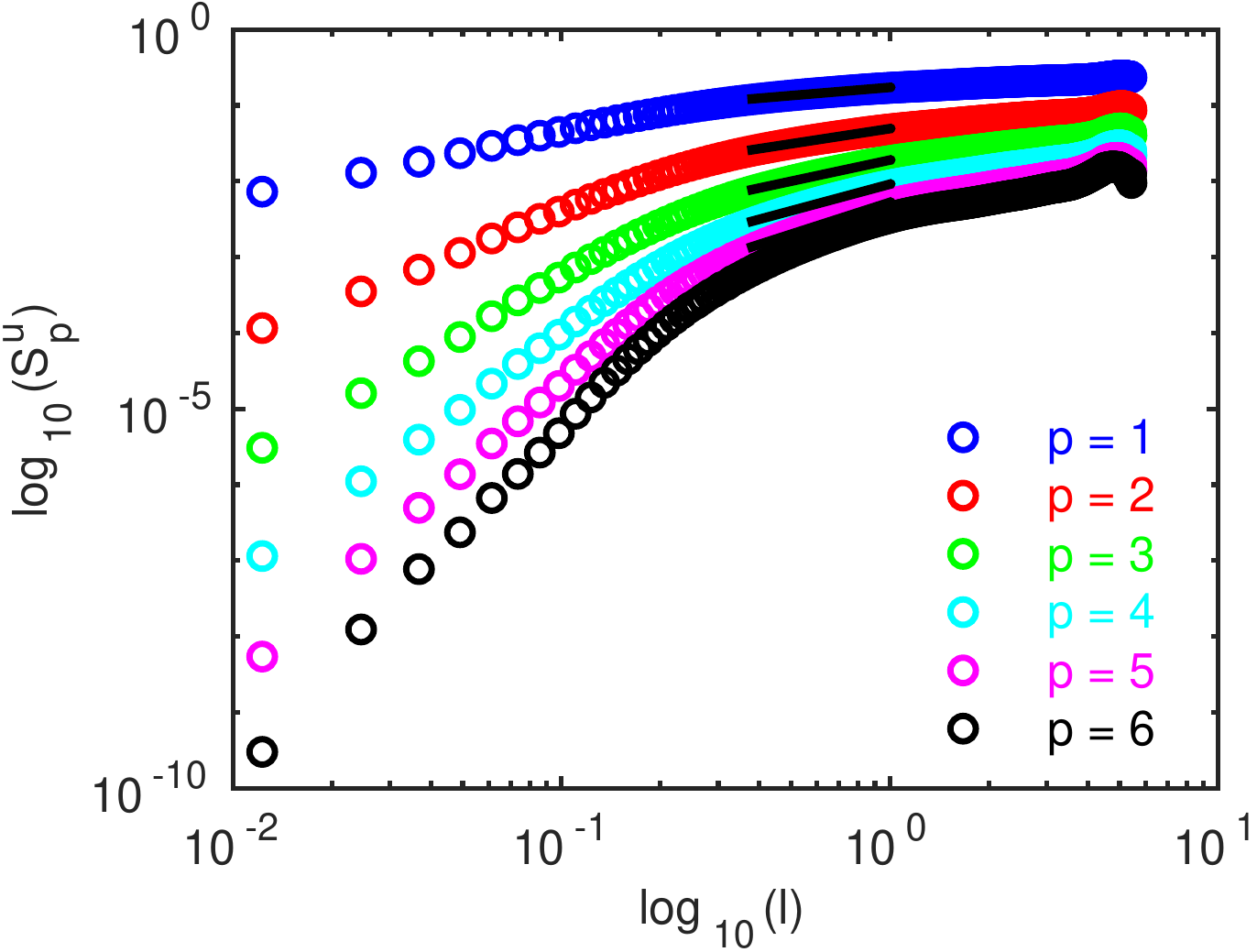} &
\includegraphics [scale=0.5]{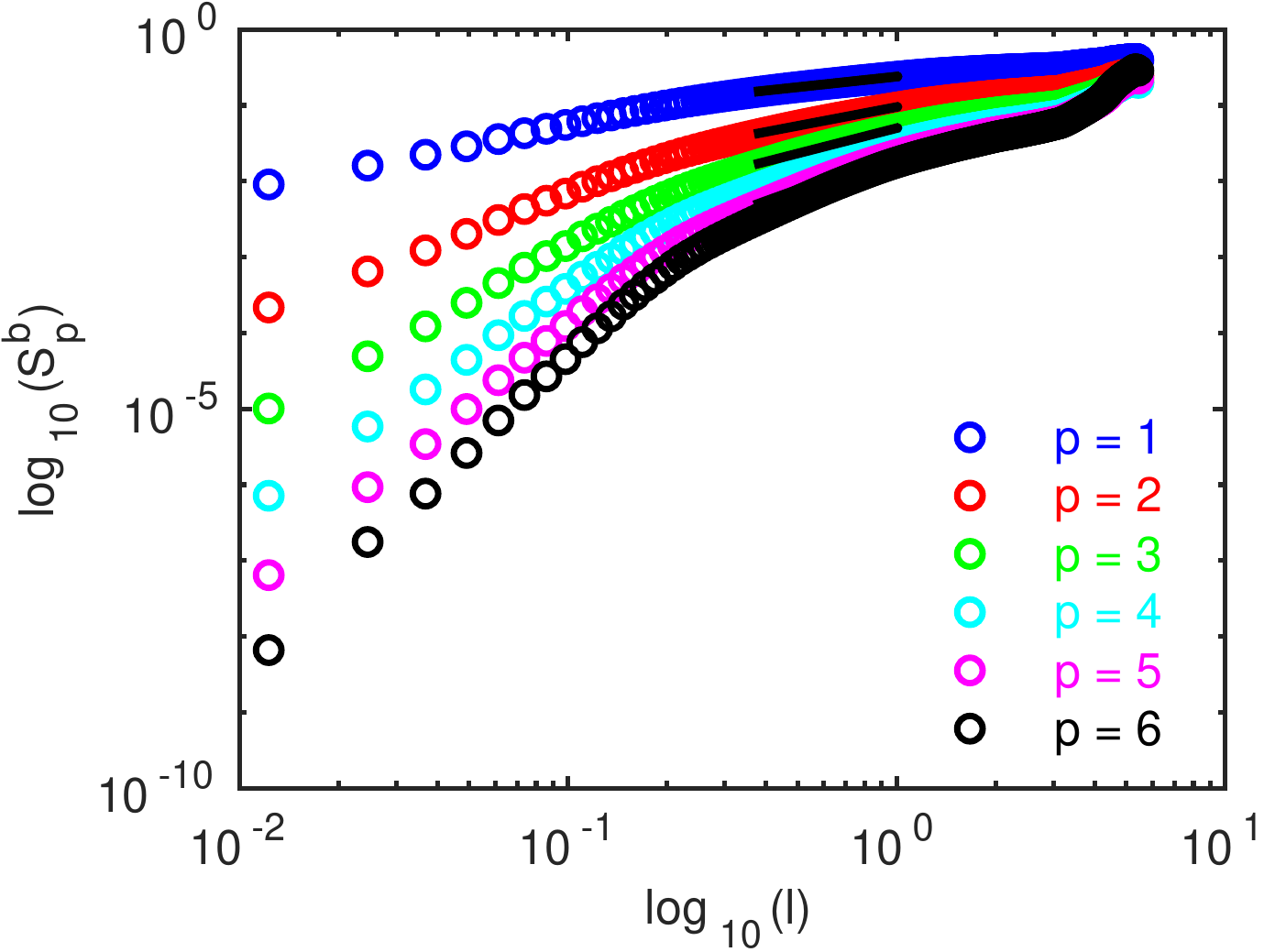}
\end{tabular}
\begin{tabular}{c c}
\textbf{(c)} & \textbf{(d)} \\
\includegraphics [scale=0.5]{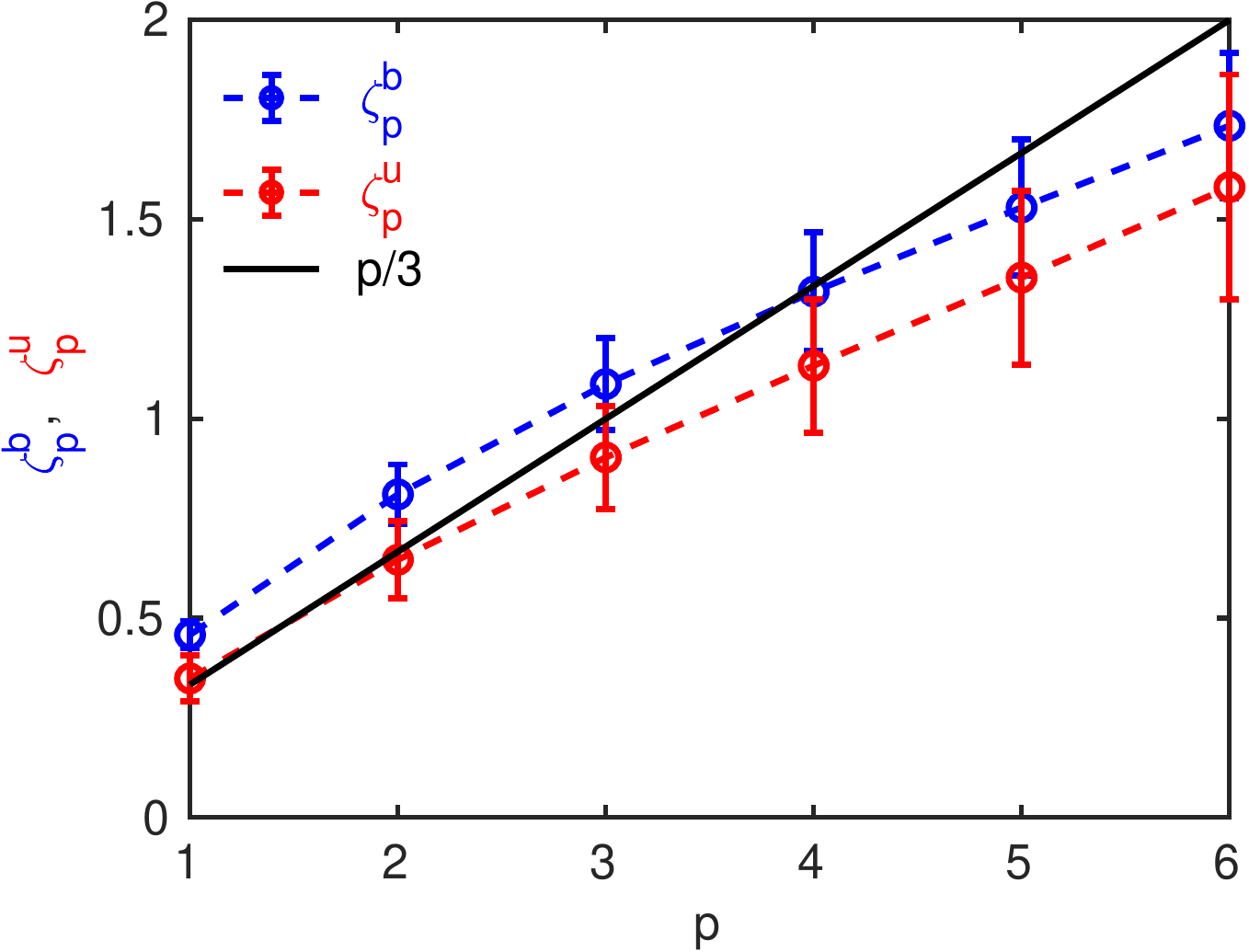} &
\includegraphics [scale=0.5]{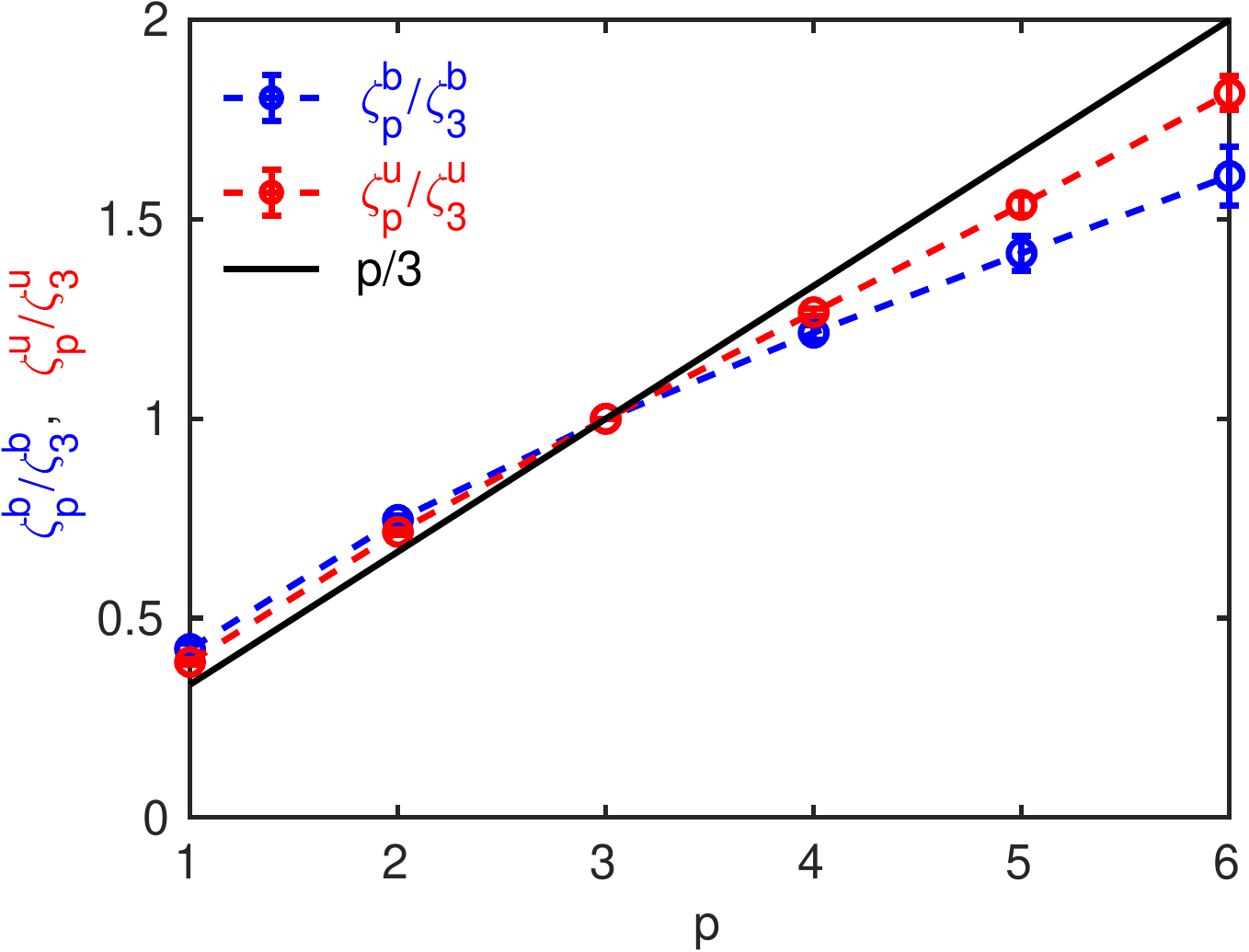} 
\end{tabular}
\vskip -.1cm
\caption{Log-log (base 10) plots versus $l$, at $t_{c}$, for $Run3$ and
(a) velocity and (b) magnetic structure functions (Eq.~\ref{eq:strucfun});
the order $p$ goes from $1-6$; we indicate by straight, black lines the regions 
that we use to obtain estimates for the  multiscaling 
exponents (Eqs.~\ref{eq:zetap}-\ref{eq:zetap12}). Plots versus the order $p$ of 
(c) the multiscaling exponents (Eqs.~\ref{eq:zetap}-\ref{eq:zetap12}) and
(d) their ratios; for reference we show the linear K41 scaling of exponents with $p/3$.}
\label{fig:8}
\end{figure*}

\begin{figure*}[t]
\begin{tabular}{c c}
\textbf{(a)} & \textbf{(b)} \\
\includegraphics [scale=0.5]{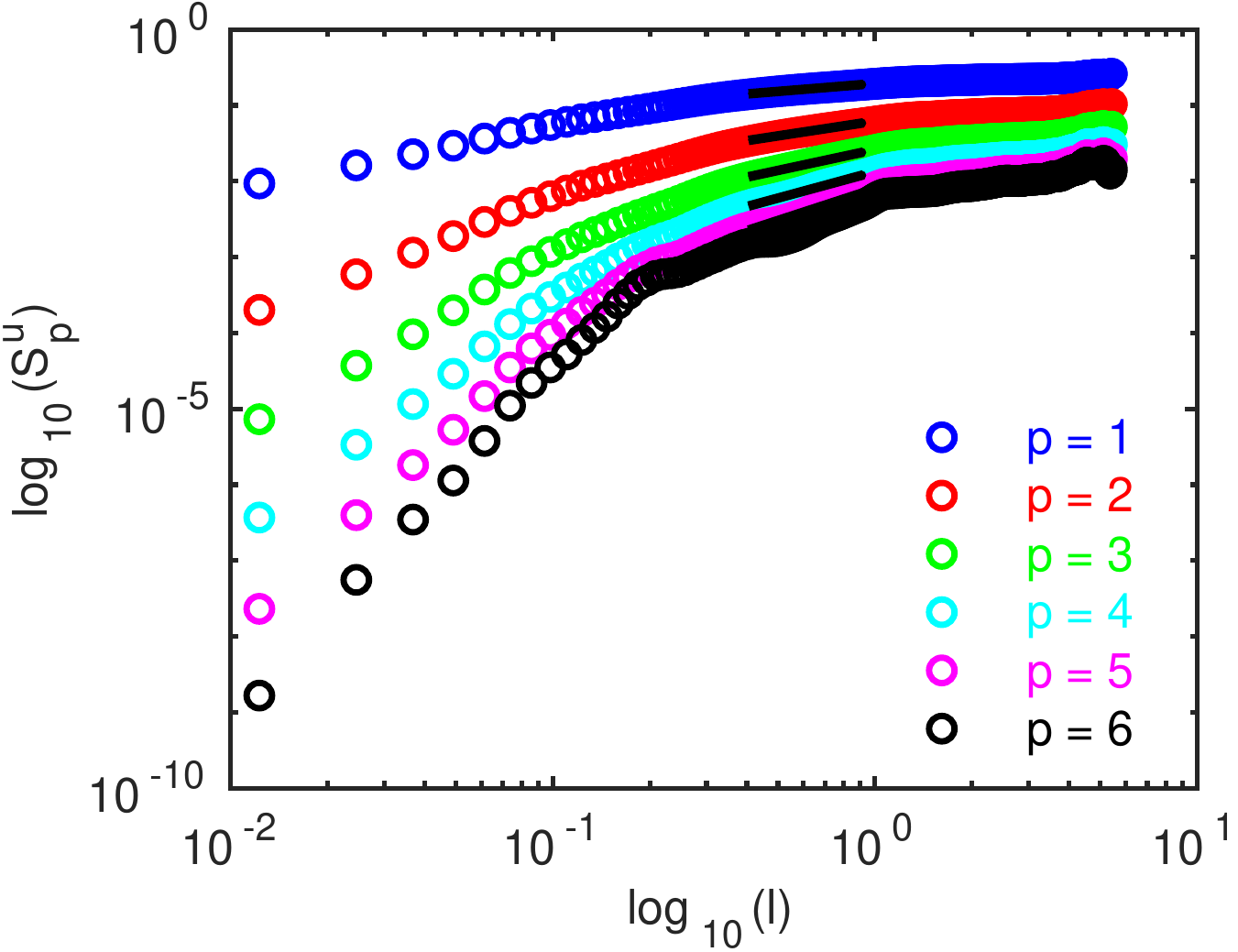} &
\includegraphics [scale=0.5]{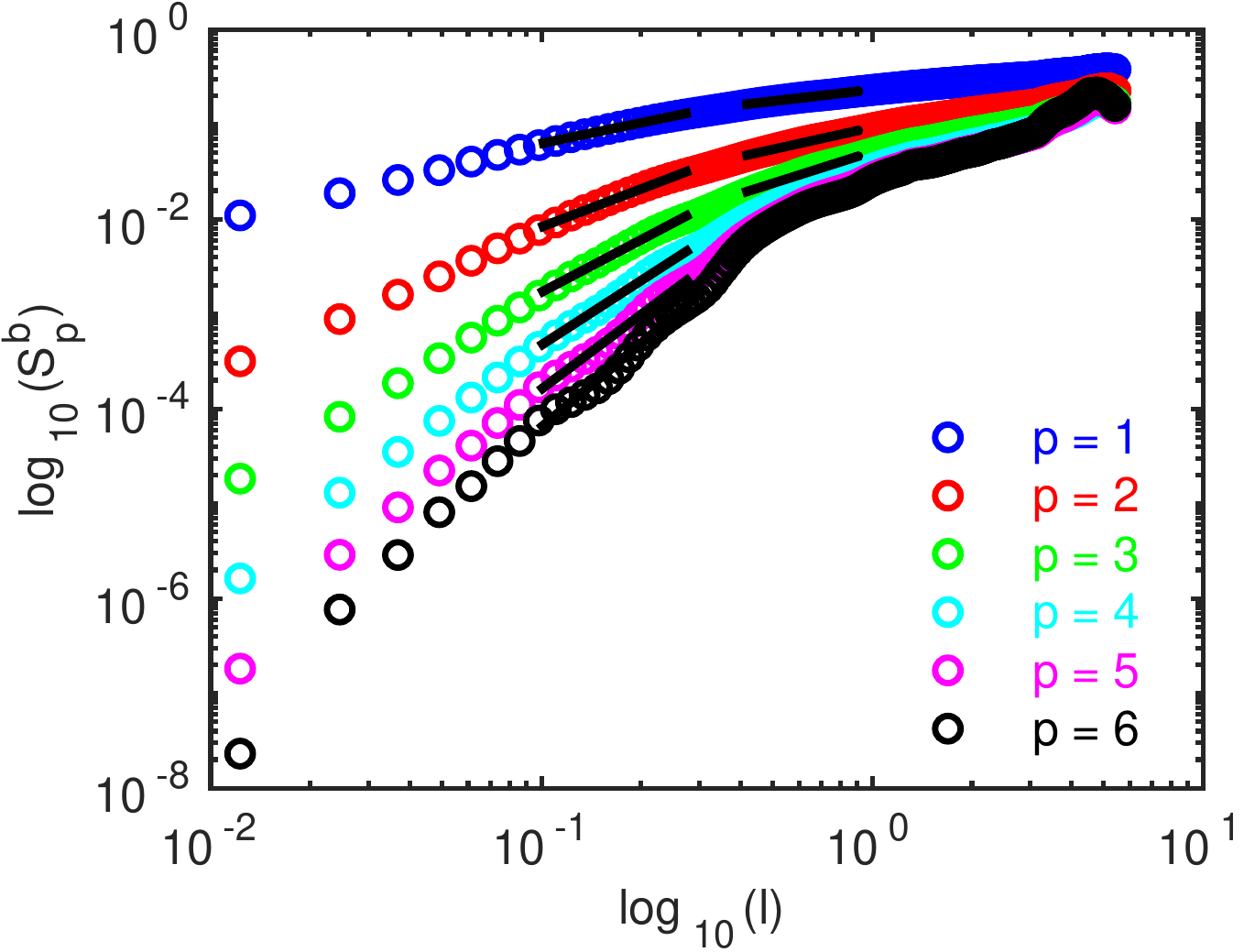}
\end{tabular}
\begin{tabular}{c c}
\textbf{(c)} & \textbf{(d)} \\
\includegraphics [scale=0.5]{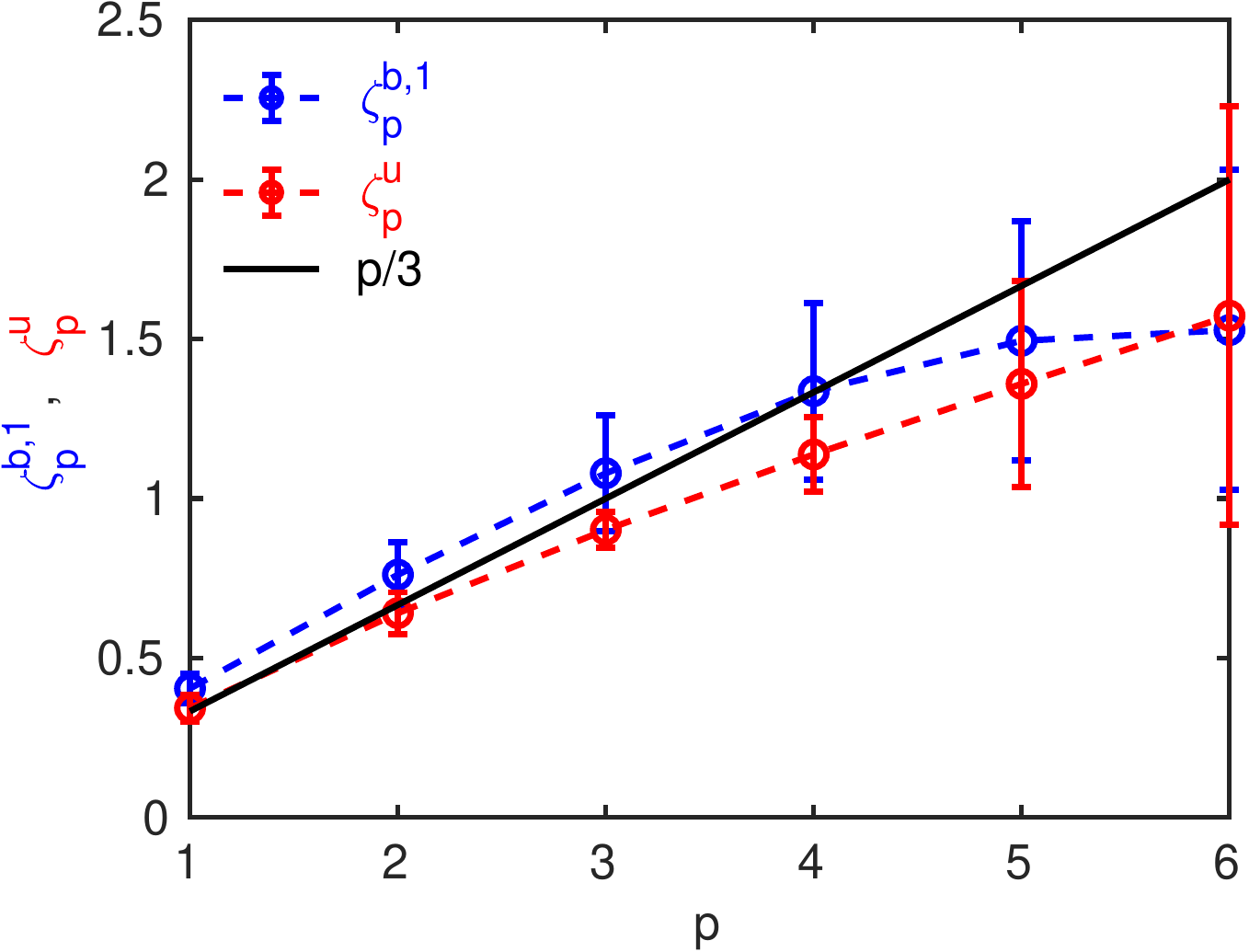} &
\includegraphics [scale=0.5]{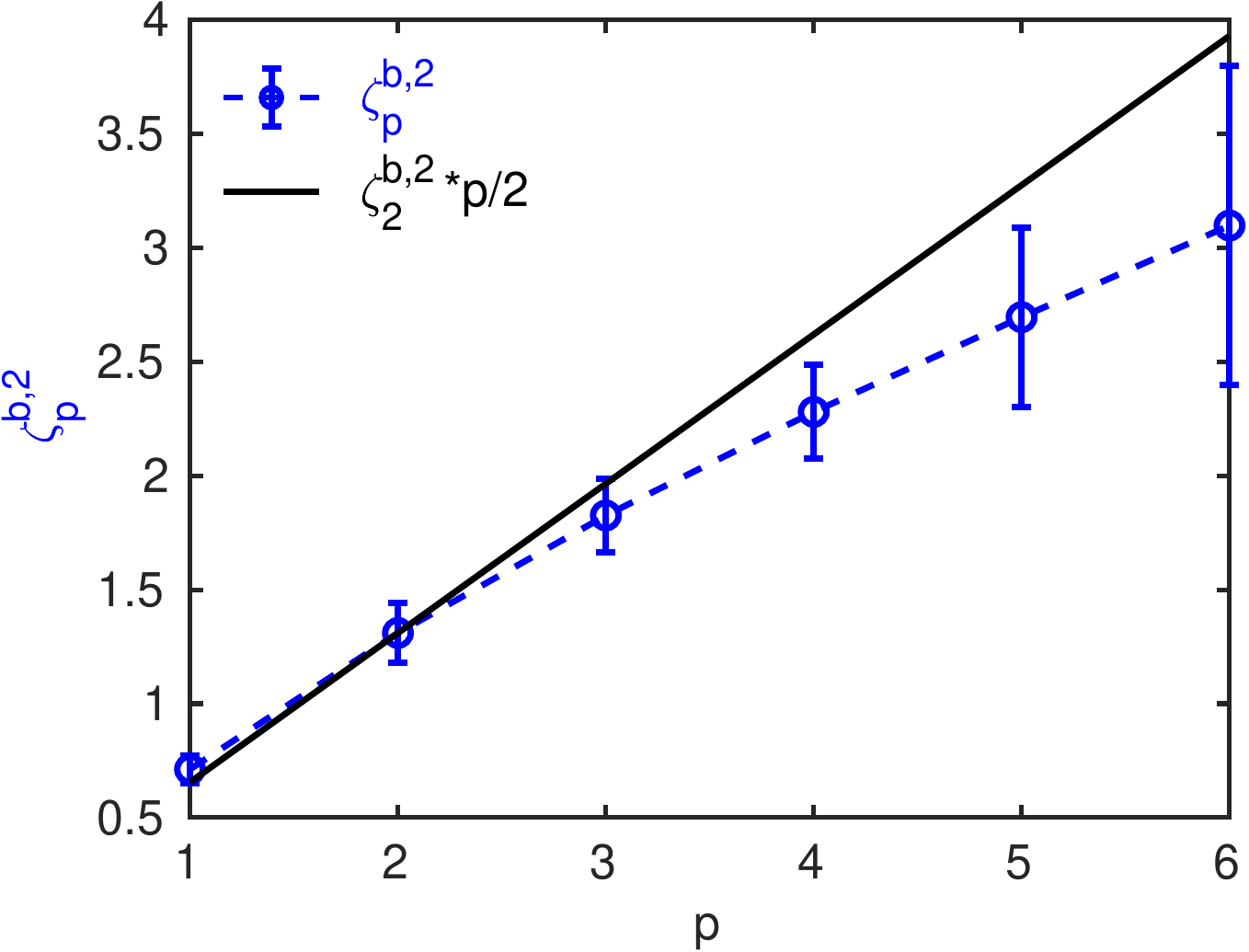}
\end{tabular}
\begin{tabular}{c c}
\textbf{(e)} & \textbf{(f)} \\
\includegraphics [scale=0.5]{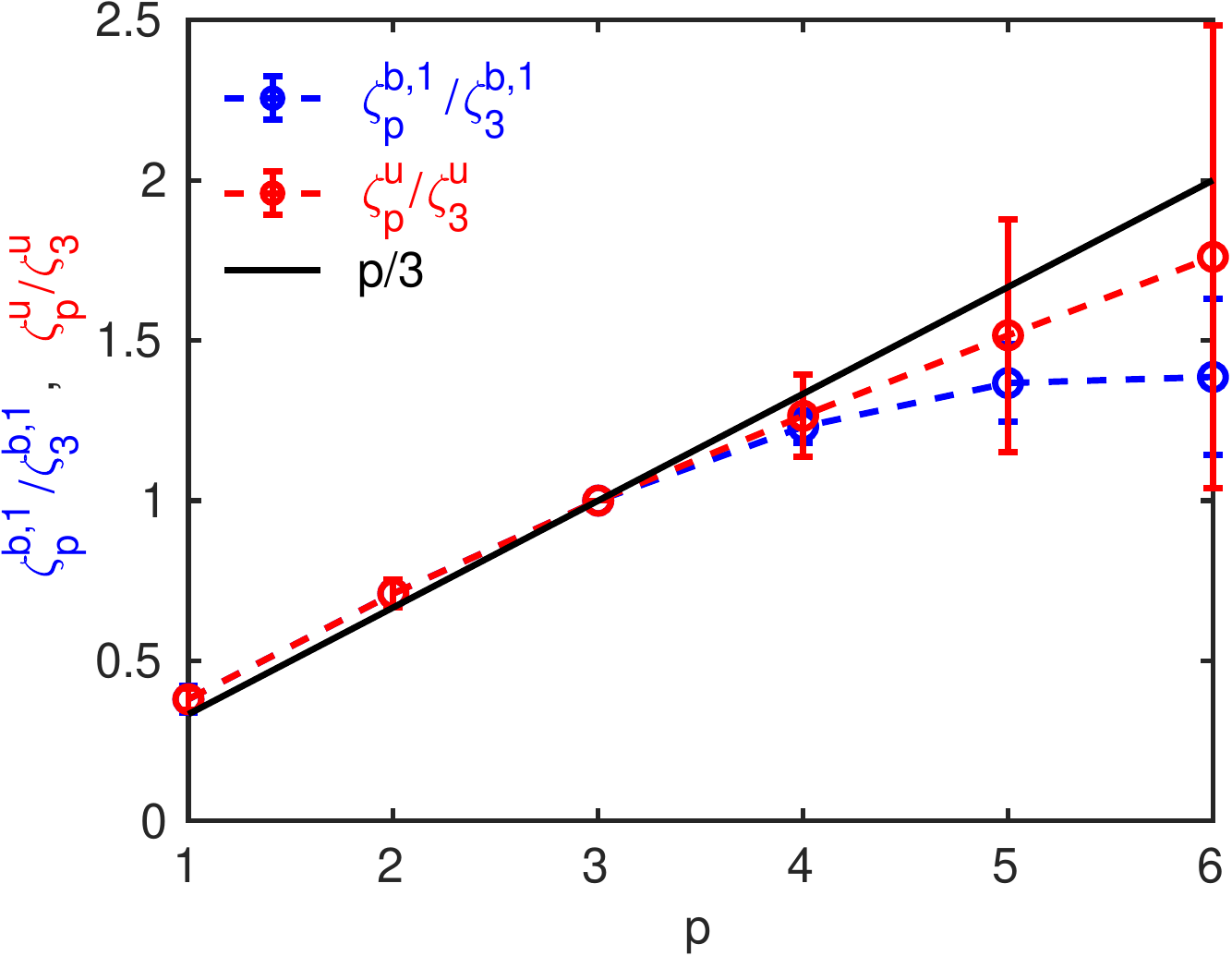} &
\includegraphics [scale=0.5]{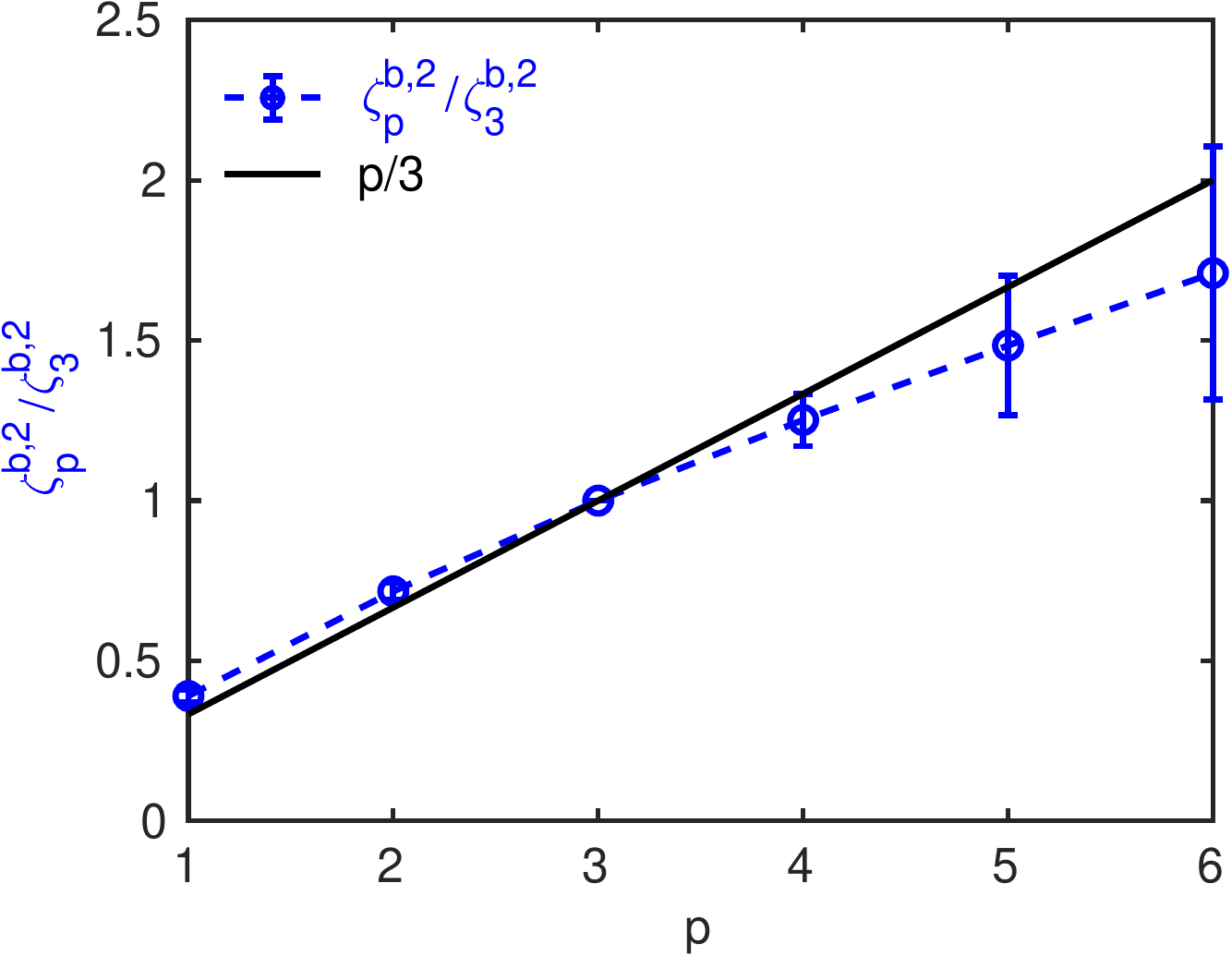} 
\end{tabular}
\vskip -.1cm
\caption{Log-log (base 10) plots versus $l$, at $t_{c}$, for $Run4$ and
(a) velocity and (b) magnetic structure functions (Eq.~\ref{eq:strucfun});
the order $p$ goes from $1-6$; we indicate by straight, black lines the regions 
that we use to obtain estimates for the  multiscaling 
exponents (Eqs.~\ref{eq:zetap}-\ref{eq:zetap12}). Plots versus the order $p$ of 
the multiscaling exponents (Eqs.~\ref{eq:zetap}-\ref{eq:zetap12})  for (c)
the inertial range and (d) the intermediate-dissipation range.
Plots versus the order $p$ of 
the ratios of multiscaling exponents (Eqs.~\ref{eq:zetap}-\ref{eq:zetap12}) 
for (e) the inertial range and (f) the intermediate-dissipation range;
for reference we show simple scaling predictions.}
\label{fig:9}
\end{figure*}

\begin{figure*}[t]
\begin{tabular}{c c c}
\textbf{(a)} & \textbf{(b)} & \textbf{(c)} \\
\includegraphics [scale=0.4]{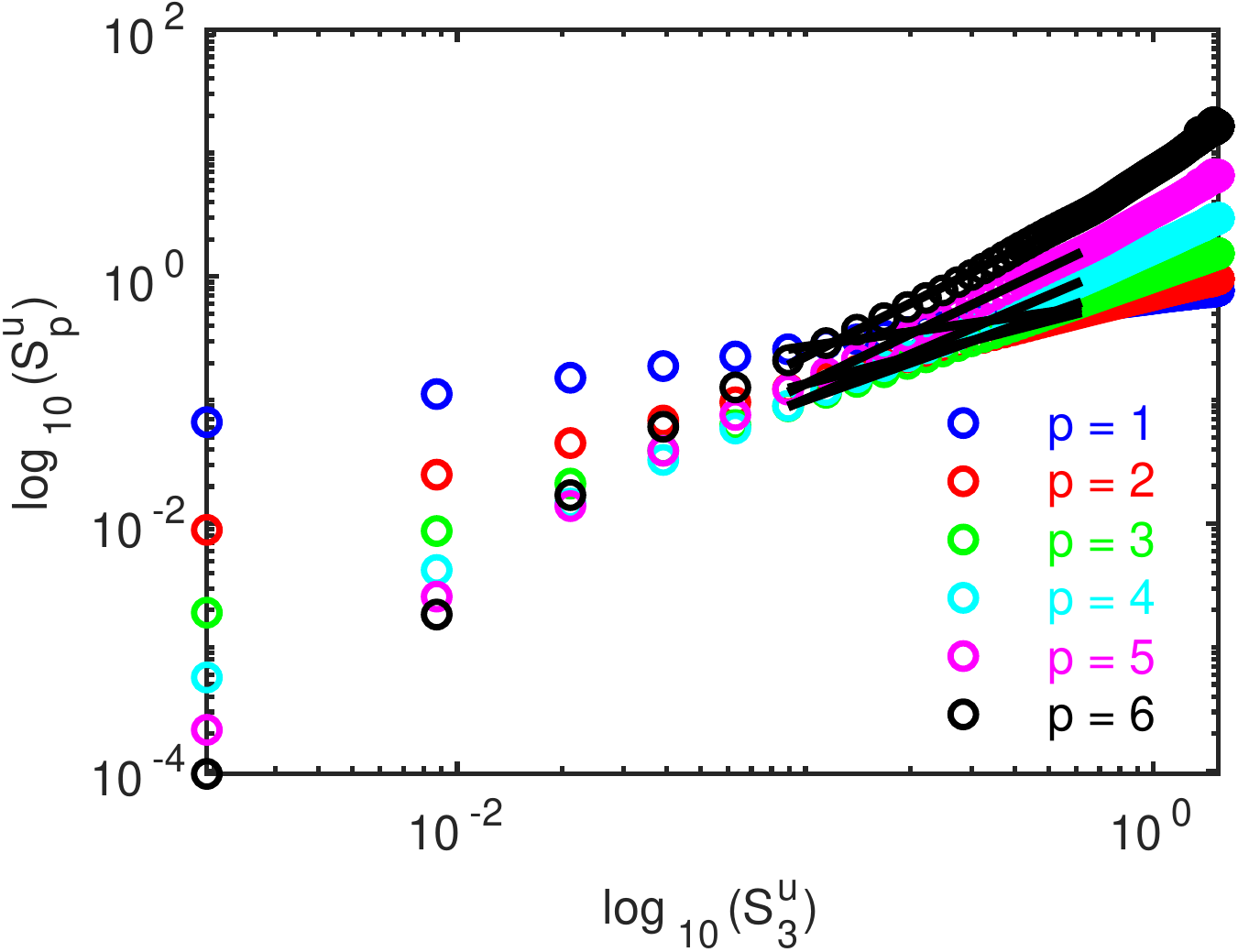} &
\includegraphics [scale=0.4]{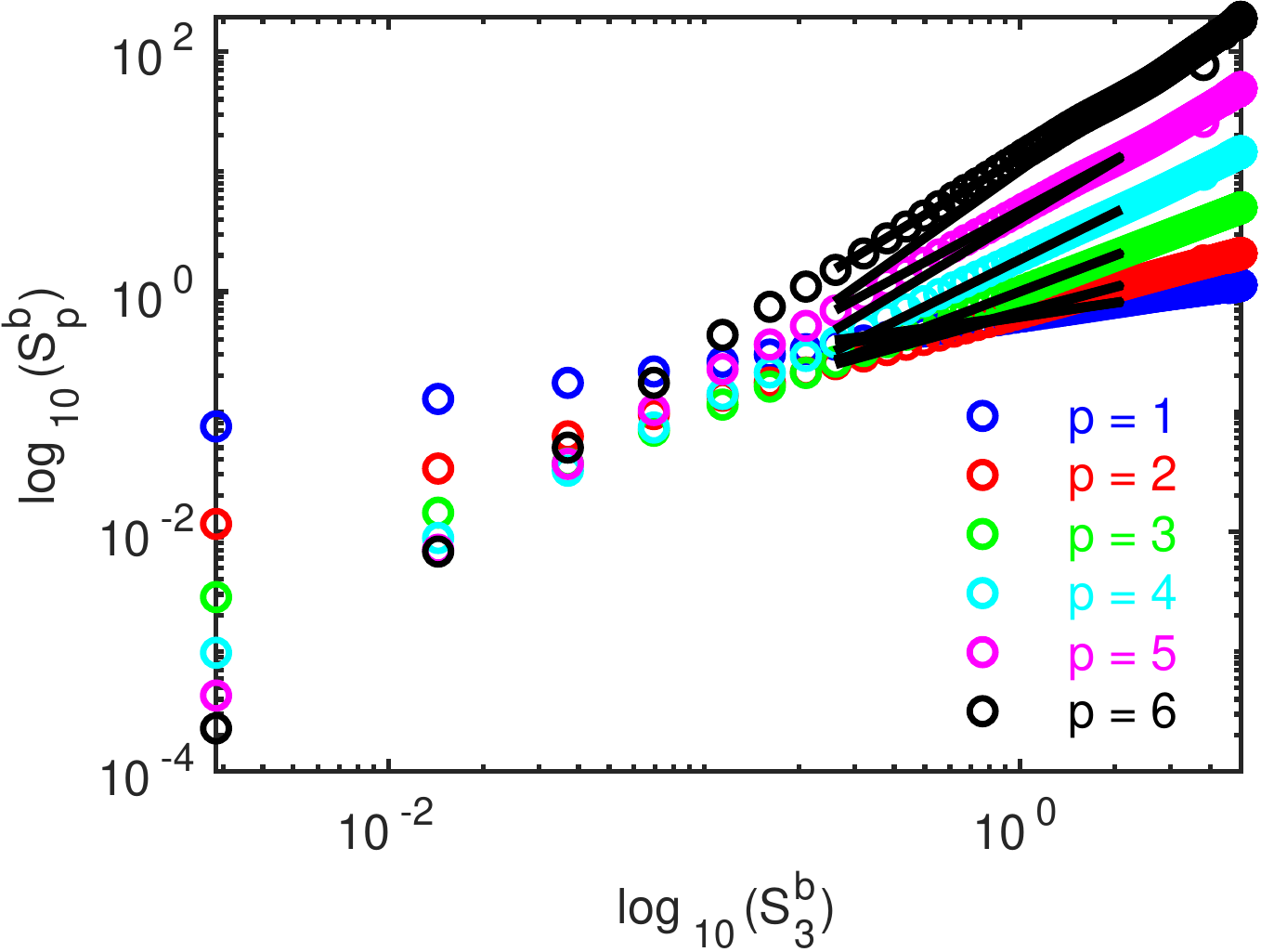} &
\includegraphics [scale=0.4]{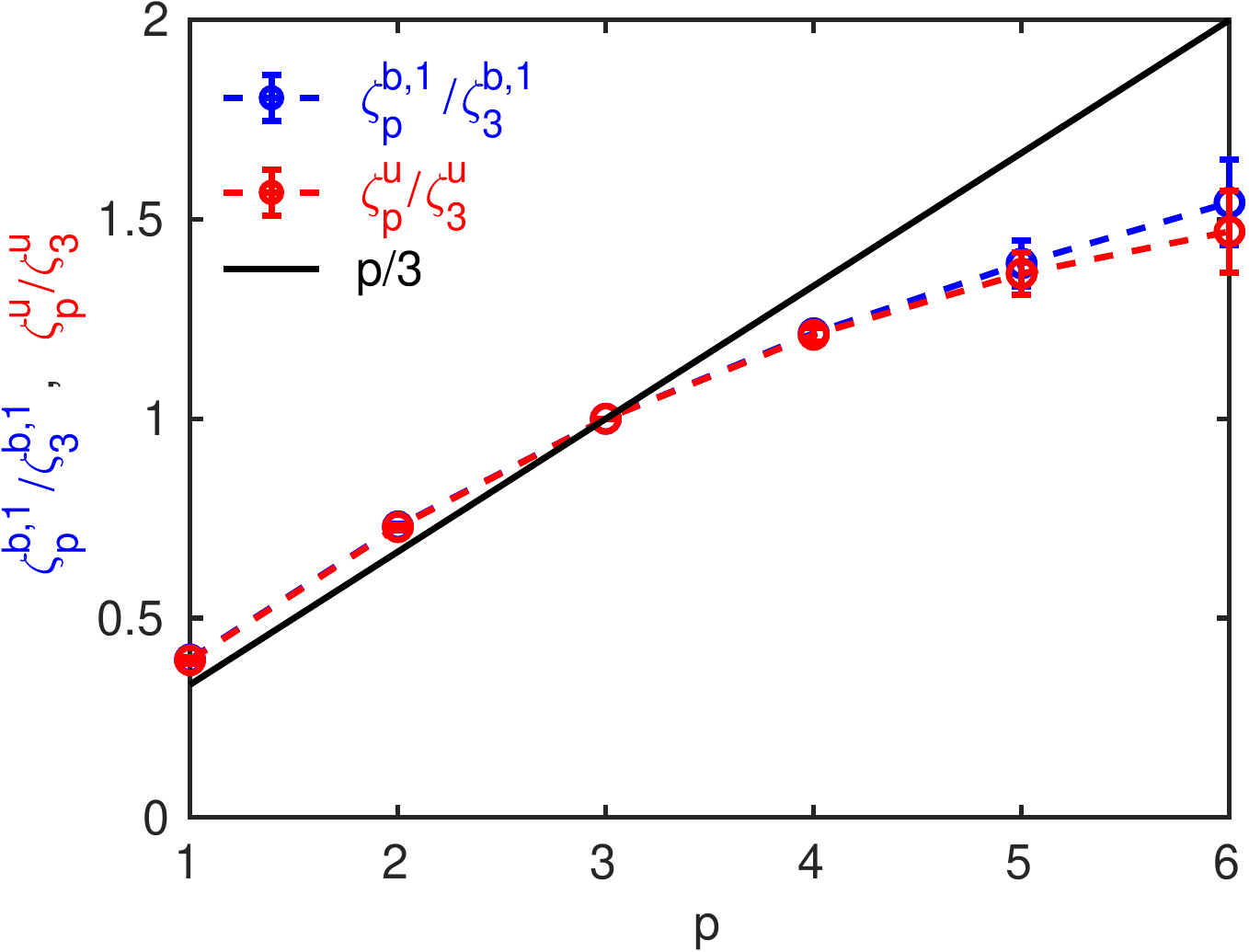} 
\end{tabular}
\vskip -.1cm
\caption{Log-log (base 10) ESS plots (see text), at $t_{c}$, for $Run5a$ and
(a) velocity and (b) magnetic structure functions (Eq.~\ref{eq:strucfun});
the order $p$ goes from $1-6$; we indicate by straight, black lines the regions 
that we use to obtain estimates for the ratios of multiscaling 
exponents (Eqs.~\ref{eq:zetap}-\ref{eq:zetap12}). Plots versus the order $p$ of 
(c) the ratios of multiscaling exponents (Eqs.~\ref{eq:zetap}-\ref{eq:zetap12}); 
for reference we show the K41 scaling of exponents with $p/3$.}
\label{fig:10}
\end{figure*}

\begin{figure*}[t]
\begin{tabular}{c c}
\textbf{(a)} & \textbf{(b)} \\
\includegraphics [scale=0.5]{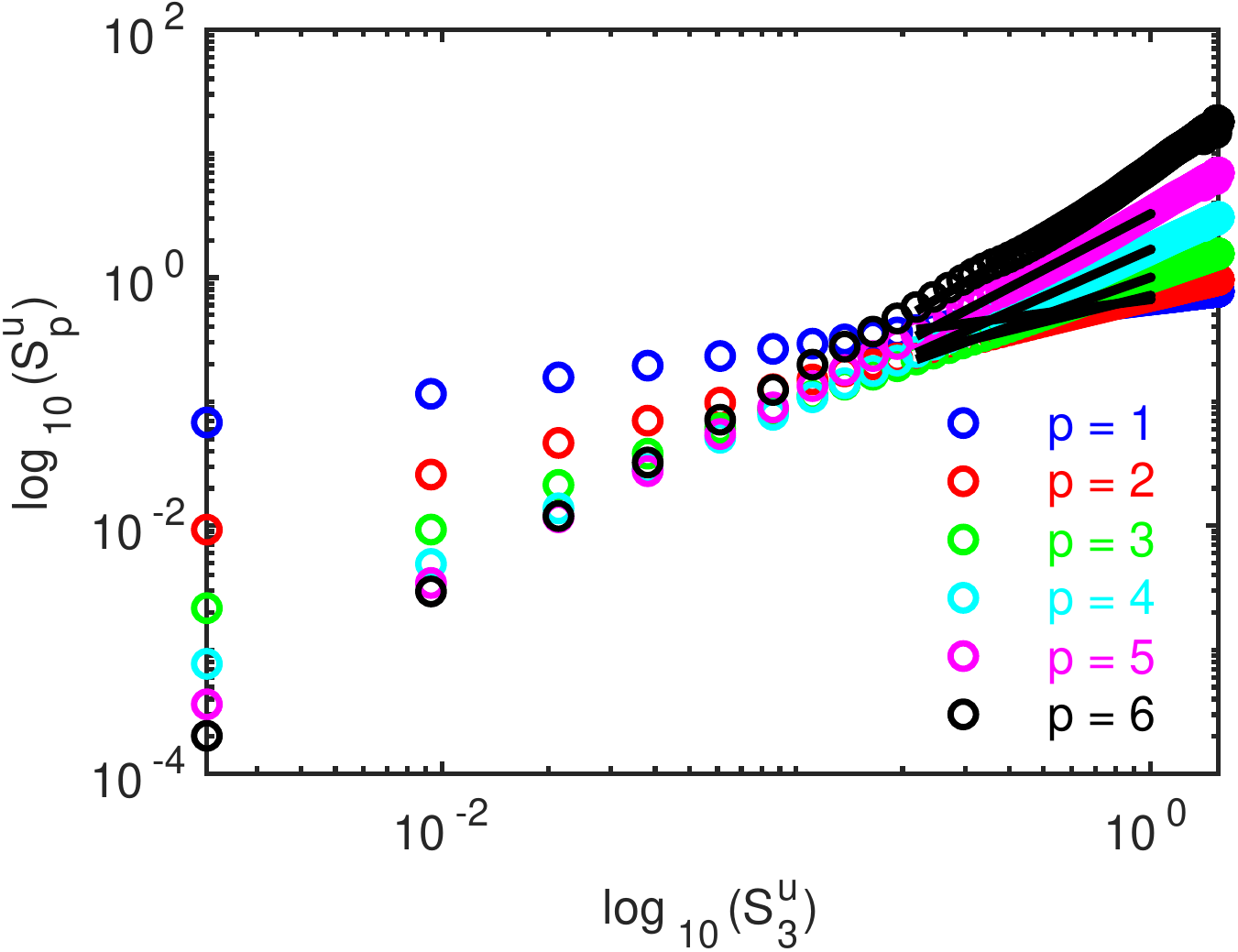} &
\includegraphics [scale=0.5]{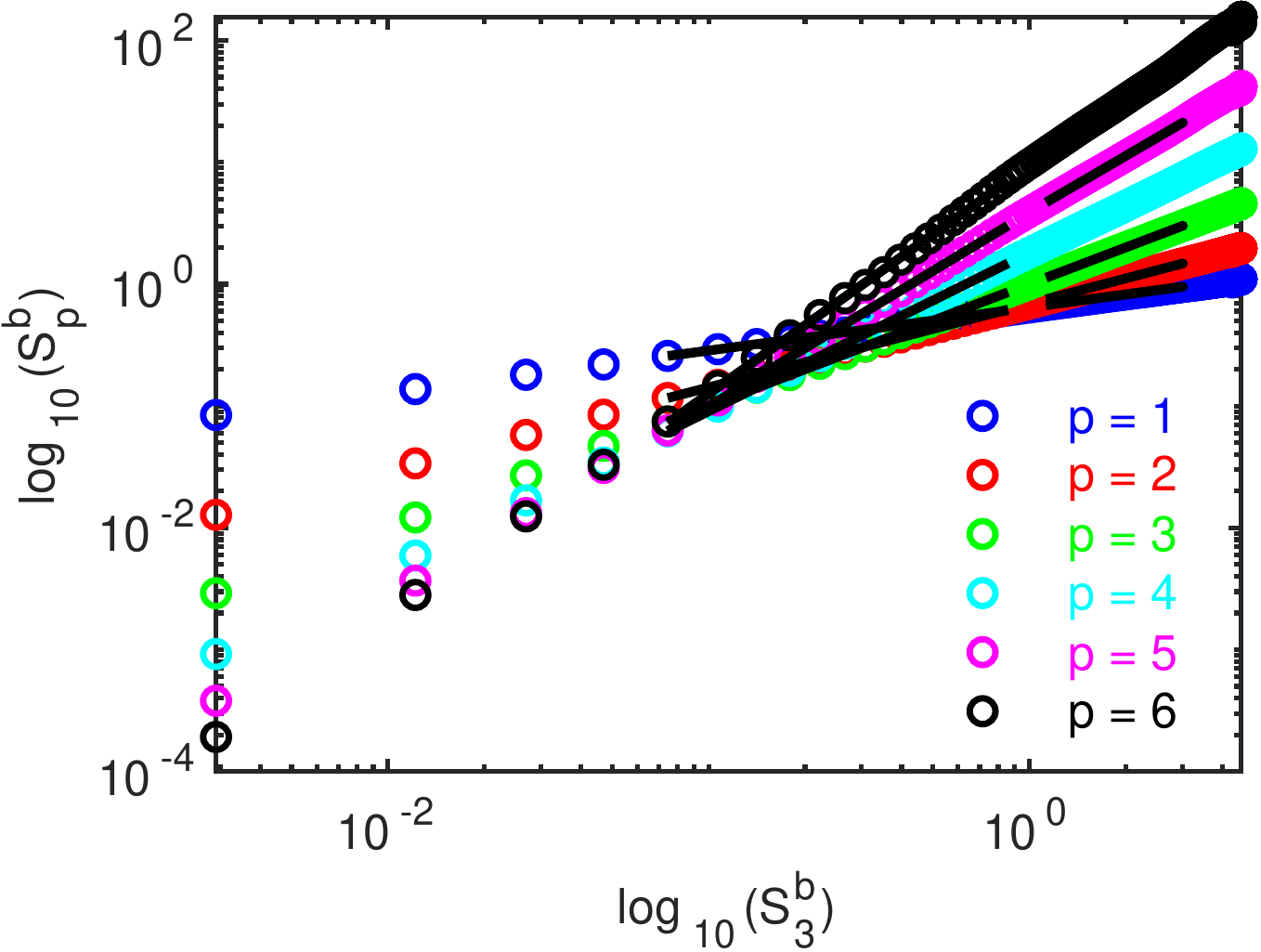}
\end{tabular}
\begin{tabular}{c c}
\textbf{(c)} & \textbf{(d)} \\
\includegraphics [scale=0.5]{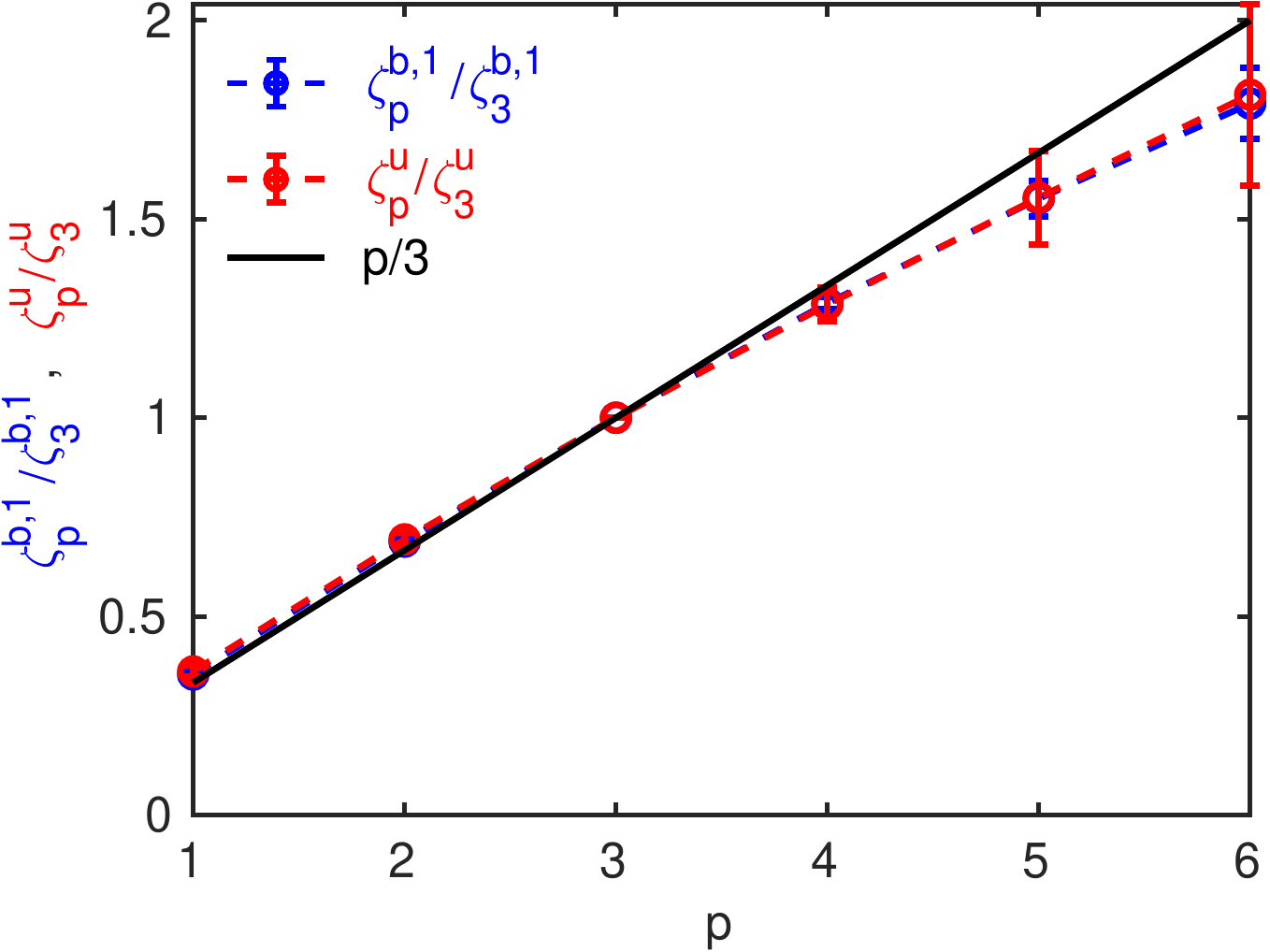} &
\includegraphics [scale=0.5]{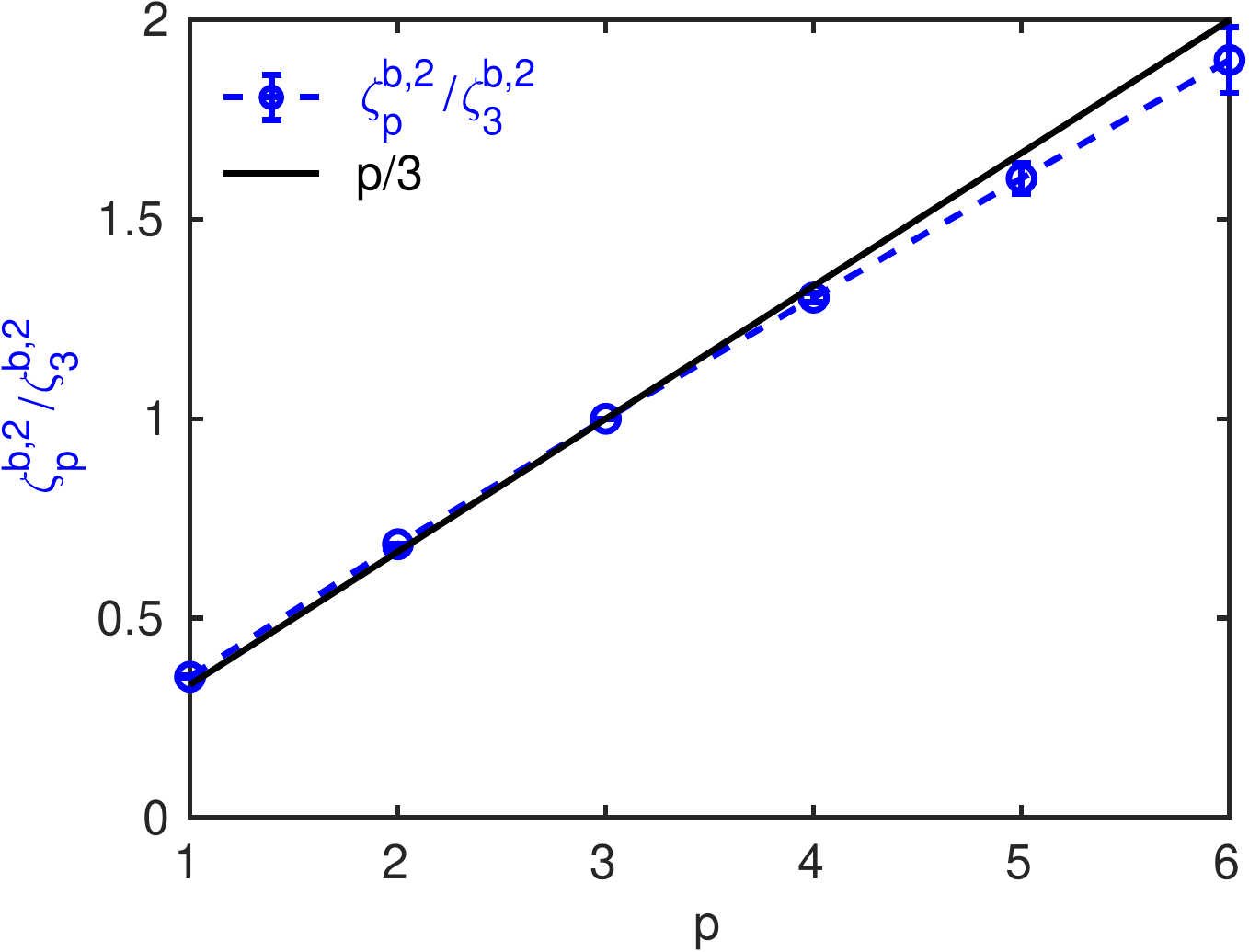} 
\end{tabular}
\vskip -.1cm

\caption{Log-log (base 10) ESS plots (see text), at $t_{c}$, for $Run5b$ and
(a) velocity and (b) magnetic structure functions (Eq.~\ref{eq:strucfun});
the order $p$ goes from $1-6$; we indicate by straight, black lines the regions 
that we use to obtain estimates for the ratios of multiscaling 
exponents (Eqs.~\ref{eq:zetap}-\ref{eq:zetap12}). Plots versus the order $p$ of 
the ratios of multiscaling exponents (Eqs.~\ref{eq:zetap}-\ref{eq:zetap12}) for (c)
the inertial range and (d) the intermediate-dissipation range; 
for reference we show the K41 scaling of exponents with $p/3$.}
\label{fig:11}
\end{figure*}

\begin{figure*}[t]
\begin{tabular}{c c}
\textbf{(a)} & \textbf{(b)} \\
\includegraphics [scale=0.5]{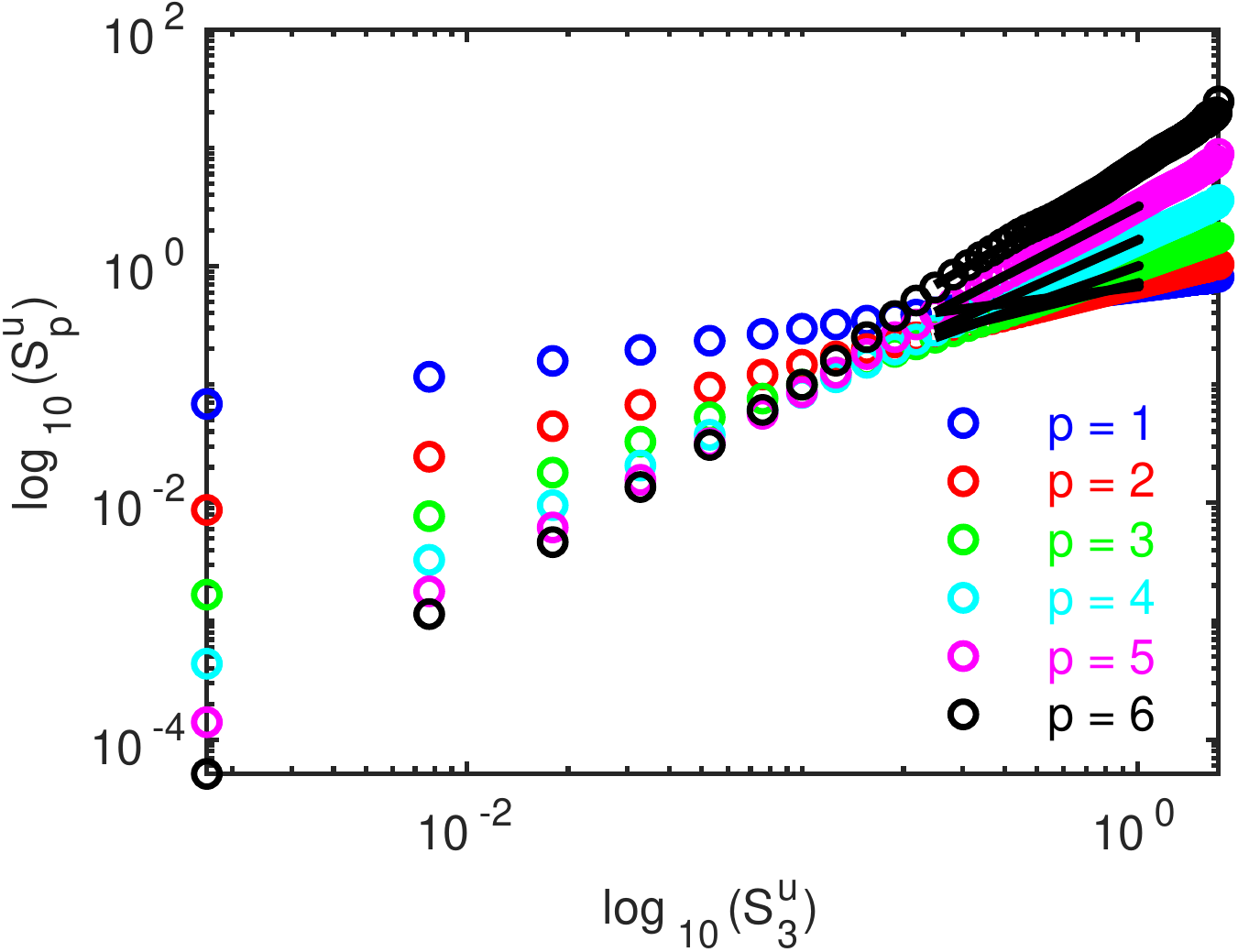} &
\includegraphics [scale=0.5]{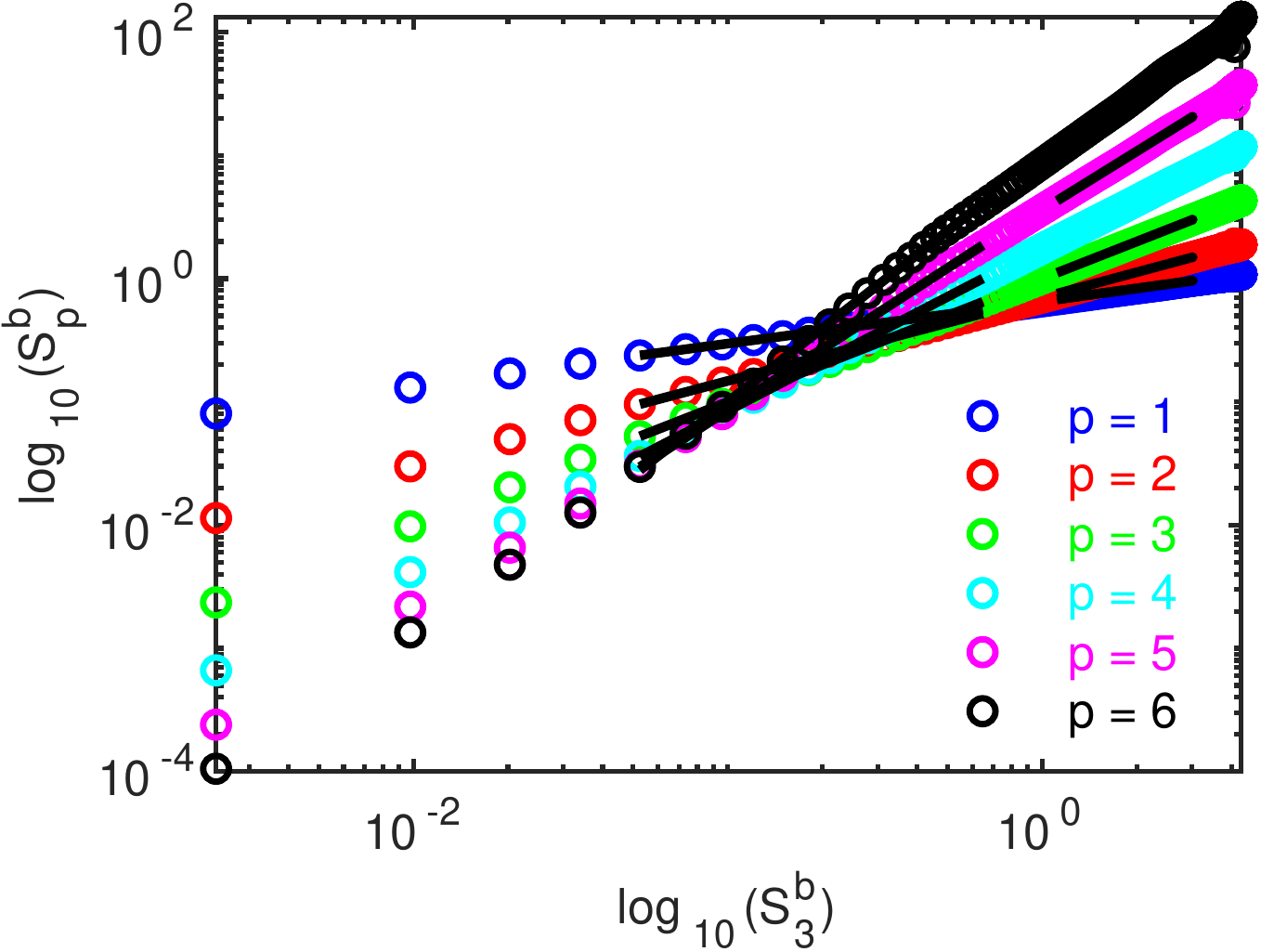}
\end{tabular}
\begin{tabular}{c c}
\textbf{(c)} & \textbf{(d)} \\
\includegraphics [scale=0.5]{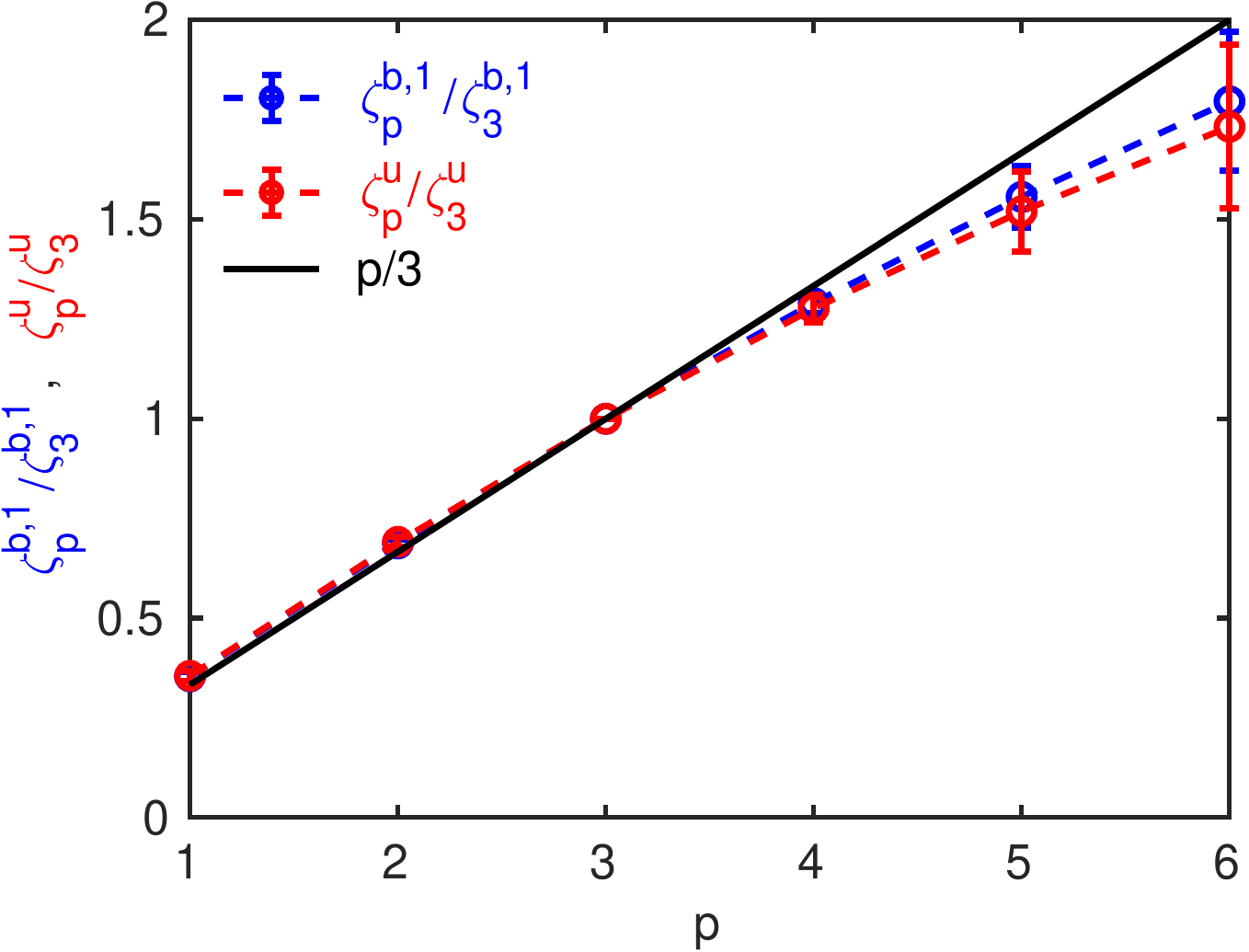} &
\includegraphics [scale=0.5]{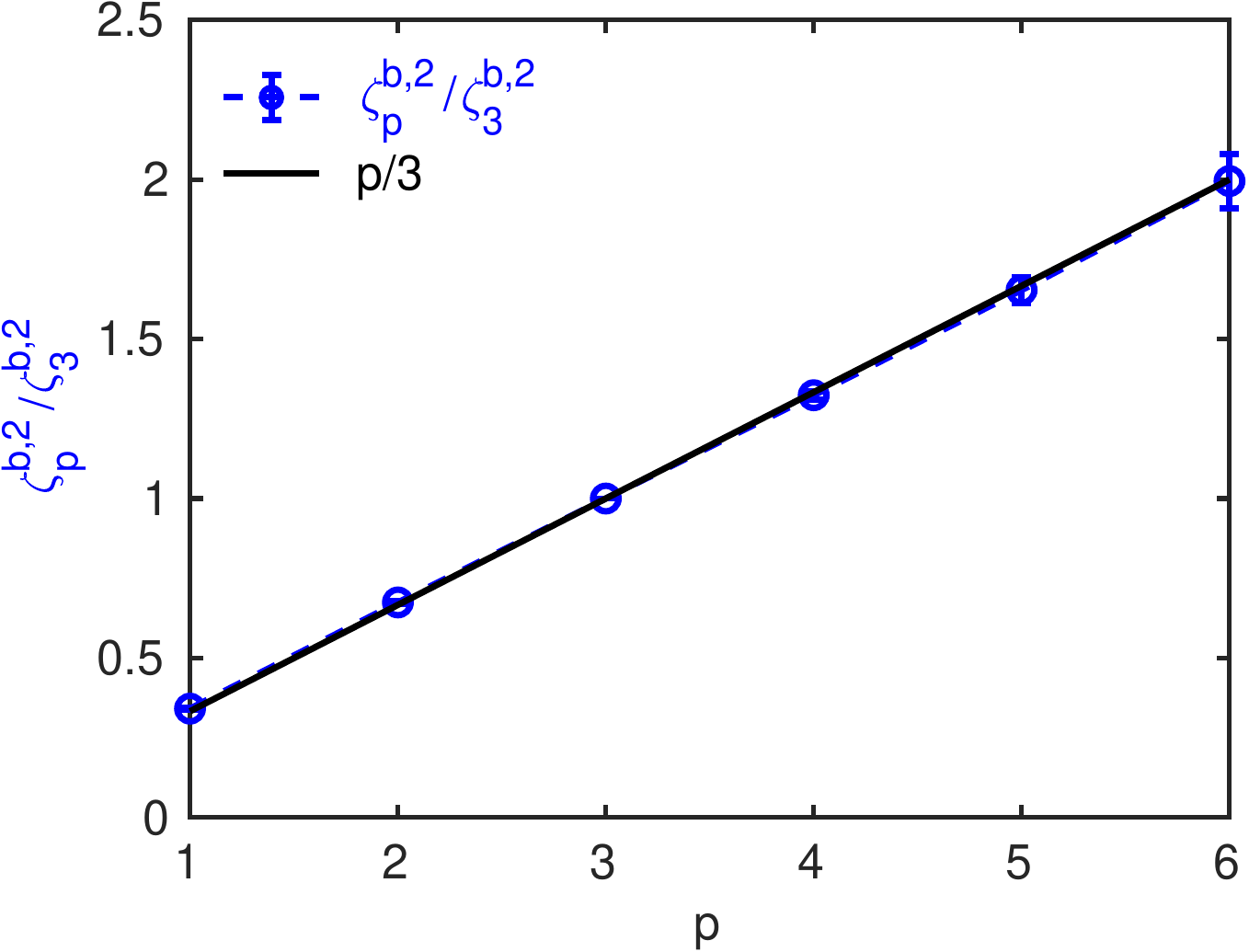} 
\end{tabular}
\vskip -.1cm
\caption{Log-log (base 10) ESS plots (see text), at $t_{c}$, for $Run5c$ and
(a) velocity and (b) magnetic structure functions (Eq.~\ref{eq:strucfun});
the order $p$ goes from $1-6$; we indicate by straight, black lines the regions 
that we use to obtain estimates for the ratios of multiscaling 
exponents (Eqs.~\ref{eq:zetap}-\ref{eq:zetap12}). Plots versus the order $p$ of 
the ratios of multiscaling exponents (Eqs.~\ref{eq:zetap}-\ref{eq:zetap12}) for (c)
the inertial range and (d) the intermediate-dissipation range; 
for reference we show the K41 scaling of exponents with $p/3$.}
\label{fig:12}
\end{figure*}

We turn now to structure functions (Eq.~\ref{eq:strucfun}), multiscaling
exponents (Eqs.~\ref{eq:zetap}-\ref{eq:zetap12}), and the ratios of these exponents.
We present plots of such structure functions in Figs.~\ref{fig:8}-\ref{fig:12}.
To extract multiscaling exponents (Eqs.~\ref{eq:zetap}-\ref{eq:zetap12}) directly, we
use log-log (base 10) plots of structure functions (Eq.~\ref{eq:strucfun}) versus the separation $l$. We obtain such plots, at $t_{c}$, for $Run3$ and
$Run4$ in, respectively, Figs.~\ref{fig:8}(a)-(b) and Figs.~\ref{fig:9}(a)-(b),
for order $p$ from $1-6$; we indicate by straight, black lines the regions that
we use to obtain estimates for the  multiscaling
exponents (Eqs.~\ref{eq:zetap}-\ref{eq:zetap12}); we get these exponents and their
error bars by using a local-slope analysis~\cite{sahoo2011systematics}.  We
present plots, versus $p$, of the resulting multiscaling
exponents (Eqs.~\ref{eq:zetap}-\ref{eq:zetap12}), for $Run3$ and $Run4$ in,
respectively, Figs.~\ref{fig:8}(c) and Figs.~\ref{fig:9}(c)-(d). If the scaling
range extends over a limited range of scales, this range can often be extended
by using the extended-self-similarity (ESS)
procedure~\cite{benzi1993extended,chakraborty2010extended}. In this procedure,
we use log-log (base 10) plots of the structure function of order $p$, versus,
e.g., the $p=3$ structure function; straight-line regions in such plots yield
the ratios of the order $p$ and order $p=3$  multiscaling
exponents (Eqs.~\ref{eq:zetap}-\ref{eq:zetap12}). [This has proved to be especially
useful in incompressible-fluid turbulence, where the third-order exponent is
known to be $1$.] We give such ESS plots for structure functions in
Figs.~\ref{fig:10}(a)-(b) ($Run5a$), Figs.~\ref{fig:11}(a)-(b) ($Run5b$), and
Figs.~\ref{fig:12}(a)-(b) ($Run5c$); we then plot the resulting  multiscaling-exponent (Eqs.~\ref{eq:zetap}-\ref{eq:zetap12}) ratios in Figs.~\ref{fig:9}(e)-(f)
($Run4$), Fig.~\ref{fig:10}(c) ($Run5a$), Figs.~\ref{fig:11}(c)-(d) ($Run5b$),
and Figs.~\ref{fig:12}(c)-(d) ($Run5c$). We list our values for
multiscaling exponents (Eqs.~\ref{eq:zetap}-\ref{eq:zetap12}), along with error bars, 
in Tables ~\ref{table:2}-\ref{table:4}; these Tables provide a quantitative
summary of our results for these exponents. We note, at the qualitative level,
that prior experimental studies suggest that the magnetic-field structure functions 
in 3D HMHD turbulence exihibit (a) \textit{inertial} and \textit{intermediate-dissipation} 
scaling regions and (b) multiscaling is replaced by simple scaling in the second
of these ranges; our results are in consonance with these observations. \par

\begin{table*}
\hspace*{-0.5cm}
\begin{center}
\resizebox{!}{0.09\textheight}{
\begin{tabular} {c c c c c c c}
\hline
\hline
{\bf $p$} &
{\bf $\zeta_{p}^{u,Run3}$} &
{\bf $\zeta_{p}^{b,Run3}$} &
{\bf $\zeta_{p}^{u,Run4}$} &
{\bf $\zeta_{p}^{(b,1),Run4}$} &
{\bf $\zeta_{p}^{(b,2),Run4}$} \\ 
\hline
 $1$ &
 $0.35 \pm 0.06$ &
 $0.45 \pm 0.03$ &
  $0.34 \pm 0.04$ &
 $0.40 \pm 0.04$ &
 $0.71 \pm 0.06$ \\
 $2 $&
 $0.65 \pm 0.10$ &
 $0.81 \pm 0.07$ &
 $0.64 \pm 0.07$ &
 $0.76 \pm 0.10$ &
 $1.31 \pm 0.13$ \\
 $3$ &
 $0.90 \pm 0.13$ &
 $1.09 \pm 0.12$ &
 $0.90 \pm 0.06$ &
 $1.08 \pm 0.18$ &
 $1.83 \pm 0.16$ \\
 $4$ &
 $1.13 \pm 0.17$ &
 $1.32 \pm 0.15$ &
 $1.14 \pm 0.12$ &
 $1.34 \pm 0.28$ &
 $2.28 \pm 0.21$ \\
 $5$ &
 $1.35 \pm 0.22$ &
 $1.53 \pm 0.17$ &
 $1.36 \pm 0.32$ &
 $1.49 \pm 0.38$ &
 $2.69 \pm 0.39$ \\
 $6$ &
 $1.58 \pm 0.28$ &
 $1.73 \pm 0.18$ &
 $1.57 \pm 0.66$ &
 $1.53 \pm 0.50$ &
 $3.10 \pm 0.70$ \\
\hline
\hline
\end{tabular}%
}
\end{center}
\caption{List of the multiscaling exponents (Eqs.~\ref{eq:zetap}-\ref{eq:zetap12}), for $Run3$ 
and $Run4$. $Run4$ has two sets of exponents, one for the inertial range and the other 
for the intermediate-disspation range (see text).}
\label{table:2}
\end{table*}

\begin{table*}
\hspace*{-0.5cm}
\begin{center}
\resizebox{!}{0.09\textheight}{
\begin{tabular} {c c c c c c c }
\hline
\hline
{\bf $p$} &
{\bf $({\zeta_{p}^{u}}/{\zeta_{3}^{u}})^{Run3}$ } &
{\bf $({\zeta_{p}^{b}}/{\zeta_{3}^{b}})^{Run3}$ } &
{\bf $({\zeta_{p}^{u}}/{\zeta_{3}^{u}})^{Run4}$} &
{\bf $({\zeta_{p}^{b,1}}/{\zeta_{3}^{b,1}})^{Run4} $} &
{\bf $({\zeta_{p}^{b,2}}/{\zeta_{3}^{b,2}})^{Run4}$} \\ 
\hline
 $1$ &
 $0.39 \pm 0.01$ &
 $0.42 \pm 0.01$ &
 $0.38 \pm 0.03$ &
 $0.38 \pm 0.04$ &
 $0.39 \pm 0.02$ \\
 $2 $&
 $0.72 \pm 0.01$ &
 $0.75 \pm 0.01$ &
 $0.71 \pm 0.04$ &
 $0.71 \pm 0.04$ &
 $0.72 \pm 0.03$ \\
 $3$ &
 $1.0 \pm 0.00$ &
 $1.0 \pm 0.00$ &
 $1.0 \pm 0.00$ &
 $1.0 \pm 0.00$ &
 $1.0 \pm 0.00$\\
 $4$ &
 $1.27 \pm 0.01$ &
 $1.22 \pm 0.02$ &
 $1.26 \pm 0.13$ &
 $1.23 \pm 0.05$ &
 $1.25 \pm 0.08$ \\
 $5$ &
 $1.54 \pm 0.02$ &
 $1.42 \pm 0.04$ &
 $1.52 \pm 0.36$ &
 $1.37 \pm 0.12$ &
 $1.48 \pm 0.22$\\
 $6$ &
 $1.82 \pm 0.04$ &
 $1.61 \pm 0.07$ &
 $1.76 \pm 0.72$ &
 $1.39 \pm 0.24$ &
 $1.71 \pm 0.39$\\
\hline
\hline
\end{tabular}%
}
\end{center}
\caption{List of the ratios of multiscaling exponents (Eqs.~\ref{eq:zetap}-\ref{eq:zetap12}), 
for $Run3$ and $Run4$. $Run4$ has two sets of exponents, one for the inertial range and 
the other for the intermediate-disspation range (see text).}
\label{table:3}
\end{table*}
\begin{table*}
\hspace*{-0.5cm}
\begin{center}
\resizebox{!}{0.062\textheight}{
\begin{tabular} {c c c c c c c c c}
\hline
\hline
{\bf $p$} &
{\bf $({\zeta_{p}^{u}}/{\zeta_{3}^{u}})^{Run5a}$} &
{\bf $({\zeta_{p}^{b}}/{\zeta_{3}^{b}})^{Run5a}$} &
{\bf $({\zeta_{p}^{u}}/{\zeta_{3}^{u}})^{Run5b}$} &
{\bf $({\zeta_{p}^{b,1}}/{\zeta_{3}^{b,1}})^{Run5b}$} &
{\bf $({\zeta_{p}^{b,2}}/{\zeta_{3}^{b,2}})^{Run5b}$} &
{\bf $({\zeta_{p}^{u}}/{\zeta_{3}^{u}})^{Run5c}$} &
{\bf $({\zeta_{p}^{b,1}}/{\zeta_{3}^{b,1}})^{Run5c}$} &
{\bf $({\zeta_{p}^{b,2}}/{\zeta_{3}^{b,2}})^{Run5c}$} \\
\hline
 $1$ &
 $0.39 \pm 0.01$ &
 $0.40 \pm 0.01$ &
 $0.36 \pm 0.01$ &
 $0.35 \pm 0.01$ &
 $0.35 \pm 0.01$ &
 $0.36 \pm 0.01$ &
 $0.35 \pm 0.01$ &
 $0.34 \pm 0.01$ \\
 $2 $&
 $0.73 \pm 0.01$ &
 $0.73 \pm 0.01$ &
 $0.70 \pm 0.02$ &
 $0.69 \pm 0.01$ &
 $0.68 \pm 0.01$ &
 $0.69 \pm 0.01$ &
 $0.69 \pm 0.01$ &
 $0.67 \pm 0.01$ \\
 $3$ &
 $1.0 \pm 0.00$ &
 $1.0 \pm 0.00$ &
 $1.0 \pm 0.00$ &
 $1.0 \pm 0.00$ &
 $1.0 \pm 0.00$ &
 $1.0 \pm 0.00$ &
 $1.0 \pm 0.00$ &
 $1.0 \pm 0.00$\\
 $4$ &
 $1.21 \pm 0.02$ &
 $1.21 \pm 0.02$ &
 $1.28 \pm 0.04$ &
 $1.29 \pm 0.02$ &
 $1.30 \pm 0.01$ &
 $1.28 \pm 0.03$ &
 $1.29 \pm 0.03$ &
 $1.32 \pm 0.01$ \\
 $5$ &
 $1.36 \pm 0.05$ &
 $1.39 \pm 0.06$ &
 $1.55 \pm 0.12$ &
 $1.55 \pm 0.04$ &
 $1.60 \pm 0.04$ &
 $1.52 \pm 0.10$ &
 $1.56 \pm 0.08$ &
 $1.65 \pm 0.04$\\
 $6$ &
 $1.47 \pm 0.10$ &
 $1.54 \pm 0.11$ &
 $1.81 \pm 0.23$ &
 $1.79 \pm 0.09$ &
 $1.89 \pm 0.08$ &
 $1.73 \pm 0.20$ &
 $1.80 \pm 0.17$ &
 $1.99 \pm 0.09$\\
\hline
\hline
\end{tabular}%
}
\end{center}
\caption{List of the ratios of multiscaling exponents (Eqs.~\ref{eq:zetap}-\ref{eq:zetap12}), 
obtained via the ESS procedure (see text), for $Run5a$, $Run5b$, and $Run5c$. 
$Run5b$ and $Run5c$ have two sets of exponents, one for the inertial range and the other 
for the intermediate-disspation range (see text).}
\label{table:4}
\end{table*}
%

%
%


%
%
%

\begin{figure*}[t]
\centering 
\begin{tabular}{c c c}
\textbf{(a)} & \textbf{(b)} & \textbf{(c)} \\
\includegraphics [scale=0.4]{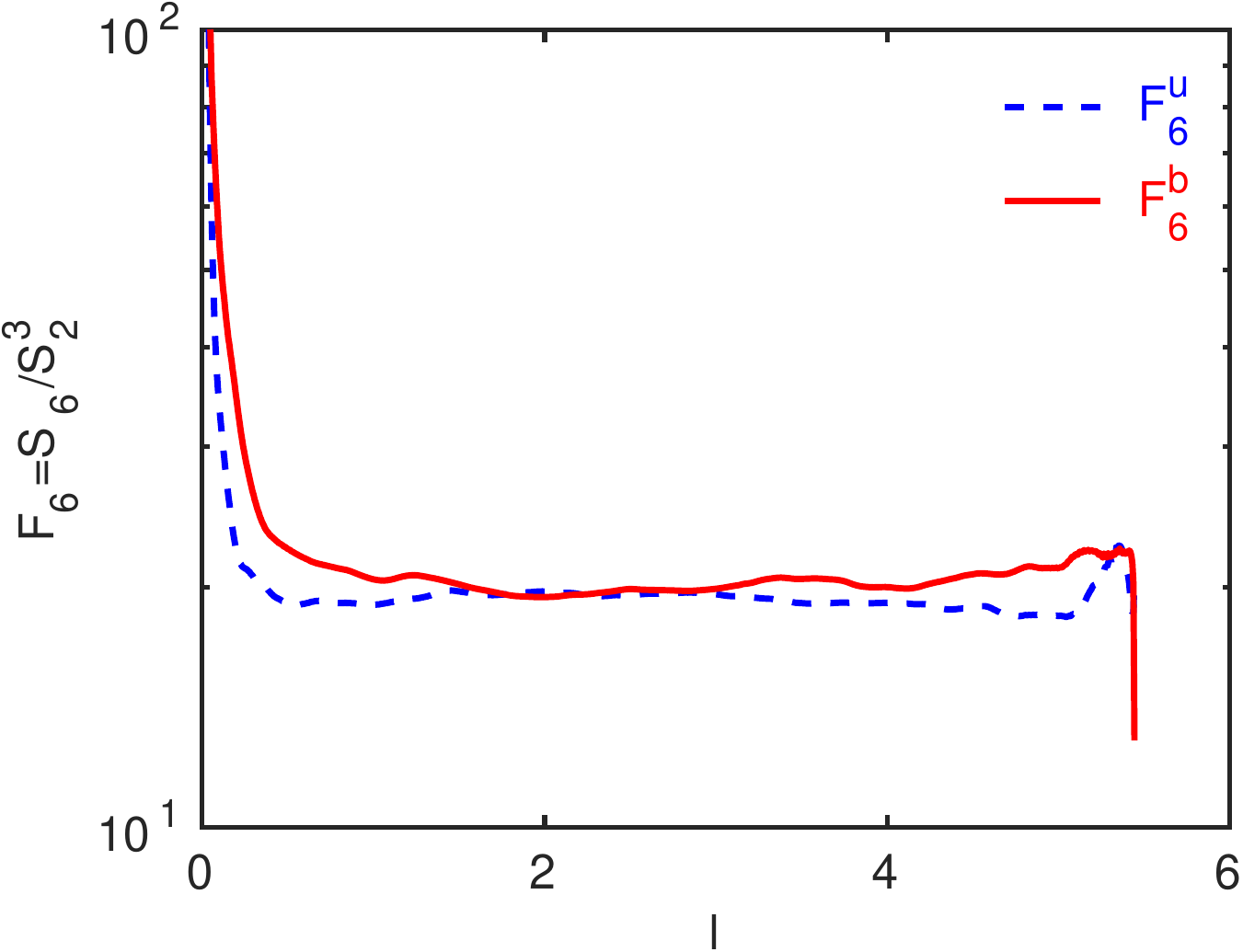} &
\includegraphics [scale=0.4]{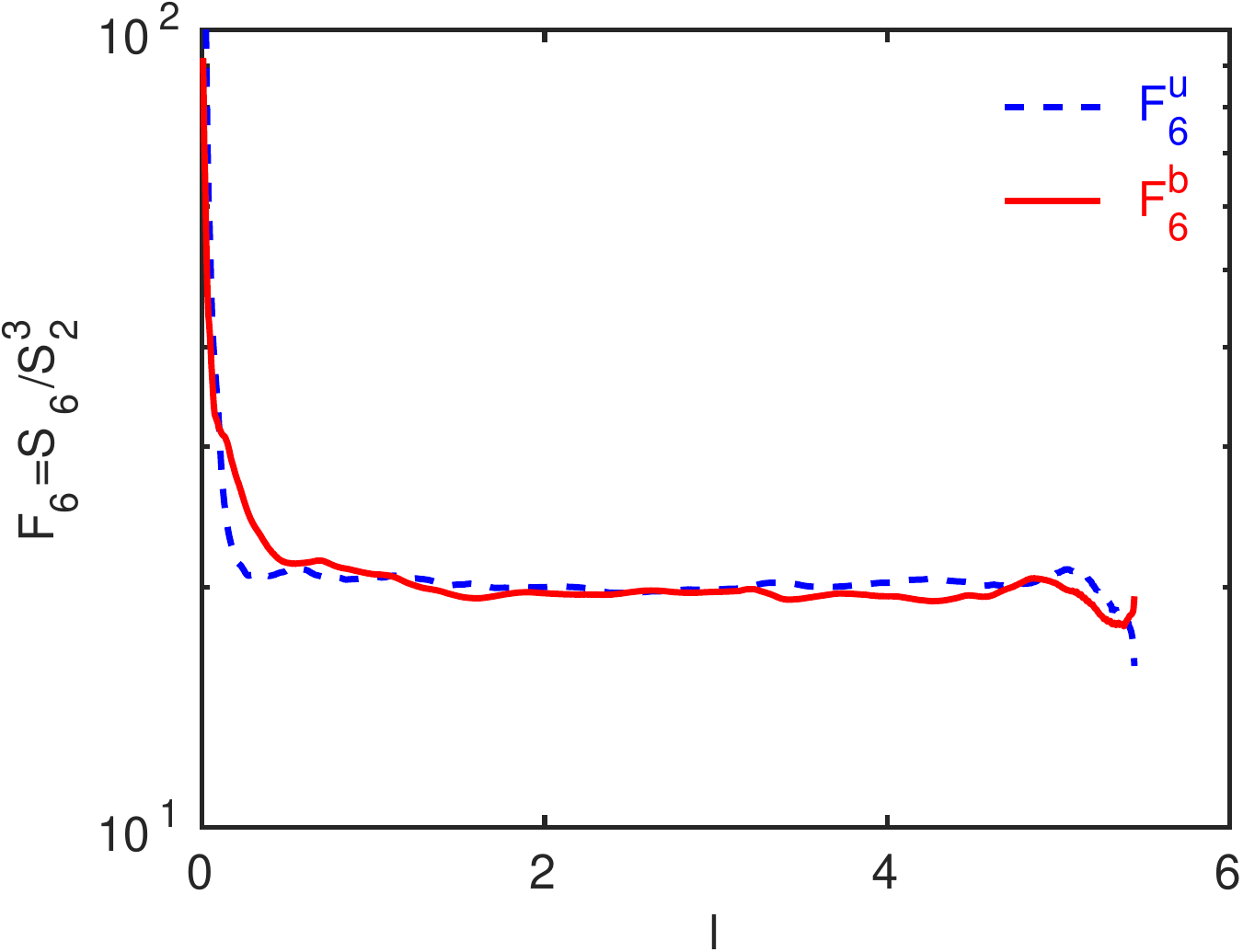} &
\includegraphics [scale=0.4]{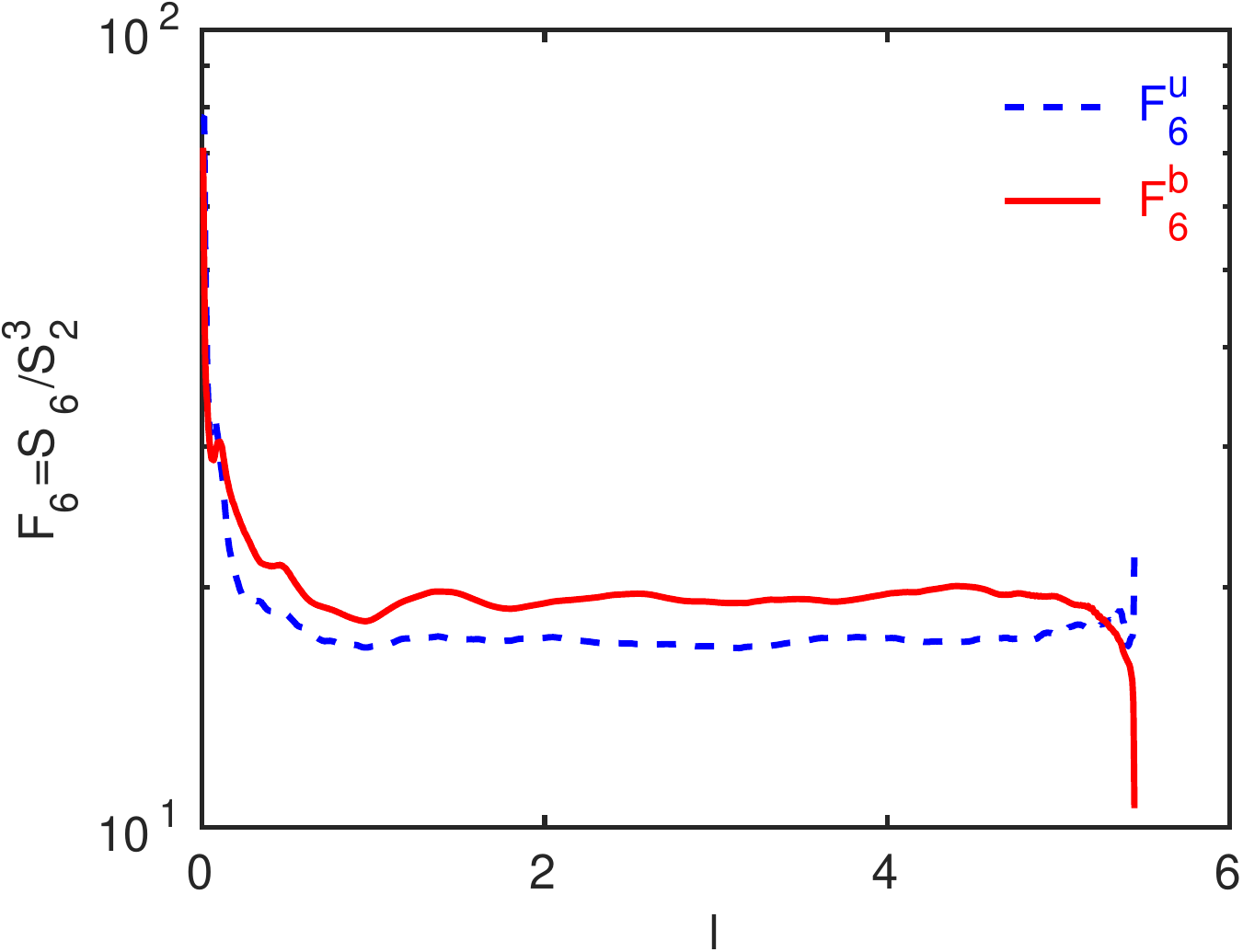}
\end{tabular}
\vskip -.1cm
\caption{Semilog plots of the hyperflatnesses (Eq.~\ref{eq:f6}), for the velocity and the 
magnetic fields, in (a) $Run5a$, (b) $Run5b$, and (c) $Run5c$.} 
\label{fig:13}
\end{figure*}

The length-scale dependence of the hyperflatness (Eq.~\ref{eq:f6}) is
often used to characterise small-scale intermittency. We present semilog plots 
of the hyperflatnesses (Eq.~\ref{eq:f6}), for the velocity and the magnetic 
fields, in $Run5a$, $Run5b$, and $Run5c$ Figs.~\ref{fig:13} (a), (b), and (c),
respectively. In these plots, we observe that these hyperflatnesses are more-or-less
flat, over a large range of $l$, but they increase rapidly, at small $l$;
this is the hallmark of small-scale intermittency.
%
%
%
\section{Conclusions}
\label{sec:conclusion}

Our study, which has been motivated by the uncertainties in spectral exponents
and the statistical properties of 3D HMHD turbulence, is a systematic
investigation of these properties by pseudospectral DNSs. Although many
numerical studies of 3D HMHD turbulence have been carried out earlier, none of
them has compared results from two different types of initial conditions
(initial condition A (Eq.~\ref{eq:initA} and $Run2 - Run4$) and  initial
condition B (Eq.~\ref{eq:initB} and $Run5b - Run5c$)) nor studied, in one work,
the different statistical properties we consider. (For similar studies of 3D
MHD turbulence we refer the reader to
Refs.~\cite{sahoo2011systematics,dallas2013structures}. Our work provides
valuable insights into the initial-condition dependence of the spectral
exponent $\alpha_1$ and the multiscaling
exponents (Eqs.~\ref{eq:zetap}-\ref{eq:zetap12}). We find clear evidence of inertial-
and intermediate-dissipation range scaling, with the value of the spectral
exponent $\alpha$ consistent with the K41 result $5/3$.  In the
intermediate-dissipation range, the value of $\alpha_1$ is consistent with  (A)
$11/3$, for the initial condition A, and (B) $7/3$, for the initial condition
B.  These different values can be attributed, in part, to the disparities in
the $k$ dependence of $E_{b}(k)/E_{u}(k)$ for these two initial conditions.
Our computations of the PDFs of the cosines of the angles between various
fields, such as the velocity ${\bf u}$ and $\bf{\omega}$ or ${\bf b}$ and ${\bf
j}$, help us to highlight the importance of the Hall term in suppressing the
tendency of alignment (or antialignment) of these fields for both the initial
conditions (A) and (B).  We carry out a careful exploration of intermittency in
3D HMHD turbulence by calculating the PDFs of the velocity and magnetic-field
increments as a function of the separation length scale $l$. We compute the
$l$-dependence of velocity and magnetic-field structure functions and,
therefrom, their order-$p$ multiscaling
exponents (Eqs.~\ref{eq:zetap}-\ref{eq:zetap12}). In the inertial
(intermediate-dissipation) range, we find clear signatures of multiscaling
(simple scaling), in consonance with solar-wind
results~\cite{kiyani2009global}.

\begin{acknowledgments} 
RP and SKY thank Nadia Bihari Padhan for discussions. SKY thanks the UGC-D.S. Kothari 
Postdoctoral Fellowship; RP and SKY thank CSIR and DST (India) 
for support and SERC (IISc) for computational resources. This research was partially supported by JSPS KAKENHI Grant Number 17K05734 and 20H00225, Japan.
The numerical simulations were performed on {\it Plasma Simulator} of NIFS with the support and under the auspices of the NIFS Collaboration Research program(NIFS20KNSSS133), as well as on the Oakforest-PACS supercomputer of the University of Tokyo, being partially supported by the Joint Usage/Research Center for Interdisciplinary Large-Scale Information Infrastructures in Japan (jh190006-NAJ, jh200002-NAH).
\end{acknowledgments} 
\appendix*
\section{Joint Probability Distribution Functions}
\begin{figure*}[t]
\centering 
\begin{tabular}{c c}
\textbf{(a)} & \textbf{(b)} \\
\includegraphics [scale=0.5]{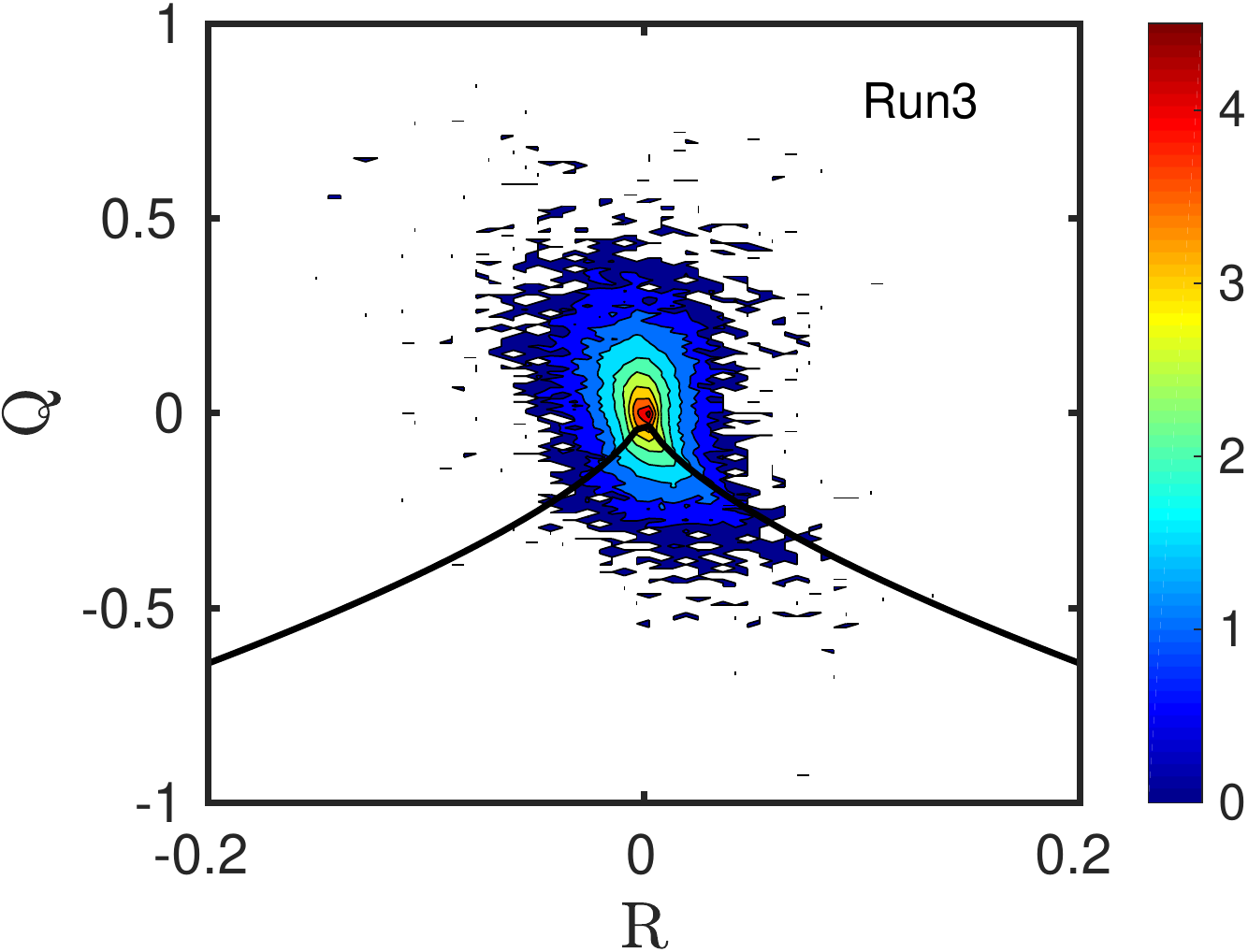} &
\includegraphics [scale=0.5]{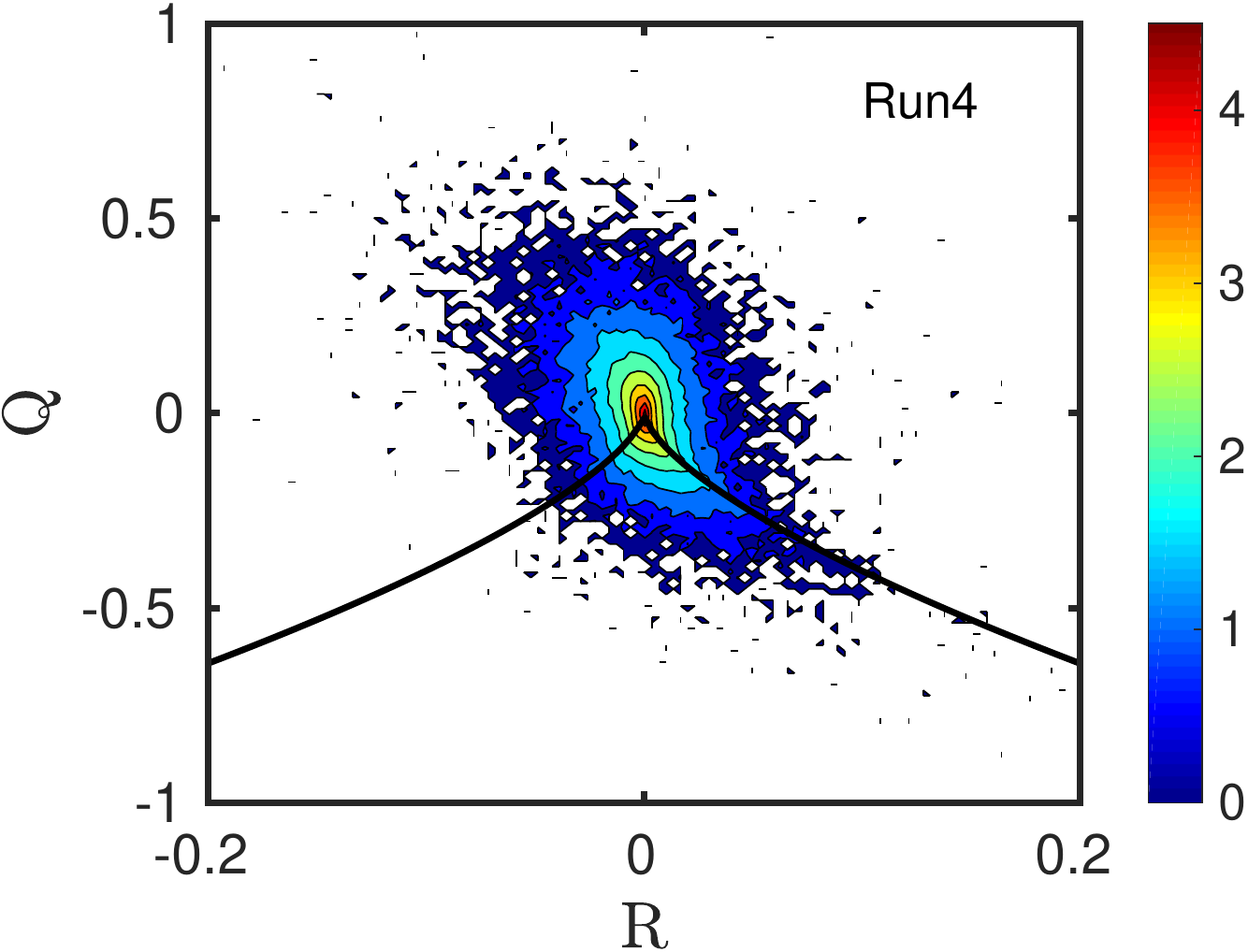}
\end{tabular}
\begin{tabular}{c c}
\textbf{(c)} & \textbf{(d)} \\
\includegraphics [scale=0.5]{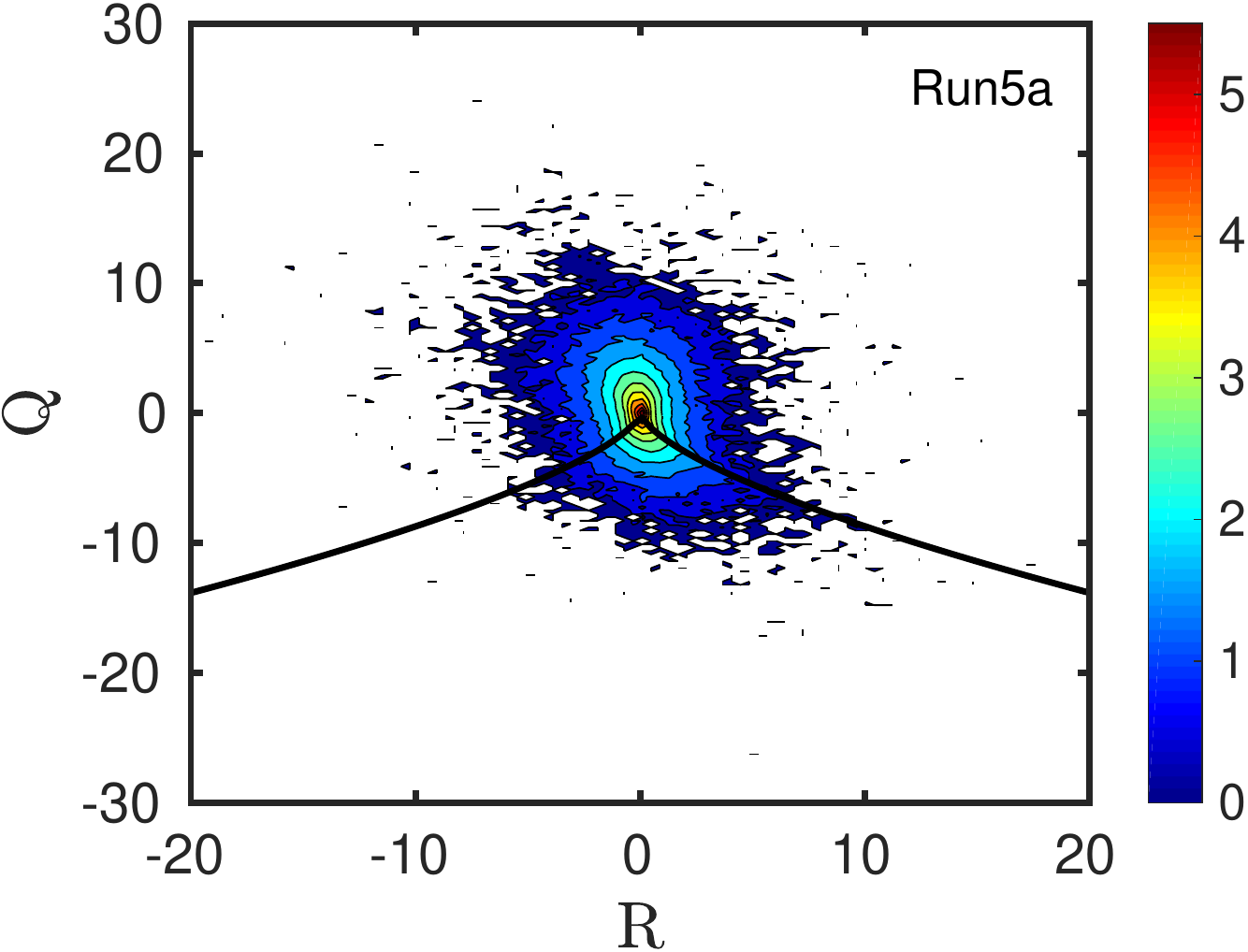} &
\includegraphics [scale=0.5]{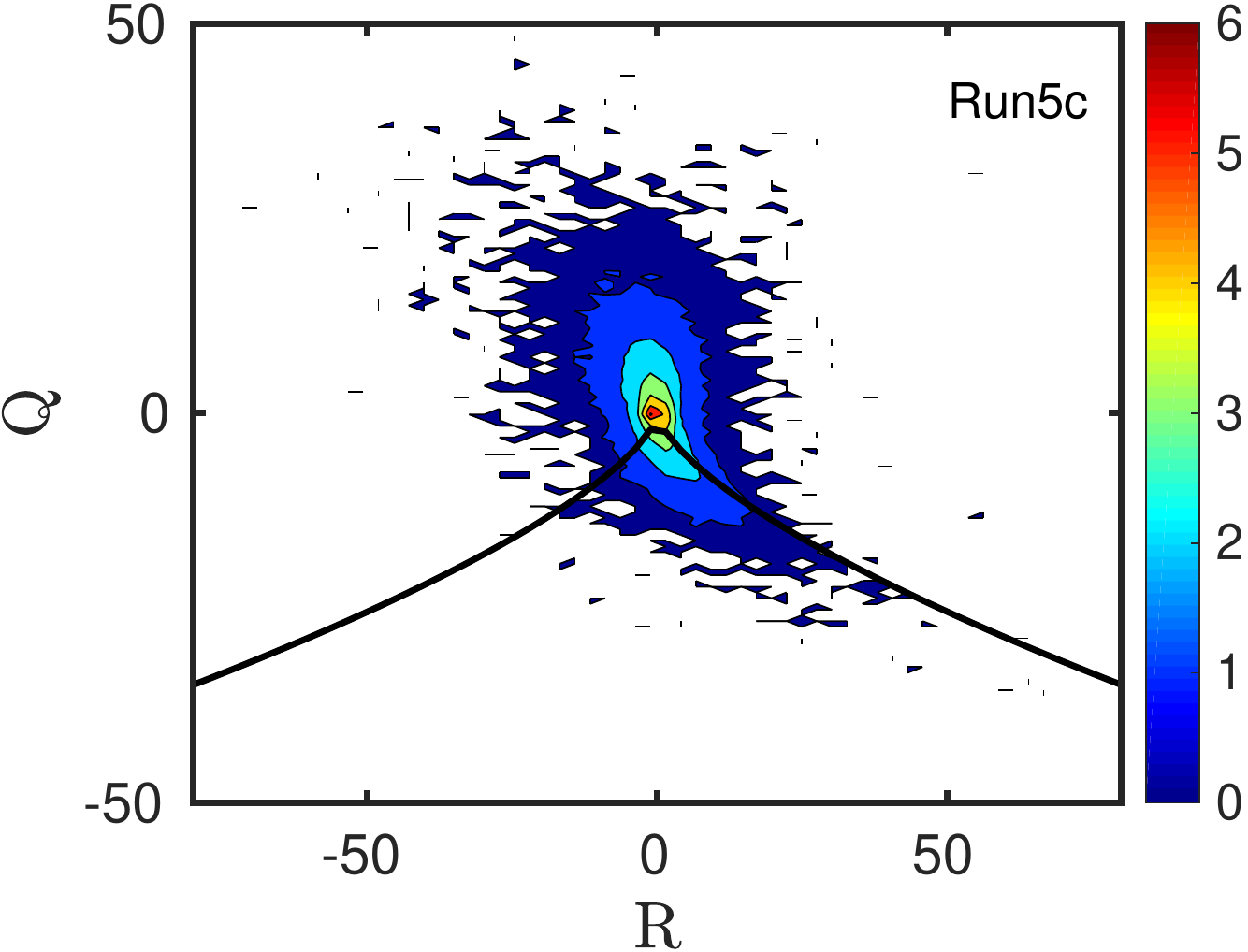} 
\end{tabular}
\vskip -.1cm
\caption{Color-contour plots of the JPDFs of $Q$ and $R$ for $\left(a\right)$ $Run3$, 
$\left(b\right)$ $Run4$, $\left(c\right)$ $Run5a$, and $\left(d\right)$ $Run5c$.}
\label{fig:14}
\end{figure*}

We display color-contour-plots of the joint probability distribution functions (JPDFs) 
of $Q$ and $R$, at the cascade-completion time $\tau_{c}$, in Fig.~\ref{fig:14} from 
$Run3$, $Run4$, $Run5a$ and $Run5c$, with $Q$ and $R$ the following, well-known 
invariants (see, e.g., Ref.~\cite{sahoo2011systematics}) for an ideal, incompressible 
fluid:  
\begin{eqnarray}
Q &=&-\frac{1}{2}tr(A^{2}); \nonumber \\ 
R &=&-\frac{1}{3}tr(A^{3});
\end{eqnarray}
here, $A$ is the velocity-derivative tensor with components 
$A_{ij}=\partial_{i}u_{j}$; the zero-discriminant line 
$D = \frac{27}{4}R^{2}+Q^{3}=0$ is shown by black curves in these plots. 
We observe that the characteristic, tear-drop shape of these JPDFS of $Q$ and $R$ 
are more prominent in 3D HMHD turbulence ($Run4$ and $Run5c$) than in their 3D MHD
counterparts ($Run3$ and $Run5a$); and the tails of these JPDFs are more elongated
in 3D HMHD turbulence ($Run4$ and $Run5c$) than in their 3D MHD
counterparts ($Run3$ and $Run5a$).
%
%
%

\begin{figure*}[t]
\centering 
\begin{tabular}{c c}
\textbf{(a)} & \textbf{(b)} \\
\includegraphics [scale=0.5]{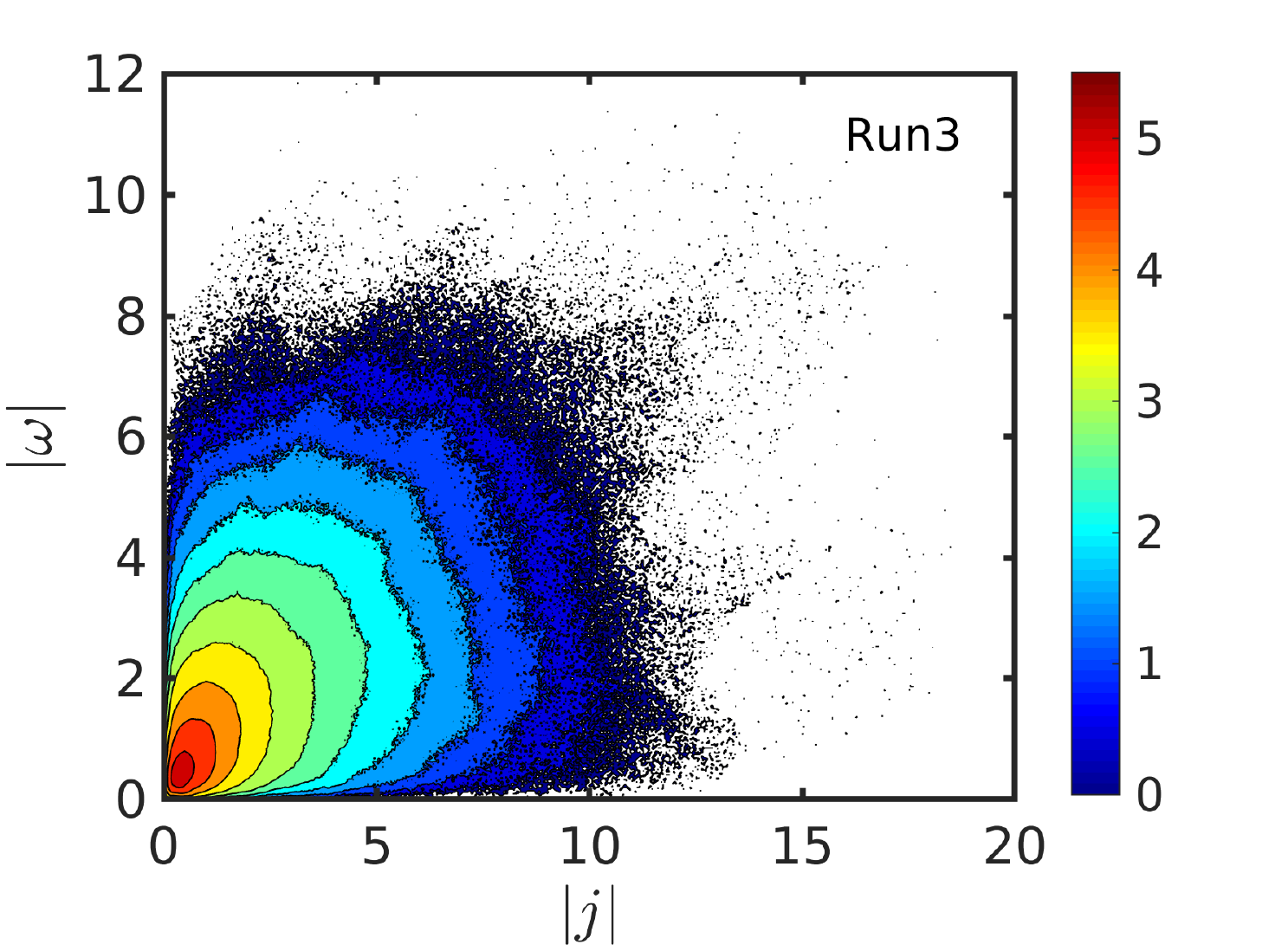} &
\includegraphics [scale=0.5]{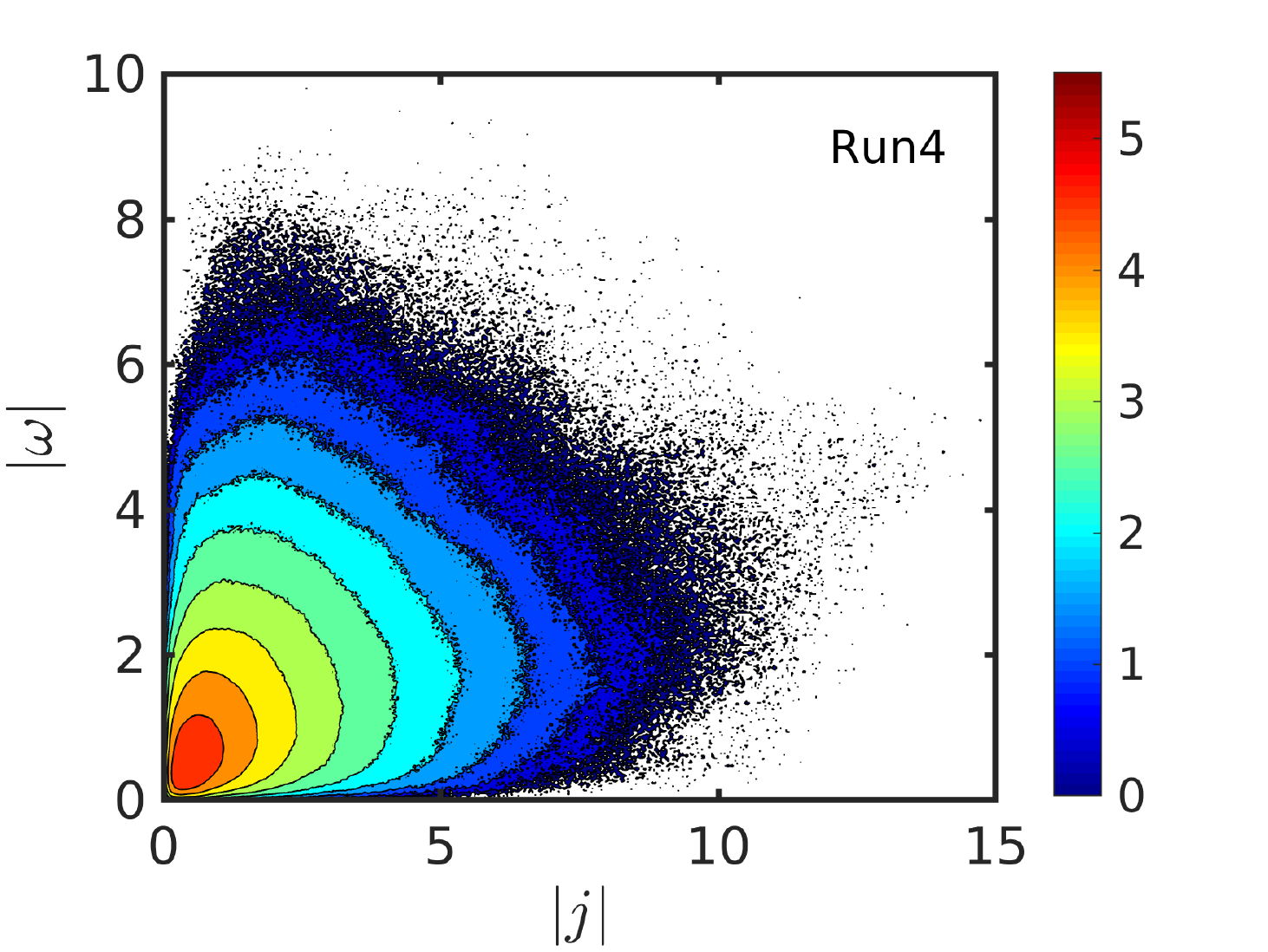}
\end{tabular}
\begin{tabular}{c c}
\textbf{(c)} & \textbf{(d)} \\
\includegraphics [scale=0.5]{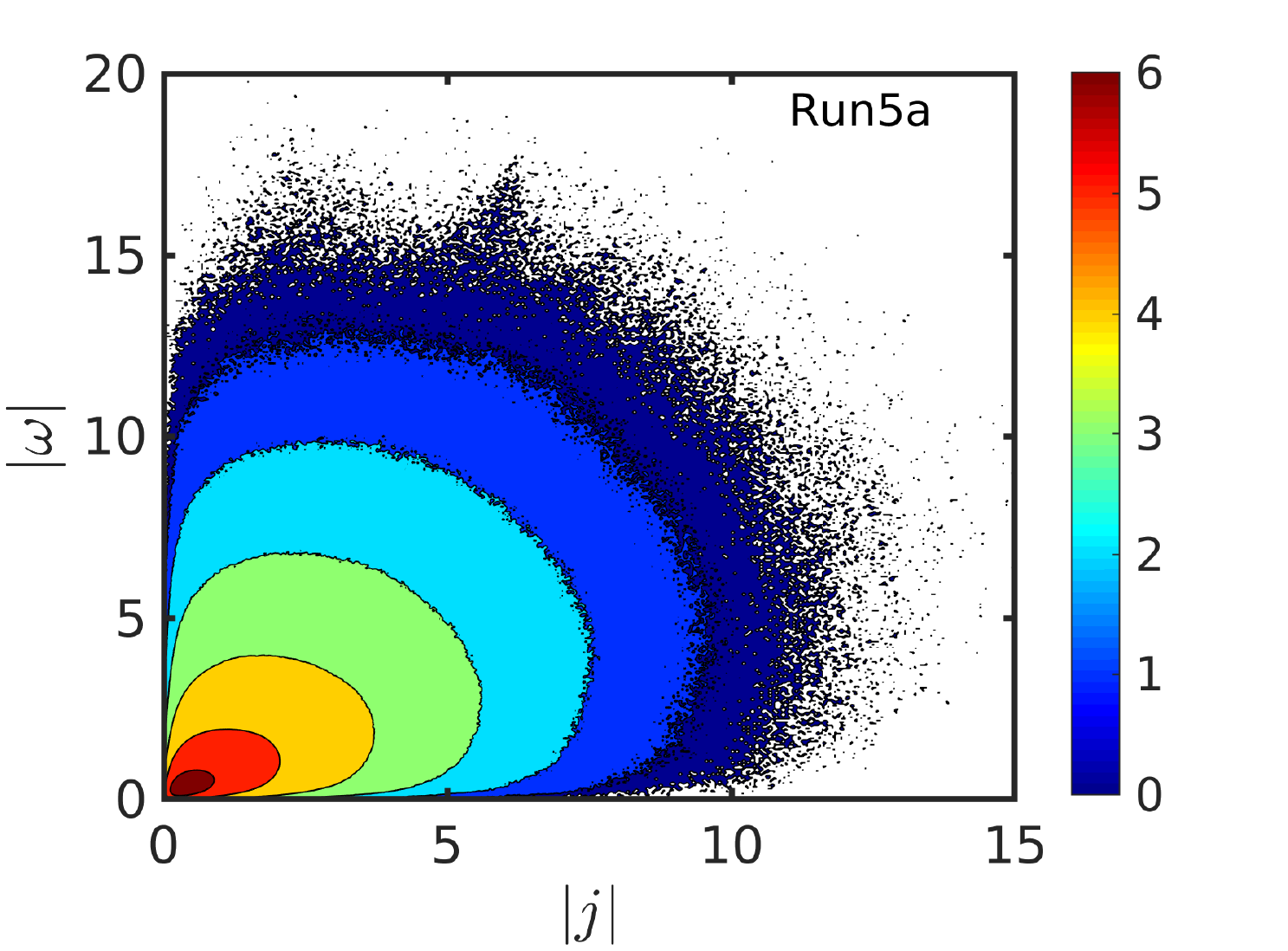} &
\includegraphics [scale=0.5]{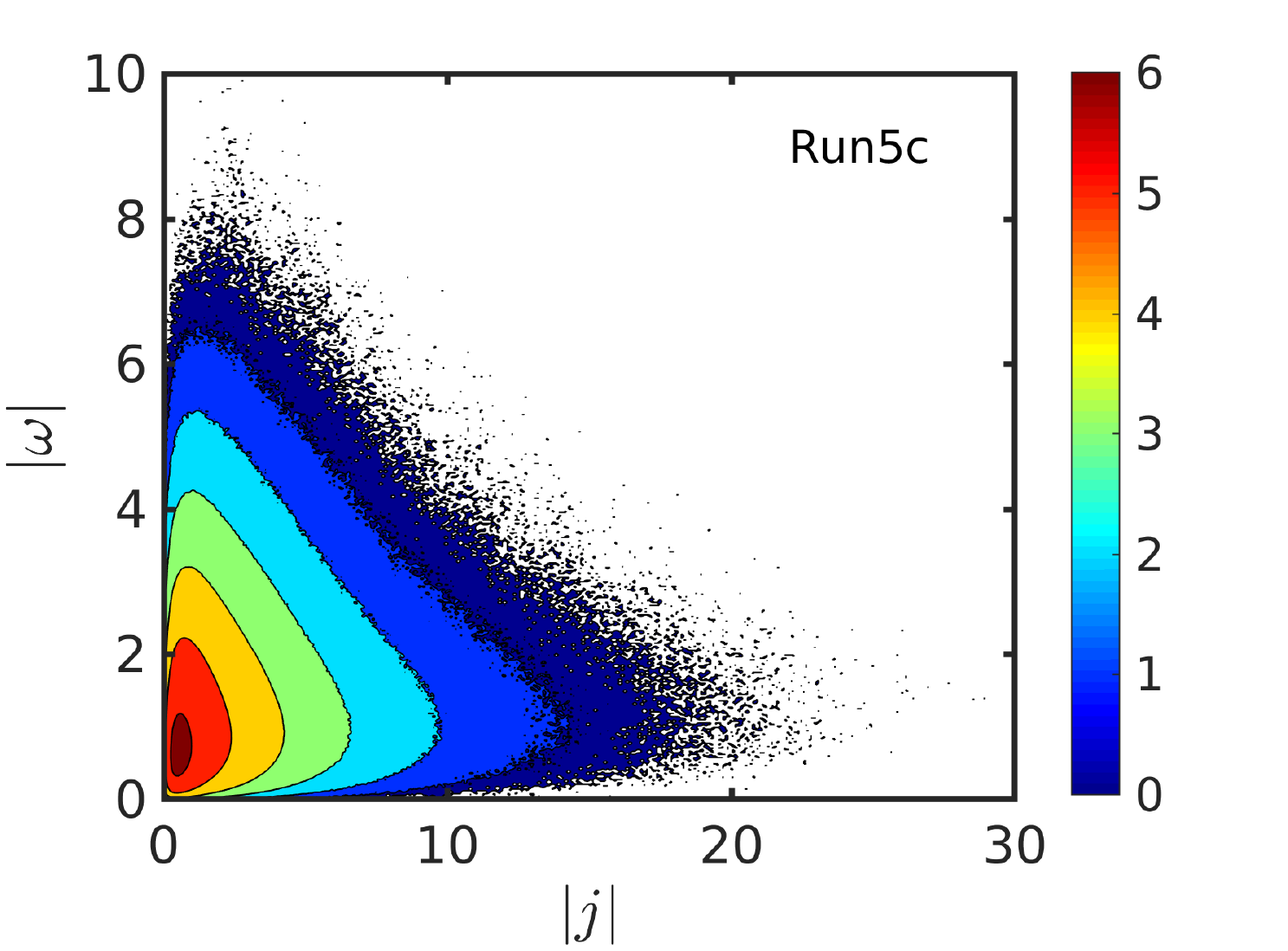} 
\end{tabular}
\vskip -.1cm
\caption{Color-contour plots of the JPDFs of $\omega$ and $j$ for $\left(a\right)$ $Run3$, 
$\left(b\right)$ $Run4$, $\left(c\right)$ $Run5a$, and $\left(d\right)$ $Run5c$.}
\label{fig:15}
\end{figure*}

In Fig.~\ref{fig:15} we show plots of JPDFs of $\omega$ and $j$, the moduli of the vorticity and the current, at the
cascade-completion time $\tau_{c}$, from $Run3$, $Run4$, $Run5a$ and $Run5c$.
We observe that in 3D MHD turbulence ($Run3$ and $Run5a$) the outer region of the JPDFs
of $\omega$ and $j$ show curved, roughly circular, contours, whereas in
in 3D HMHD turbulence ($Run4$ and $Run5c$) the counterparts of these contours are
flattened and suppressed significantly; furthermore, there is an elongation of the JPDFs 
of $\omega$ and $j$ along the axes, at low values of $\omega$ and $j$. 

%
%
\bibliography{mybib}

\begin{thebibliography}{69}%
\makeatletter
\providecommand \@ifxundefined [1]{%
 \@ifx{#1\undefined}
}%
\providecommand \@ifnum [1]{%
 \ifnum #1\expandafter \@firstoftwo
 \else \expandafter \@secondoftwo
 \fi
}%
\providecommand \@ifx [1]{%
 \ifx #1\expandafter \@firstoftwo
 \else \expandafter \@secondoftwo
 \fi
}%
\providecommand \natexlab [1]{#1}%
\providecommand \enquote  [1]{``#1''}%
\providecommand \bibnamefont  [1]{#1}%
\providecommand \bibfnamefont [1]{#1}%
\providecommand \citenamefont [1]{#1}%
\providecommand \href@noop [0]{\@secondoftwo}%
\providecommand \href [0]{\begingroup \@sanitize@url \@href}%
\providecommand \@href[1]{\@@startlink{#1}\@@href}%
\providecommand \@@href[1]{\endgroup#1\@@endlink}%
\providecommand \@sanitize@url [0]{\catcode `\\12\catcode `\$12\catcode
  `\&12\catcode `\#12\catcode `\^12\catcode `\_12\catcode `\%12\relax}%
\providecommand \@@startlink[1]{}%
\providecommand \@@endlink[0]{}%
\providecommand \url  [0]{\begingroup\@sanitize@url \@url }%
\providecommand \@url [1]{\endgroup\@href {#1}{\urlprefix }}%
\providecommand \urlprefix  [0]{URL }%
\providecommand \Eprint [0]{\href }%
\providecommand \doibase [0]{https://doi.org/}%
\providecommand \selectlanguage [0]{\@gobble}%
\providecommand \bibinfo  [0]{\@secondoftwo}%
\providecommand \bibfield  [0]{\@secondoftwo}%
\providecommand \translation [1]{[#1]}%
\providecommand \BibitemOpen [0]{}%
\providecommand \bibitemStop [0]{}%
\providecommand \bibitemNoStop [0]{.\EOS\space}%
\providecommand \EOS [0]{\spacefactor3000\relax}%
\providecommand \BibitemShut  [1]{\csname bibitem#1\endcsname}%
\let\auto@bib@innerbib\@empty
\bibitem [{\citenamefont {Choudhuri}\ \emph {et~al.}(1998)\citenamefont
  {Choudhuri} \emph {et~al.}}]{choudhuri1998physics}%
  \BibitemOpen
  \bibfield  {author} {\bibinfo {author} {\bibfnamefont {A.~R.}\ \bibnamefont
  {Choudhuri}} \emph {et~al.},\ }\href@noop {} {\emph {\bibinfo {title} {The
  physics of fluids and plasmas: an introduction for astrophysicists}}}\
  (\bibinfo  {publisher} {Cambridge University Press},\ \bibinfo {year}
  {1998})\BibitemShut {NoStop}%
\bibitem [{\citenamefont {Krishan}(1999)}]{krishan1999astrophysical}%
  \BibitemOpen
  \bibfield  {author} {\bibinfo {author} {\bibfnamefont {V.}~\bibnamefont
  {Krishan}},\ }\href@noop {} {\emph {\bibinfo {title} {Astrophysical plasmas
  and fluids}}},\ Vol.\ \bibinfo {volume} {235}\ (\bibinfo  {publisher}
  {Springer Science \& Business Media},\ \bibinfo {year} {1999})\BibitemShut
  {NoStop}%
\bibitem [{\citenamefont {R{\"u}diger}\ and\ \citenamefont
  {Hollerbach}(2006)}]{rudiger2006magnetic}%
  \BibitemOpen
  \bibfield  {author} {\bibinfo {author} {\bibfnamefont {G.}~\bibnamefont
  {R{\"u}diger}}\ and\ \bibinfo {author} {\bibfnamefont {R.}~\bibnamefont
  {Hollerbach}},\ }\href@noop {} {\emph {\bibinfo {title} {The magnetic
  universe: geophysical and astrophysical dynamo theory}}}\ (\bibinfo
  {publisher} {John Wiley \& Sons},\ \bibinfo {year} {2006})\BibitemShut
  {NoStop}%
\bibitem [{\citenamefont {Goedbloed}\ \emph {et~al.}(2004)\citenamefont
  {Goedbloed}, \citenamefont {Goedbloed},\ and\ \citenamefont
  {Poedts}}]{goedbloed2004principles}%
  \BibitemOpen
  \bibfield  {author} {\bibinfo {author} {\bibfnamefont {J.~H.}\ \bibnamefont
  {Goedbloed}}, \bibinfo {author} {\bibfnamefont {J.}~\bibnamefont
  {Goedbloed}},\ and\ \bibinfo {author} {\bibfnamefont {S.}~\bibnamefont
  {Poedts}},\ }\href@noop {} {\emph {\bibinfo {title} {Principles of
  magnetohydrodynamics: with applications to laboratory and astrophysical
  plasmas}}}\ (\bibinfo  {publisher} {Cambridge university press},\ \bibinfo
  {year} {2004})\BibitemShut {NoStop}%
\bibitem [{\citenamefont {Biskamp}(2003)}]{biskamp2003magnetohydrodynamic}%
  \BibitemOpen
  \bibfield  {author} {\bibinfo {author} {\bibfnamefont {D.}~\bibnamefont
  {Biskamp}},\ }\href@noop {} {\emph {\bibinfo {title} {Magnetohydrodynamic
  turbulence}}}\ (\bibinfo  {publisher} {Cambridge University Press},\ \bibinfo
  {year} {2003})\BibitemShut {NoStop}%
\bibitem [{\citenamefont {Verma}(2004)}]{verma2004statistical}%
  \BibitemOpen
  \bibfield  {author} {\bibinfo {author} {\bibfnamefont {M.~K.}\ \bibnamefont
  {Verma}},\ }\bibfield  {title} {\bibinfo {title} {Statistical theory of
  magnetohydrodynamic turbulence: recent results},\ }\href@noop {} {\bibfield
  {journal} {\bibinfo  {journal} {Physics Reports}\ }\textbf {\bibinfo {volume}
  {401}},\ \bibinfo {pages} {229} (\bibinfo {year} {2004})}\BibitemShut
  {NoStop}%
\bibitem [{\citenamefont {Davidson}(2002)}]{davidson2002introduction}%
  \BibitemOpen
  \bibfield  {author} {\bibinfo {author} {\bibfnamefont {P.~A.}\ \bibnamefont
  {Davidson}},\ }\href@noop {} {\bibinfo {title} {An introduction to
  magnetohydrodynamics}} (\bibinfo {year} {2002})\BibitemShut {NoStop}%
\bibitem [{\citenamefont {Sahoo}\ \emph {et~al.}(2011)\citenamefont {Sahoo},
  \citenamefont {Perlekar},\ and\ \citenamefont
  {Pandit}}]{sahoo2011systematics}%
  \BibitemOpen
  \bibfield  {author} {\bibinfo {author} {\bibfnamefont {G.}~\bibnamefont
  {Sahoo}}, \bibinfo {author} {\bibfnamefont {P.}~\bibnamefont {Perlekar}},\
  and\ \bibinfo {author} {\bibfnamefont {R.}~\bibnamefont {Pandit}},\
  }\bibfield  {title} {\bibinfo {title} {Systematics of the
  magnetic-prandtl-number dependence of homogeneous, isotropic
  magnetohydrodynamic turbulence},\ }\href@noop {} {\bibfield  {journal}
  {\bibinfo  {journal} {New Journal of Physics}\ }\textbf {\bibinfo {volume}
  {13}},\ \bibinfo {pages} {013036} (\bibinfo {year} {2011})}\BibitemShut
  {NoStop}%
\bibitem [{\citenamefont {Basu}\ \emph {et~al.}(2014)\citenamefont {Basu},
  \citenamefont {Naji},\ and\ \citenamefont {Pandit}}]{basu2014structure}%
  \BibitemOpen
  \bibfield  {author} {\bibinfo {author} {\bibfnamefont {A.}~\bibnamefont
  {Basu}}, \bibinfo {author} {\bibfnamefont {A.}~\bibnamefont {Naji}},\ and\
  \bibinfo {author} {\bibfnamefont {R.}~\bibnamefont {Pandit}},\ }\bibfield
  {title} {\bibinfo {title} {Structure-function hierarchies and von
  k{\'a}rm{\'a}n--howarth relations for turbulence in magnetohydrodynamical
  equations},\ }\href@noop {} {\bibfield  {journal} {\bibinfo  {journal}
  {Physical Review E}\ }\textbf {\bibinfo {volume} {89}},\ \bibinfo {pages}
  {012117} (\bibinfo {year} {2014})}\BibitemShut {NoStop}%
\bibitem [{\citenamefont {Moffatt}\ and\ \citenamefont
  {Dormy}(2019)}]{moffatt2019self}%
  \BibitemOpen
  \bibfield  {author} {\bibinfo {author} {\bibfnamefont {K.}~\bibnamefont
  {Moffatt}}\ and\ \bibinfo {author} {\bibfnamefont {E.}~\bibnamefont
  {Dormy}},\ }\href@noop {} {\emph {\bibinfo {title} {Self-exciting fluid
  dynamos}}},\ Vol.~\bibinfo {volume} {59}\ (\bibinfo  {publisher} {Cambridge
  University Press},\ \bibinfo {year} {2019})\BibitemShut {NoStop}%
\bibitem [{\citenamefont {Roberts}\ and\ \citenamefont
  {Glatzmaier}(2000)}]{roberts2000geodynamo}%
  \BibitemOpen
  \bibfield  {author} {\bibinfo {author} {\bibfnamefont {P.~H.}\ \bibnamefont
  {Roberts}}\ and\ \bibinfo {author} {\bibfnamefont {G.~A.}\ \bibnamefont
  {Glatzmaier}},\ }\bibfield  {title} {\bibinfo {title} {Geodynamo theory and
  simulations},\ }\href@noop {} {\bibfield  {journal} {\bibinfo  {journal}
  {Reviews of modern physics}\ }\textbf {\bibinfo {volume} {72}},\ \bibinfo
  {pages} {1081} (\bibinfo {year} {2000})}\BibitemShut {NoStop}%
\bibitem [{\citenamefont {Shew}\ and\ \citenamefont
  {Lathrop}(2005)}]{shew2005liquid}%
  \BibitemOpen
  \bibfield  {author} {\bibinfo {author} {\bibfnamefont {W.~L.}\ \bibnamefont
  {Shew}}\ and\ \bibinfo {author} {\bibfnamefont {D.~P.}\ \bibnamefont
  {Lathrop}},\ }\bibfield  {title} {\bibinfo {title} {Liquid sodium model of
  geophysical core convection},\ }\href@noop {} {\bibfield  {journal} {\bibinfo
   {journal} {Physics of the Earth and Planetary Interiors}\ }\textbf {\bibinfo
  {volume} {153}},\ \bibinfo {pages} {136} (\bibinfo {year}
  {2005})}\BibitemShut {NoStop}%
\bibitem [{\citenamefont {Monchaux}\ \emph {et~al.}(2007)\citenamefont
  {Monchaux}, \citenamefont {Berhanu}, \citenamefont {Bourgoin}, \citenamefont
  {Moulin}, \citenamefont {Odier}, \citenamefont {Pinton}, \citenamefont
  {Volk}, \citenamefont {Fauve}, \citenamefont {Mordant}, \citenamefont
  {P{\'e}tr{\'e}lis} \emph {et~al.}}]{monchaux2007generation}%
  \BibitemOpen
  \bibfield  {author} {\bibinfo {author} {\bibfnamefont {R.}~\bibnamefont
  {Monchaux}}, \bibinfo {author} {\bibfnamefont {M.}~\bibnamefont {Berhanu}},
  \bibinfo {author} {\bibfnamefont {M.}~\bibnamefont {Bourgoin}}, \bibinfo
  {author} {\bibfnamefont {M.}~\bibnamefont {Moulin}}, \bibinfo {author}
  {\bibfnamefont {P.}~\bibnamefont {Odier}}, \bibinfo {author} {\bibfnamefont
  {J.-F.}\ \bibnamefont {Pinton}}, \bibinfo {author} {\bibfnamefont
  {R.}~\bibnamefont {Volk}}, \bibinfo {author} {\bibfnamefont {S.}~\bibnamefont
  {Fauve}}, \bibinfo {author} {\bibfnamefont {N.}~\bibnamefont {Mordant}},
  \bibinfo {author} {\bibfnamefont {F.}~\bibnamefont {P{\'e}tr{\'e}lis}}, \emph
  {et~al.},\ }\bibfield  {title} {\bibinfo {title} {Generation of a magnetic
  field by dynamo action in a turbulent flow of liquid sodium},\ }\href@noop {}
  {\bibfield  {journal} {\bibinfo  {journal} {Physical review letters}\
  }\textbf {\bibinfo {volume} {98}},\ \bibinfo {pages} {044502} (\bibinfo
  {year} {2007})}\BibitemShut {NoStop}%
\bibitem [{\citenamefont {Salem}\ \emph {et~al.}(2009)\citenamefont {Salem},
  \citenamefont {Mangeney}, \citenamefont {Bale},\ and\ \citenamefont
  {Veltri}}]{salem2009solar}%
  \BibitemOpen
  \bibfield  {author} {\bibinfo {author} {\bibfnamefont {C.}~\bibnamefont
  {Salem}}, \bibinfo {author} {\bibfnamefont {A.}~\bibnamefont {Mangeney}},
  \bibinfo {author} {\bibfnamefont {S.}~\bibnamefont {Bale}},\ and\ \bibinfo
  {author} {\bibfnamefont {P.}~\bibnamefont {Veltri}},\ }\bibfield  {title}
  {\bibinfo {title} {Solar wind magnetohydrodynamics turbulence: anomalous
  scaling and role of intermittency},\ }\href@noop {} {\bibfield  {journal}
  {\bibinfo  {journal} {The Astrophysical Journal}\ }\textbf {\bibinfo {volume}
  {702}},\ \bibinfo {pages} {537} (\bibinfo {year} {2009})}\BibitemShut
  {NoStop}%
\bibitem [{\citenamefont {Podesta}\ \emph {et~al.}(2007)\citenamefont
  {Podesta}, \citenamefont {Roberts},\ and\ \citenamefont
  {Goldstein}}]{podesta2007spectral}%
  \BibitemOpen
  \bibfield  {author} {\bibinfo {author} {\bibfnamefont {J.}~\bibnamefont
  {Podesta}}, \bibinfo {author} {\bibfnamefont {D.}~\bibnamefont {Roberts}},\
  and\ \bibinfo {author} {\bibfnamefont {M.}~\bibnamefont {Goldstein}},\
  }\bibfield  {title} {\bibinfo {title} {Spectral exponents of kinetic and
  magnetic energy spectra in solar wind turbulence},\ }\href@noop {} {\bibfield
   {journal} {\bibinfo  {journal} {The Astrophysical Journal}\ }\textbf
  {\bibinfo {volume} {664}},\ \bibinfo {pages} {543} (\bibinfo {year}
  {2007})}\BibitemShut {NoStop}%
\bibitem [{\citenamefont {Podesta}\ \emph {et~al.}(2009)\citenamefont
  {Podesta}, \citenamefont {Chandran}, \citenamefont {Bhattacharjee},
  \citenamefont {Roberts},\ and\ \citenamefont {Goldstein}}]{podesta2009scale}%
  \BibitemOpen
  \bibfield  {author} {\bibinfo {author} {\bibfnamefont {J.}~\bibnamefont
  {Podesta}}, \bibinfo {author} {\bibfnamefont {B.~D.}\ \bibnamefont
  {Chandran}}, \bibinfo {author} {\bibfnamefont {A.}~\bibnamefont
  {Bhattacharjee}}, \bibinfo {author} {\bibfnamefont {D.}~\bibnamefont
  {Roberts}},\ and\ \bibinfo {author} {\bibfnamefont {M.}~\bibnamefont
  {Goldstein}},\ }\bibfield  {title} {\bibinfo {title} {Scale-dependent angle
  of alignment between velocity and magnetic field fluctuations in solar wind
  turbulence},\ }\href@noop {} {\bibfield  {journal} {\bibinfo  {journal}
  {Journal of Geophysical Research: Space Physics}\ }\textbf {\bibinfo {volume}
  {114}} (\bibinfo {year} {2009})}\BibitemShut {NoStop}%
\bibitem [{\citenamefont {Weygand}\ \emph {et~al.}(2007)\citenamefont
  {Weygand}, \citenamefont {Matthaeus}, \citenamefont {Dasso}, \citenamefont
  {Kivelson},\ and\ \citenamefont {Walker}}]{weygand2007taylor}%
  \BibitemOpen
  \bibfield  {author} {\bibinfo {author} {\bibfnamefont {J.~M.}\ \bibnamefont
  {Weygand}}, \bibinfo {author} {\bibfnamefont {W.}~\bibnamefont {Matthaeus}},
  \bibinfo {author} {\bibfnamefont {S.}~\bibnamefont {Dasso}}, \bibinfo
  {author} {\bibfnamefont {M.}~\bibnamefont {Kivelson}},\ and\ \bibinfo
  {author} {\bibfnamefont {R.}~\bibnamefont {Walker}},\ }\bibfield  {title}
  {\bibinfo {title} {Taylor scale and effective magnetic reynolds number
  determination from plasma sheet and solar wind magnetic field fluctuations},\
  }\href@noop {} {\bibfield  {journal} {\bibinfo  {journal} {Journal of
  Geophysical Research: Space Physics}\ }\textbf {\bibinfo {volume} {112}}
  (\bibinfo {year} {2007})}\BibitemShut {NoStop}%
\bibitem [{\citenamefont {Pouquet}\ \emph
  {et~al.}(2020{\natexlab{a}})\citenamefont {Pouquet}, \citenamefont
  {Rosenberg},\ and\ \citenamefont {Stawarz}}]{pouquet2020interplay}%
  \BibitemOpen
  \bibfield  {author} {\bibinfo {author} {\bibfnamefont {A.}~\bibnamefont
  {Pouquet}}, \bibinfo {author} {\bibfnamefont {D.}~\bibnamefont {Rosenberg}},\
  and\ \bibinfo {author} {\bibfnamefont {J.~E.}\ \bibnamefont {Stawarz}},\
  }\bibfield  {title} {\bibinfo {title} {Interplay between turbulence and
  waves:large-scale helical transfer, and small-scale dissipation and mixing in
  fluid and hall-mhd turbulence},\ }\href@noop {} {\bibfield  {journal}
  {\bibinfo  {journal} {Rendiconti Lincei. Scienze Fisiche e Naturali}\
  }\textbf {\bibinfo {volume} {31}},\ \bibinfo {pages} {949} (\bibinfo {year}
  {2020}{\natexlab{a}})}\BibitemShut {NoStop}%
\bibitem [{\citenamefont {Pouquet}\ \emph
  {et~al.}(2020{\natexlab{b}})\citenamefont {Pouquet}, \citenamefont
  {Stawarz},\ and\ \citenamefont {Rosenberg}}]{pouquet2020coupling}%
  \BibitemOpen
  \bibfield  {author} {\bibinfo {author} {\bibfnamefont {A.}~\bibnamefont
  {Pouquet}}, \bibinfo {author} {\bibfnamefont {J.~E.}\ \bibnamefont
  {Stawarz}},\ and\ \bibinfo {author} {\bibfnamefont {D.}~\bibnamefont
  {Rosenberg}},\ }\bibfield  {title} {\bibinfo {title} {Coupling large eddies
  and waves in turbulence: Case study of magnetic helicity at the ion inertial
  scale},\ }\href@noop {} {\bibfield  {journal} {\bibinfo  {journal}
  {Atmosphere}\ }\textbf {\bibinfo {volume} {11}},\ \bibinfo {pages} {203}
  (\bibinfo {year} {2020}{\natexlab{b}})}\BibitemShut {NoStop}%
\bibitem [{\citenamefont {Falceta-Gon{\c{c}}alves}\ \emph
  {et~al.}(2014)\citenamefont {Falceta-Gon{\c{c}}alves}, \citenamefont {Kowal},
  \citenamefont {Falgarone},\ and\ \citenamefont
  {Chian}}]{falceta2014turbulence}%
  \BibitemOpen
  \bibfield  {author} {\bibinfo {author} {\bibfnamefont {D.}~\bibnamefont
  {Falceta-Gon{\c{c}}alves}}, \bibinfo {author} {\bibfnamefont
  {G.}~\bibnamefont {Kowal}}, \bibinfo {author} {\bibfnamefont
  {E.}~\bibnamefont {Falgarone}},\ and\ \bibinfo {author} {\bibfnamefont
  {A.-L.}\ \bibnamefont {Chian}},\ }\bibfield  {title} {\bibinfo {title}
  {Turbulence in the interstellar medium},\ }\href@noop {} {\bibfield
  {journal} {\bibinfo  {journal} {Nonlinear Processes in Geophysics}\ }\textbf
  {\bibinfo {volume} {21}},\ \bibinfo {pages} {587} (\bibinfo {year}
  {2014})}\BibitemShut {NoStop}%
\bibitem [{\citenamefont {Lighthill}(1960)}]{lighthill1960studies}%
  \BibitemOpen
  \bibfield  {author} {\bibinfo {author} {\bibfnamefont {M.~J.}\ \bibnamefont
  {Lighthill}},\ }\bibfield  {title} {\bibinfo {title} {Studies on
  magneto-hydrodynamic waves and other anisotropic wave motions},\ }\href@noop
  {} {\bibfield  {journal} {\bibinfo  {journal} {Philosophical Transactions of
  the Royal Society of London. Series A, Mathematical and Physical Sciences}\
  }\textbf {\bibinfo {volume} {252}},\ \bibinfo {pages} {397} (\bibinfo {year}
  {1960})}\BibitemShut {NoStop}%
\bibitem [{\citenamefont {G{\'o}mez}\ \emph {et~al.}(2010)\citenamefont
  {G{\'o}mez}, \citenamefont {Mininni},\ and\ \citenamefont
  {Dmitruk}}]{gomez2010hall}%
  \BibitemOpen
  \bibfield  {author} {\bibinfo {author} {\bibfnamefont {D.~O.}\ \bibnamefont
  {G{\'o}mez}}, \bibinfo {author} {\bibfnamefont {P.~D.}\ \bibnamefont
  {Mininni}},\ and\ \bibinfo {author} {\bibfnamefont {P.}~\bibnamefont
  {Dmitruk}},\ }\bibfield  {title} {\bibinfo {title} {Hall-magnetohydrodynamic
  small-scale dynamos},\ }\href@noop {} {\bibfield  {journal} {\bibinfo
  {journal} {Physical Review E}\ }\textbf {\bibinfo {volume} {82}},\ \bibinfo
  {pages} {036406} (\bibinfo {year} {2010})}\BibitemShut {NoStop}%
\bibitem [{\citenamefont {Hnat}\ \emph {et~al.}(2005)\citenamefont {Hnat},
  \citenamefont {Chapman},\ and\ \citenamefont
  {Rowlands}}]{hnat2005compressibility}%
  \BibitemOpen
  \bibfield  {author} {\bibinfo {author} {\bibfnamefont {B.}~\bibnamefont
  {Hnat}}, \bibinfo {author} {\bibfnamefont {S.~C.}\ \bibnamefont {Chapman}},\
  and\ \bibinfo {author} {\bibfnamefont {G.}~\bibnamefont {Rowlands}},\
  }\bibfield  {title} {\bibinfo {title} {Compressibility in solar wind plasma
  turbulence},\ }\href@noop {} {\bibfield  {journal} {\bibinfo  {journal}
  {Physical review letters}\ }\textbf {\bibinfo {volume} {94}},\ \bibinfo
  {pages} {204502} (\bibinfo {year} {2005})}\BibitemShut {NoStop}%
\bibitem [{\citenamefont {Horbury}\ \emph {et~al.}(2008)\citenamefont
  {Horbury}, \citenamefont {Forman},\ and\ \citenamefont
  {Oughton}}]{horbury2008anisotropic}%
  \BibitemOpen
  \bibfield  {author} {\bibinfo {author} {\bibfnamefont {T.~S.}\ \bibnamefont
  {Horbury}}, \bibinfo {author} {\bibfnamefont {M.}~\bibnamefont {Forman}},\
  and\ \bibinfo {author} {\bibfnamefont {S.}~\bibnamefont {Oughton}},\
  }\bibfield  {title} {\bibinfo {title} {Anisotropic scaling of
  magnetohydrodynamic turbulence},\ }\href@noop {} {\bibfield  {journal}
  {\bibinfo  {journal} {Physical Review Letters}\ }\textbf {\bibinfo {volume}
  {101}},\ \bibinfo {pages} {175005} (\bibinfo {year} {2008})}\BibitemShut
  {NoStop}%
\bibitem [{\citenamefont {Martin}\ \emph {et~al.}(2012)\citenamefont {Martin},
  \citenamefont {Dmitruk},\ and\ \citenamefont {Gomez}}]{martin2012energy}%
  \BibitemOpen
  \bibfield  {author} {\bibinfo {author} {\bibfnamefont {L.~N.}\ \bibnamefont
  {Martin}}, \bibinfo {author} {\bibfnamefont {P.}~\bibnamefont {Dmitruk}},\
  and\ \bibinfo {author} {\bibfnamefont {D.~O.}\ \bibnamefont {Gomez}},\
  }\bibfield  {title} {\bibinfo {title} {Energy spectrum, dissipation, and
  spatial structures in reduced hall magnetohydrodynamic},\ }\href@noop {}
  {\bibfield  {journal} {\bibinfo  {journal} {Physics of Plasmas}\ }\textbf
  {\bibinfo {volume} {19}},\ \bibinfo {pages} {052305} (\bibinfo {year}
  {2012})}\BibitemShut {NoStop}%
\bibitem [{\citenamefont {Mininni}\ \emph {et~al.}(2002)\citenamefont
  {Mininni}, \citenamefont {G{\'o}mez},\ and\ \citenamefont
  {Mahajan}}]{mininni2002dynamo}%
  \BibitemOpen
  \bibfield  {author} {\bibinfo {author} {\bibfnamefont {P.~D.}\ \bibnamefont
  {Mininni}}, \bibinfo {author} {\bibfnamefont {D.~O.}\ \bibnamefont
  {G{\'o}mez}},\ and\ \bibinfo {author} {\bibfnamefont {S.~M.}\ \bibnamefont
  {Mahajan}},\ }\bibfield  {title} {\bibinfo {title} {Dynamo action in hall
  magnetohydrodynamics},\ }\href@noop {} {\bibfield  {journal} {\bibinfo
  {journal} {The Astrophysical Journal Letters}\ }\textbf {\bibinfo {volume}
  {567}},\ \bibinfo {pages} {L81} (\bibinfo {year} {2002})}\BibitemShut
  {NoStop}%
\bibitem [{\citenamefont {Mininni}\ \emph {et~al.}(2003)\citenamefont
  {Mininni}, \citenamefont {G{\'o}mez},\ and\ \citenamefont
  {Mahajan}}]{mininni2003dynamo}%
  \BibitemOpen
  \bibfield  {author} {\bibinfo {author} {\bibfnamefont {P.~D.}\ \bibnamefont
  {Mininni}}, \bibinfo {author} {\bibfnamefont {D.~O.}\ \bibnamefont
  {G{\'o}mez}},\ and\ \bibinfo {author} {\bibfnamefont {S.~M.}\ \bibnamefont
  {Mahajan}},\ }\bibfield  {title} {\bibinfo {title} {Dynamo action in
  magnetohydrodynamics and hall-magnetohydrodynamics},\ }\href@noop {}
  {\bibfield  {journal} {\bibinfo  {journal} {The Astrophysical Journal}\
  }\textbf {\bibinfo {volume} {587}},\ \bibinfo {pages} {472} (\bibinfo {year}
  {2003})}\BibitemShut {NoStop}%
\bibitem [{\citenamefont {Mininni}\ \emph {et~al.}(2005)\citenamefont
  {Mininni}, \citenamefont {G{\'o}mez},\ and\ \citenamefont
  {Mahajan}}]{mininni2005direct}%
  \BibitemOpen
  \bibfield  {author} {\bibinfo {author} {\bibfnamefont {P.~D.}\ \bibnamefont
  {Mininni}}, \bibinfo {author} {\bibfnamefont {D.~O.}\ \bibnamefont
  {G{\'o}mez}},\ and\ \bibinfo {author} {\bibfnamefont {S.~M.}\ \bibnamefont
  {Mahajan}},\ }\bibfield  {title} {\bibinfo {title} {Direct simulations of
  helical hall-mhd turbulence and dynamo action},\ }\href@noop {} {\bibfield
  {journal} {\bibinfo  {journal} {The Astrophysical Journal}\ }\textbf
  {\bibinfo {volume} {619}},\ \bibinfo {pages} {1019} (\bibinfo {year}
  {2005})}\BibitemShut {NoStop}%
\bibitem [{\citenamefont {Mininni}\ \emph {et~al.}(2007)\citenamefont
  {Mininni}, \citenamefont {Alexakis},\ and\ \citenamefont
  {Pouquet}}]{mininni2007energy}%
  \BibitemOpen
  \bibfield  {author} {\bibinfo {author} {\bibfnamefont {P.~D.}\ \bibnamefont
  {Mininni}}, \bibinfo {author} {\bibfnamefont {A.}~\bibnamefont {Alexakis}},\
  and\ \bibinfo {author} {\bibfnamefont {A.}~\bibnamefont {Pouquet}},\
  }\bibfield  {title} {\bibinfo {title} {Energy transfer in hall-mhd
  turbulence: cascades, backscatter, and dynamo action},\ }\href@noop {}
  {\bibfield  {journal} {\bibinfo  {journal} {Journal of plasma physics}\
  }\textbf {\bibinfo {volume} {73}},\ \bibinfo {pages} {377} (\bibinfo {year}
  {2007})}\BibitemShut {NoStop}%
\bibitem [{\citenamefont {Boffetta}\ \emph {et~al.}(1999)\citenamefont
  {Boffetta}, \citenamefont {Celani}, \citenamefont {Crisanti},\ and\
  \citenamefont {Prandi}}]{boffetta1999intermittency}%
  \BibitemOpen
  \bibfield  {author} {\bibinfo {author} {\bibfnamefont {G.}~\bibnamefont
  {Boffetta}}, \bibinfo {author} {\bibfnamefont {A.}~\bibnamefont {Celani}},
  \bibinfo {author} {\bibfnamefont {A.}~\bibnamefont {Crisanti}},\ and\
  \bibinfo {author} {\bibfnamefont {R.}~\bibnamefont {Prandi}},\ }\bibfield
  {title} {\bibinfo {title} {Intermittency of two-dimensional decaying electron
  magnetohydrodynamic turbulence},\ }\href@noop {} {\bibfield  {journal}
  {\bibinfo  {journal} {Physical Review E}\ }\textbf {\bibinfo {volume} {59}},\
  \bibinfo {pages} {3724} (\bibinfo {year} {1999})}\BibitemShut {NoStop}%
\bibitem [{\citenamefont {Chapman}\ \emph {et~al.}(2008)\citenamefont
  {Chapman}, \citenamefont {Hnat},\ and\ \citenamefont
  {Kiyani}}]{chapman2008solar}%
  \BibitemOpen
  \bibfield  {author} {\bibinfo {author} {\bibfnamefont {S.}~\bibnamefont
  {Chapman}}, \bibinfo {author} {\bibfnamefont {B.}~\bibnamefont {Hnat}},\ and\
  \bibinfo {author} {\bibfnamefont {K.}~\bibnamefont {Kiyani}},\ }\bibfield
  {title} {\bibinfo {title} {Solar cycle dependence of scaling in solar wind
  fluctuations},\ }\href@noop {} {\bibfield  {journal} {\bibinfo  {journal}
  {Nonlinear Processes in Geophysics}\ }\textbf {\bibinfo {volume} {15}},\
  \bibinfo {pages} {445} (\bibinfo {year} {2008})}\BibitemShut {NoStop}%
\bibitem [{\citenamefont {Shaikh}\ and\ \citenamefont
  {Shukla}(2009)}]{shaikh20093d}%
  \BibitemOpen
  \bibfield  {author} {\bibinfo {author} {\bibfnamefont {D.}~\bibnamefont
  {Shaikh}}\ and\ \bibinfo {author} {\bibfnamefont {P.}~\bibnamefont
  {Shukla}},\ }\bibfield  {title} {\bibinfo {title} {3d simulations of
  fluctuation spectra in the hall-mhd plasma},\ }\href@noop {} {\bibfield
  {journal} {\bibinfo  {journal} {Physical Review Letters}\ }\textbf {\bibinfo
  {volume} {102}},\ \bibinfo {pages} {045004} (\bibinfo {year}
  {2009})}\BibitemShut {NoStop}%
\bibitem [{\citenamefont {Hori}\ and\ \citenamefont
  {Miura}(2008)}]{hori2008spectrum}%
  \BibitemOpen
  \bibfield  {author} {\bibinfo {author} {\bibfnamefont {D.}~\bibnamefont
  {Hori}}\ and\ \bibinfo {author} {\bibfnamefont {H.}~\bibnamefont {Miura}},\
  }\bibfield  {title} {\bibinfo {title} {Spectrum properties of hall mhd
  turbulence},\ }\href@noop {} {\bibfield  {journal} {\bibinfo  {journal}
  {Plasma and Fusion Research}\ }\textbf {\bibinfo {volume} {3}},\ \bibinfo
  {pages} {S1053} (\bibinfo {year} {2008})}\BibitemShut {NoStop}%
\bibitem [{\citenamefont {Miura}\ and\ \citenamefont
  {Araki}(2014)}]{miura2014structure}%
  \BibitemOpen
  \bibfield  {author} {\bibinfo {author} {\bibfnamefont {H.}~\bibnamefont
  {Miura}}\ and\ \bibinfo {author} {\bibfnamefont {K.}~\bibnamefont {Araki}},\
  }\bibfield  {title} {\bibinfo {title} {Structure transitions induced by the
  hall term in homogeneous and isotropic magnetohydrodynamic turbulence},\
  }\href@noop {} {\bibfield  {journal} {\bibinfo  {journal} {Physics of
  Plasmas}\ }\textbf {\bibinfo {volume} {21}},\ \bibinfo {pages} {072313}
  (\bibinfo {year} {2014})}\BibitemShut {NoStop}%
\bibitem [{\citenamefont {Miura}\ \emph {et~al.}(2016)\citenamefont {Miura},
  \citenamefont {Araki},\ and\ \citenamefont {Hamba}}]{miura2016hall}%
  \BibitemOpen
  \bibfield  {author} {\bibinfo {author} {\bibfnamefont {H.}~\bibnamefont
  {Miura}}, \bibinfo {author} {\bibfnamefont {K.}~\bibnamefont {Araki}},\ and\
  \bibinfo {author} {\bibfnamefont {F.}~\bibnamefont {Hamba}},\ }\bibfield
  {title} {\bibinfo {title} {Hall effects and sub-grid-scale modeling in
  magnetohydrodynamic turbulence simulations},\ }\href@noop {} {\bibfield
  {journal} {\bibinfo  {journal} {Journal of Computational Physics}\ }\textbf
  {\bibinfo {volume} {316}},\ \bibinfo {pages} {385} (\bibinfo {year}
  {2016})}\BibitemShut {NoStop}%
\bibitem [{\citenamefont {Miura}\ and\ \citenamefont
  {Araki}(2012)}]{miura2012coarse}%
  \BibitemOpen
  \bibfield  {author} {\bibinfo {author} {\bibfnamefont {H.}~\bibnamefont
  {Miura}}\ and\ \bibinfo {author} {\bibfnamefont {K.}~\bibnamefont {Araki}},\
  }\bibfield  {title} {\bibinfo {title} {Coarse-graining study of homogeneous
  and isotropic hall magnetohydrodynamics turbulence},\ }\href@noop {}
  {\bibfield  {journal} {\bibinfo  {journal} {Plasma Physics and Controlled
  Fusion}\ }\textbf {\bibinfo {volume} {55}},\ \bibinfo {pages} {014012}
  (\bibinfo {year} {2012})}\BibitemShut {NoStop}%
\bibitem [{\citenamefont {Miura}\ \emph {et~al.}(2019)\citenamefont {Miura},
  \citenamefont {Yang},\ and\ \citenamefont {Gotoh}}]{miura2019hall}%
  \BibitemOpen
  \bibfield  {author} {\bibinfo {author} {\bibfnamefont {H.}~\bibnamefont
  {Miura}}, \bibinfo {author} {\bibfnamefont {J.}~\bibnamefont {Yang}},\ and\
  \bibinfo {author} {\bibfnamefont {T.}~\bibnamefont {Gotoh}},\ }\bibfield
  {title} {\bibinfo {title} {Hall magnetohydrodynamic turbulence with a
  magnetic prandtl number larger than unity},\ }\href@noop {} {\bibfield
  {journal} {\bibinfo  {journal} {Physical Review E}\ }\textbf {\bibinfo
  {volume} {100}},\ \bibinfo {pages} {063207} (\bibinfo {year}
  {2019})}\BibitemShut {NoStop}%
\bibitem [{\citenamefont {Banerjee}\ \emph {et~al.}(2013)\citenamefont
  {Banerjee}, \citenamefont {Ray}, \citenamefont {Sahoo},\ and\ \citenamefont
  {Pandit}}]{banerjee2013multiscaling}%
  \BibitemOpen
  \bibfield  {author} {\bibinfo {author} {\bibfnamefont {D.}~\bibnamefont
  {Banerjee}}, \bibinfo {author} {\bibfnamefont {S.~S.}\ \bibnamefont {Ray}},
  \bibinfo {author} {\bibfnamefont {G.}~\bibnamefont {Sahoo}},\ and\ \bibinfo
  {author} {\bibfnamefont {R.}~\bibnamefont {Pandit}},\ }\bibfield  {title}
  {\bibinfo {title} {Multiscaling in hall-magnetohydrodynamic turbulence:
  insights from a shell model},\ }\href@noop {} {\bibfield  {journal} {\bibinfo
   {journal} {Physical review letters}\ }\textbf {\bibinfo {volume} {111}},\
  \bibinfo {pages} {174501} (\bibinfo {year} {2013})}\BibitemShut {NoStop}%
\bibitem [{\citenamefont {Galtier}\ and\ \citenamefont
  {Buchlin}(2007)}]{galtier2007multiscale}%
  \BibitemOpen
  \bibfield  {author} {\bibinfo {author} {\bibfnamefont {S.}~\bibnamefont
  {Galtier}}\ and\ \bibinfo {author} {\bibfnamefont {E.}~\bibnamefont
  {Buchlin}},\ }\bibfield  {title} {\bibinfo {title} {Multiscale
  hall-magnetohydrodynamic turbulence in the solar wind},\ }\href@noop {}
  {\bibfield  {journal} {\bibinfo  {journal} {The Astrophysical Journal}\
  }\textbf {\bibinfo {volume} {656}},\ \bibinfo {pages} {560} (\bibinfo {year}
  {2007})}\BibitemShut {NoStop}%
\bibitem [{\citenamefont {Galtier}(2008)}]{galtier2008karman}%
  \BibitemOpen
  \bibfield  {author} {\bibinfo {author} {\bibfnamefont {S.}~\bibnamefont
  {Galtier}},\ }\bibfield  {title} {\bibinfo {title} {von
  k{\'a}rm{\'a}n--howarth equations for hall magnetohydrodynamic flows},\
  }\href@noop {} {\bibfield  {journal} {\bibinfo  {journal} {Physical Review
  E}\ }\textbf {\bibinfo {volume} {77}},\ \bibinfo {pages} {015302} (\bibinfo
  {year} {2008})}\BibitemShut {NoStop}%
\bibitem [{\citenamefont {Meyrand}\ and\ \citenamefont
  {Galtier}(2012)}]{meyrand2012spontaneous}%
  \BibitemOpen
  \bibfield  {author} {\bibinfo {author} {\bibfnamefont {R.}~\bibnamefont
  {Meyrand}}\ and\ \bibinfo {author} {\bibfnamefont {S.}~\bibnamefont
  {Galtier}},\ }\bibfield  {title} {\bibinfo {title} {Spontaneous chiral
  symmetry breaking of hall magnetohydrodynamic turbulence},\ }\href@noop {}
  {\bibfield  {journal} {\bibinfo  {journal} {Physical review letters}\
  }\textbf {\bibinfo {volume} {109}},\ \bibinfo {pages} {194501} (\bibinfo
  {year} {2012})}\BibitemShut {NoStop}%
\bibitem [{\citenamefont {Meyrand}\ \emph {et~al.}(2018)\citenamefont
  {Meyrand}, \citenamefont {Kiyani}, \citenamefont {G{\"u}rcan},\ and\
  \citenamefont {Galtier}}]{meyrand2018coexistence}%
  \BibitemOpen
  \bibfield  {author} {\bibinfo {author} {\bibfnamefont {R.}~\bibnamefont
  {Meyrand}}, \bibinfo {author} {\bibfnamefont {K.~H.}\ \bibnamefont {Kiyani}},
  \bibinfo {author} {\bibfnamefont {{\"O}.~D.}\ \bibnamefont {G{\"u}rcan}},\
  and\ \bibinfo {author} {\bibfnamefont {S.}~\bibnamefont {Galtier}},\
  }\bibfield  {title} {\bibinfo {title} {Coexistence of weak and strong wave
  turbulence in incompressible hall magnetohydrodynamics},\ }\href@noop {}
  {\bibfield  {journal} {\bibinfo  {journal} {Physical Review X}\ }\textbf
  {\bibinfo {volume} {8}},\ \bibinfo {pages} {031066} (\bibinfo {year}
  {2018})}\BibitemShut {NoStop}%
\bibitem [{\citenamefont {Krishan}\ and\ \citenamefont
  {Mahajan}(2004)}]{krishan2004magnetic}%
  \BibitemOpen
  \bibfield  {author} {\bibinfo {author} {\bibfnamefont {V.}~\bibnamefont
  {Krishan}}\ and\ \bibinfo {author} {\bibfnamefont {S.}~\bibnamefont
  {Mahajan}},\ }\bibfield  {title} {\bibinfo {title} {Magnetic fluctuations and
  hall magnetohydrodynamic turbulence in the solar wind},\ }\href@noop {}
  {\bibfield  {journal} {\bibinfo  {journal} {Journal of Geophysical Research:
  Space Physics}\ }\textbf {\bibinfo {volume} {109}} (\bibinfo {year}
  {2004})}\BibitemShut {NoStop}%
\bibitem [{\citenamefont {G{\'o}mez}\ \emph {et~al.}(2005)\citenamefont
  {G{\'o}mez}, \citenamefont {Mininni},\ and\ \citenamefont
  {Dmitruk}}]{gomez2005mhd}%
  \BibitemOpen
  \bibfield  {author} {\bibinfo {author} {\bibfnamefont {D.~O.}\ \bibnamefont
  {G{\'o}mez}}, \bibinfo {author} {\bibfnamefont {P.~D.}\ \bibnamefont
  {Mininni}},\ and\ \bibinfo {author} {\bibfnamefont {P.}~\bibnamefont
  {Dmitruk}},\ }\bibfield  {title} {\bibinfo {title} {Mhd simulations and
  astrophysical applications},\ }\href@noop {} {\bibfield  {journal} {\bibinfo
  {journal} {Advances in Space Research}\ }\textbf {\bibinfo {volume} {35}},\
  \bibinfo {pages} {899} (\bibinfo {year} {2005})}\BibitemShut {NoStop}%
\bibitem [{\citenamefont {Hori}\ \emph {et~al.}(2005)\citenamefont {Hori},
  \citenamefont {Furukawa}, \citenamefont {Ohsaki},\ and\ \citenamefont
  {Yoshida}}]{hori2005shell}%
  \BibitemOpen
  \bibfield  {author} {\bibinfo {author} {\bibfnamefont {D.}~\bibnamefont
  {Hori}}, \bibinfo {author} {\bibfnamefont {M.}~\bibnamefont {Furukawa}},
  \bibinfo {author} {\bibfnamefont {S.}~\bibnamefont {Ohsaki}},\ and\ \bibinfo
  {author} {\bibfnamefont {Z.}~\bibnamefont {Yoshida}},\ }\bibfield  {title}
  {\bibinfo {title} {A shell model for the hall mhd system},\ }\href@noop {}
  {\bibfield  {journal} {\bibinfo  {journal} {Journal of Plasma and Fusion
  Research}\ }\textbf {\bibinfo {volume} {81}},\ \bibinfo {pages} {141}
  (\bibinfo {year} {2005})}\BibitemShut {NoStop}%
\bibitem [{\citenamefont {Chae}\ \emph {et~al.}(2014)\citenamefont {Chae},
  \citenamefont {Degond},\ and\ \citenamefont {Liu}}]{chae2014well}%
  \BibitemOpen
  \bibfield  {author} {\bibinfo {author} {\bibfnamefont {D.}~\bibnamefont
  {Chae}}, \bibinfo {author} {\bibfnamefont {P.}~\bibnamefont {Degond}},\ and\
  \bibinfo {author} {\bibfnamefont {J.-G.}\ \bibnamefont {Liu}},\ }\bibfield
  {title} {\bibinfo {title} {Well-posedness for hall-magnetohydrodynamics},\
  }in\ \href@noop {} {\emph {\bibinfo {booktitle} {Annales de l'IHP Analyse non
  lin{\'e}aire}}},\ Vol.~\bibinfo {volume} {31}\ (\bibinfo {year} {2014})\ pp.\
  \bibinfo {pages} {555--565}\BibitemShut {NoStop}%
\bibitem [{\citenamefont {Alghamdi}\ \emph {et~al.}(2018)\citenamefont
  {Alghamdi}, \citenamefont {Gala},\ and\ \citenamefont
  {Ragusa}}]{alghamdi2018regularity}%
  \BibitemOpen
  \bibfield  {author} {\bibinfo {author} {\bibfnamefont {A.}~\bibnamefont
  {Alghamdi}}, \bibinfo {author} {\bibfnamefont {S.}~\bibnamefont {Gala}},\
  and\ \bibinfo {author} {\bibfnamefont {M.}~\bibnamefont {Ragusa}},\
  }\bibfield  {title} {\bibinfo {title} {A regularity criterion of smooth
  solution for the 3d viscous hall-mhd equations},\ }\href@noop {} {\bibfield
  {journal} {\bibinfo  {journal} {Aims Mathematics}\ }\textbf {\bibinfo
  {volume} {3}},\ \bibinfo {pages} {565} (\bibinfo {year} {2018})}\BibitemShut
  {NoStop}%
\bibitem [{\citenamefont {Goldstein}\ \emph {et~al.}(1995)\citenamefont
  {Goldstein}, \citenamefont {Roberts},\ and\ \citenamefont
  {Matthaeus}}]{goldstein1995magnetohydrodynamic}%
  \BibitemOpen
  \bibfield  {author} {\bibinfo {author} {\bibfnamefont {M.~L.}\ \bibnamefont
  {Goldstein}}, \bibinfo {author} {\bibfnamefont {D.}~\bibnamefont {Roberts}},\
  and\ \bibinfo {author} {\bibfnamefont {W.}~\bibnamefont {Matthaeus}},\
  }\bibfield  {title} {\bibinfo {title} {Magnetohydrodynamic turbulence in the
  solar wind},\ }\href@noop {} {\bibfield  {journal} {\bibinfo  {journal}
  {Annual review of astronomy and astrophysics}\ }\textbf {\bibinfo {volume}
  {33}},\ \bibinfo {pages} {283} (\bibinfo {year} {1995})}\BibitemShut
  {NoStop}%
\bibitem [{\citenamefont {Verma}(1996)}]{verma1996nonclassical}%
  \BibitemOpen
  \bibfield  {author} {\bibinfo {author} {\bibfnamefont {M.~K.}\ \bibnamefont
  {Verma}},\ }\bibfield  {title} {\bibinfo {title} {Nonclassical viscosity and
  resistivity of the solar wind plasma},\ }\href@noop {} {\bibfield  {journal}
  {\bibinfo  {journal} {Journal of Geophysical Research: Space Physics}\
  }\textbf {\bibinfo {volume} {101}},\ \bibinfo {pages} {27543} (\bibinfo
  {year} {1996})}\BibitemShut {NoStop}%
\bibitem [{\citenamefont {Bruno}\ and\ \citenamefont
  {Carbone}(2013)}]{bruno2013}%
  \BibitemOpen
  \bibfield  {author} {\bibinfo {author} {\bibfnamefont {R.}~\bibnamefont
  {Bruno}}\ and\ \bibinfo {author} {\bibfnamefont {V.}~\bibnamefont
  {Carbone}},\ }\bibfield  {title} {\bibinfo {title} {The solar wind as a
  turbulence laboratory},\ }\href@noop {} {\bibfield  {journal} {\bibinfo
  {journal} {Living Rev. Sol. Phys.}\ }\textbf {\bibinfo {volume} {10}},\
  \bibinfo {pages} {2} (\bibinfo {year} {2013})}\BibitemShut {NoStop}%
\bibitem [{\citenamefont {Bruno}\ and\ \citenamefont
  {Carbone}(2016)}]{bruno2016turbulence}%
  \BibitemOpen
  \bibfield  {author} {\bibinfo {author} {\bibfnamefont {R.}~\bibnamefont
  {Bruno}}\ and\ \bibinfo {author} {\bibfnamefont {V.}~\bibnamefont
  {Carbone}},\ }\href@noop {} {\emph {\bibinfo {title} {Turbulence in the solar
  wind}}},\ Vol.\ \bibinfo {volume} {928}\ (\bibinfo  {publisher} {Springer},\
  \bibinfo {year} {2016})\BibitemShut {NoStop}%
\bibitem [{\citenamefont {Kiyani}\ \emph {et~al.}(2009)\citenamefont {Kiyani},
  \citenamefont {Chapman}, \citenamefont {Khotyaintsev}, \citenamefont
  {Dunlop},\ and\ \citenamefont {Sahraoui}}]{kiyani2009global}%
  \BibitemOpen
  \bibfield  {author} {\bibinfo {author} {\bibfnamefont {K.}~\bibnamefont
  {Kiyani}}, \bibinfo {author} {\bibfnamefont {S.}~\bibnamefont {Chapman}},
  \bibinfo {author} {\bibfnamefont {Y.~V.}\ \bibnamefont {Khotyaintsev}},
  \bibinfo {author} {\bibfnamefont {M.}~\bibnamefont {Dunlop}},\ and\ \bibinfo
  {author} {\bibfnamefont {F.}~\bibnamefont {Sahraoui}},\ }\bibfield  {title}
  {\bibinfo {title} {Global scale-invariant dissipation in collisionless plasma
  turbulence},\ }\href@noop {} {\bibfield  {journal} {\bibinfo  {journal}
  {Physical review letters}\ }\textbf {\bibinfo {volume} {103}},\ \bibinfo
  {pages} {075006} (\bibinfo {year} {2009})}\BibitemShut {NoStop}%
\bibitem [{\citenamefont {Matthaeus}\ and\ \citenamefont
  {Goldstein}(1982)}]{matthaeus1982measurement}%
  \BibitemOpen
  \bibfield  {author} {\bibinfo {author} {\bibfnamefont {W.~H.}\ \bibnamefont
  {Matthaeus}}\ and\ \bibinfo {author} {\bibfnamefont {M.~L.}\ \bibnamefont
  {Goldstein}},\ }\bibfield  {title} {\bibinfo {title} {Measurement of the
  rugged invariants of magnetohydrodynamic turbulence in the solar wind},\
  }\href@noop {} {\bibfield  {journal} {\bibinfo  {journal} {Journal of
  Geophysical Research: Space Physics}\ }\textbf {\bibinfo {volume} {87}},\
  \bibinfo {pages} {6011} (\bibinfo {year} {1982})}\BibitemShut {NoStop}%
\bibitem [{\citenamefont {Matthaeus}\ \emph {et~al.}(2015)\citenamefont
  {Matthaeus}, \citenamefont {Wan}, \citenamefont {Servidio}, \citenamefont
  {Greco}, \citenamefont {Osman}, \citenamefont {Oughton},\ and\ \citenamefont
  {Dmitruk}}]{matthaeus2015intermittency}%
  \BibitemOpen
  \bibfield  {author} {\bibinfo {author} {\bibfnamefont {W.~H.}\ \bibnamefont
  {Matthaeus}}, \bibinfo {author} {\bibfnamefont {M.}~\bibnamefont {Wan}},
  \bibinfo {author} {\bibfnamefont {S.}~\bibnamefont {Servidio}}, \bibinfo
  {author} {\bibfnamefont {A.}~\bibnamefont {Greco}}, \bibinfo {author}
  {\bibfnamefont {K.~T.}\ \bibnamefont {Osman}}, \bibinfo {author}
  {\bibfnamefont {S.}~\bibnamefont {Oughton}},\ and\ \bibinfo {author}
  {\bibfnamefont {P.}~\bibnamefont {Dmitruk}},\ }\bibfield  {title} {\bibinfo
  {title} {Intermittency, nonlinear dynamics and dissipation in the solar wind
  and astrophysical plasmas},\ }\href@noop {} {\bibfield  {journal} {\bibinfo
  {journal} {Philosophical Transactions of the Royal Society A: Mathematical,
  Physical and Engineering Sciences}\ }\textbf {\bibinfo {volume} {373}},\
  \bibinfo {pages} {20140154} (\bibinfo {year} {2015})}\BibitemShut {NoStop}%
\bibitem [{\citenamefont {Alexandrova}\ \emph {et~al.}(2013)\citenamefont
  {Alexandrova}, \citenamefont {Chen}, \citenamefont {Sorriso-Valvo},
  \citenamefont {Horbury},\ and\ \citenamefont {Bale}}]{alexandrova2013solar}%
  \BibitemOpen
  \bibfield  {author} {\bibinfo {author} {\bibfnamefont {O.}~\bibnamefont
  {Alexandrova}}, \bibinfo {author} {\bibfnamefont {C.~H.~K.}\ \bibnamefont
  {Chen}}, \bibinfo {author} {\bibfnamefont {L.}~\bibnamefont {Sorriso-Valvo}},
  \bibinfo {author} {\bibfnamefont {T.~S.}\ \bibnamefont {Horbury}},\ and\
  \bibinfo {author} {\bibfnamefont {S.~D.}\ \bibnamefont {Bale}},\ }\bibfield
  {title} {\bibinfo {title} {Solar wind turbulence and the role of ion
  instabilities},\ }\href@noop {} {\bibfield  {journal} {\bibinfo  {journal}
  {Space Science Reviews}\ }\textbf {\bibinfo {volume} {178}},\ \bibinfo
  {pages} {101} (\bibinfo {year} {2013})}\BibitemShut {NoStop}%
\bibitem [{\citenamefont {Sahraoui}\ \emph {et~al.}(2020)\citenamefont
  {Sahraoui}, \citenamefont {Hadid},\ and\ \citenamefont
  {Huang}}]{sahraoui2020magnetohydrodynamic}%
  \BibitemOpen
  \bibfield  {author} {\bibinfo {author} {\bibfnamefont {F.}~\bibnamefont
  {Sahraoui}}, \bibinfo {author} {\bibfnamefont {L.}~\bibnamefont {Hadid}},\
  and\ \bibinfo {author} {\bibfnamefont {S.}~\bibnamefont {Huang}},\ }\bibfield
   {title} {\bibinfo {title} {Magnetohydrodynamic and kinetic scale turbulence
  in the near-earth space plasmas: a (short) biased review},\ }\href@noop {}
  {\bibfield  {journal} {\bibinfo  {journal} {Reviews of Modern Plasma
  Physics}\ }\textbf {\bibinfo {volume} {4}},\ \bibinfo {pages} {1} (\bibinfo
  {year} {2020})}\BibitemShut {NoStop}%
\bibitem [{\citenamefont {Alexandrova}\ \emph {et~al.}(2008)\citenamefont
  {Alexandrova}, \citenamefont {Carbone}, \citenamefont {Veltri},\ and\
  \citenamefont {Sorriso-Valvo}}]{alexandrova2008small}%
  \BibitemOpen
  \bibfield  {author} {\bibinfo {author} {\bibfnamefont {O.}~\bibnamefont
  {Alexandrova}}, \bibinfo {author} {\bibfnamefont {V.}~\bibnamefont
  {Carbone}}, \bibinfo {author} {\bibfnamefont {P.}~\bibnamefont {Veltri}},\
  and\ \bibinfo {author} {\bibfnamefont {L.}~\bibnamefont {Sorriso-Valvo}},\
  }\bibfield  {title} {\bibinfo {title} {Small-scale energy cascade of the
  solar wind turbulence},\ }\href@noop {} {\bibfield  {journal} {\bibinfo
  {journal} {The Astrophysical Journal}\ }\textbf {\bibinfo {volume} {674}},\
  \bibinfo {pages} {1153} (\bibinfo {year} {2008})}\BibitemShut {NoStop}%
\bibitem [{\citenamefont {Alexandrova}\ \emph {et~al.}(2007)\citenamefont
  {Alexandrova}, \citenamefont {Carbone}, \citenamefont {Veltri},\ and\
  \citenamefont {Sorriso-Valvo}}]{alexandrova2007solar}%
  \BibitemOpen
  \bibfield  {author} {\bibinfo {author} {\bibfnamefont {O.}~\bibnamefont
  {Alexandrova}}, \bibinfo {author} {\bibfnamefont {V.}~\bibnamefont
  {Carbone}}, \bibinfo {author} {\bibfnamefont {P.}~\bibnamefont {Veltri}},\
  and\ \bibinfo {author} {\bibfnamefont {L.}~\bibnamefont {Sorriso-Valvo}},\
  }\bibfield  {title} {\bibinfo {title} {Solar wind cluster observations:
  Turbulent spectrum and role of hall effect},\ }\href@noop {} {\bibfield
  {journal} {\bibinfo  {journal} {Planetary and Space Science}\ }\textbf
  {\bibinfo {volume} {55}},\ \bibinfo {pages} {2224} (\bibinfo {year}
  {2007})}\BibitemShut {NoStop}%
\bibitem [{\citenamefont {Zimbardo}\ \emph {et~al.}(2010)\citenamefont
  {Zimbardo}, \citenamefont {Greco}, \citenamefont {Sorriso-Valvo},
  \citenamefont {Perri}, \citenamefont {V{\"o}r{\"o}s}, \citenamefont
  {Aburjania}, \citenamefont {Chargazia},\ and\ \citenamefont
  {Alexandrova}}]{zimbardo2010magnetic}%
  \BibitemOpen
  \bibfield  {author} {\bibinfo {author} {\bibfnamefont {G.}~\bibnamefont
  {Zimbardo}}, \bibinfo {author} {\bibfnamefont {A.}~\bibnamefont {Greco}},
  \bibinfo {author} {\bibfnamefont {L.}~\bibnamefont {Sorriso-Valvo}}, \bibinfo
  {author} {\bibfnamefont {S.}~\bibnamefont {Perri}}, \bibinfo {author}
  {\bibfnamefont {Z.}~\bibnamefont {V{\"o}r{\"o}s}}, \bibinfo {author}
  {\bibfnamefont {G.}~\bibnamefont {Aburjania}}, \bibinfo {author}
  {\bibfnamefont {K.}~\bibnamefont {Chargazia}},\ and\ \bibinfo {author}
  {\bibfnamefont {O.}~\bibnamefont {Alexandrova}},\ }\bibfield  {title}
  {\bibinfo {title} {Magnetic turbulence in the geospace environment},\
  }\href@noop {} {\bibfield  {journal} {\bibinfo  {journal} {Space science
  reviews}\ }\textbf {\bibinfo {volume} {156}},\ \bibinfo {pages} {89}
  (\bibinfo {year} {2010})}\BibitemShut {NoStop}%
\bibitem [{\citenamefont {Frish}(1995)}]{frish1995}%
  \BibitemOpen
  \bibfield  {author} {\bibinfo {author} {\bibfnamefont {U.}~\bibnamefont
  {Frish}},\ }\href@noop {} {\emph {\bibinfo {title} {Turbulence: The legacy of
  A. N. Kologorov}}}\ (\bibinfo  {publisher} {Cambridge University Press},\
  \bibinfo {address} {Cambridge, England},\ \bibinfo {year} {1995})\BibitemShut
  {NoStop}%
\bibitem [{\citenamefont {Vasquez}\ \emph {et~al.}(2007)\citenamefont
  {Vasquez}, \citenamefont {Abramenko}, \citenamefont {Haggerty},\ and\
  \citenamefont {Smith}}]{vasquez2007numerous}%
  \BibitemOpen
  \bibfield  {author} {\bibinfo {author} {\bibfnamefont {B.~J.}\ \bibnamefont
  {Vasquez}}, \bibinfo {author} {\bibfnamefont {V.~I.}\ \bibnamefont
  {Abramenko}}, \bibinfo {author} {\bibfnamefont {D.~K.}\ \bibnamefont
  {Haggerty}},\ and\ \bibinfo {author} {\bibfnamefont {C.~W.}\ \bibnamefont
  {Smith}},\ }\bibfield  {title} {\bibinfo {title} {Numerous small magnetic
  field discontinuities of bartels rotation 2286 and the potential role of
  alfv{\'e}nic turbulence},\ }\href@noop {} {\bibfield  {journal} {\bibinfo
  {journal} {Journal of Geophysical Research: Space Physics}\ }\textbf
  {\bibinfo {volume} {112}} (\bibinfo {year} {2007})}\BibitemShut {NoStop}%
\bibitem [{\citenamefont {Matthaeus}\ \emph {et~al.}(2012)\citenamefont
  {Matthaeus}, \citenamefont {Montgomery}, \citenamefont {Wan},\ and\
  \citenamefont {Servidio}}]{matthaeus2012review}%
  \BibitemOpen
  \bibfield  {author} {\bibinfo {author} {\bibfnamefont {W.~H.}\ \bibnamefont
  {Matthaeus}}, \bibinfo {author} {\bibfnamefont {D.~C.}\ \bibnamefont
  {Montgomery}}, \bibinfo {author} {\bibfnamefont {M.}~\bibnamefont {Wan}},\
  and\ \bibinfo {author} {\bibfnamefont {S.}~\bibnamefont {Servidio}},\
  }\bibfield  {title} {\bibinfo {title} {A review of relaxation and structure
  in some turbulent plasmas: magnetohydrodynamics and related models},\
  }\href@noop {} {\bibfield  {journal} {\bibinfo  {journal} {Journal of
  Turbulence}\ }\textbf {\bibinfo {volume} {13}},\ \bibinfo {pages} {N37}
  (\bibinfo {year} {2012})}\BibitemShut {NoStop}%
\bibitem [{\citenamefont {Wan}\ \emph {et~al.}(2011)\citenamefont {Wan},
  \citenamefont {Osman}, \citenamefont {Matthaeus},\ and\ \citenamefont
  {Oughton}}]{wan2011investigation}%
  \BibitemOpen
  \bibfield  {author} {\bibinfo {author} {\bibfnamefont {M.}~\bibnamefont
  {Wan}}, \bibinfo {author} {\bibfnamefont {K.~T.}\ \bibnamefont {Osman}},
  \bibinfo {author} {\bibfnamefont {W.~H.}\ \bibnamefont {Matthaeus}},\ and\
  \bibinfo {author} {\bibfnamefont {S.}~\bibnamefont {Oughton}},\ }\bibfield
  {title} {\bibinfo {title} {Investigation of intermittency in
  magnetohydrodynamics and solar wind turbulence: scale-dependent kurtosis},\
  }\href@noop {} {\bibfield  {journal} {\bibinfo  {journal} {The Astrophysical
  Journal}\ }\textbf {\bibinfo {volume} {744}},\ \bibinfo {pages} {171}
  (\bibinfo {year} {2011})}\BibitemShut {NoStop}%
\bibitem [{\citenamefont {Rodriguez~Imazio}\ \emph {et~al.}(2013)\citenamefont
  {Rodriguez~Imazio}, \citenamefont {Martin}, \citenamefont {Dmitruk},\ and\
  \citenamefont {Mininni}}]{rodriguez2013intermittency}%
  \BibitemOpen
  \bibfield  {author} {\bibinfo {author} {\bibfnamefont {P.}~\bibnamefont
  {Rodriguez~Imazio}}, \bibinfo {author} {\bibfnamefont {L.~N.}\ \bibnamefont
  {Martin}}, \bibinfo {author} {\bibfnamefont {P.}~\bibnamefont {Dmitruk}},\
  and\ \bibinfo {author} {\bibfnamefont {P.~D.}\ \bibnamefont {Mininni}},\
  }\bibfield  {title} {\bibinfo {title} {Intermittency in
  hall-magnetohydrodynamics with a strong guide field},\ }\href@noop {}
  {\bibfield  {journal} {\bibinfo  {journal} {Physics of Plasmas}\ }\textbf
  {\bibinfo {volume} {20}},\ \bibinfo {pages} {052506} (\bibinfo {year}
  {2013})}\BibitemShut {NoStop}%
\bibitem [{\citenamefont {Dallas}\ and\ \citenamefont
  {Alexakis}(2013)}]{dallas2013structures}%
  \BibitemOpen
  \bibfield  {author} {\bibinfo {author} {\bibfnamefont {V.}~\bibnamefont
  {Dallas}}\ and\ \bibinfo {author} {\bibfnamefont {A.}~\bibnamefont
  {Alexakis}},\ }\bibfield  {title} {\bibinfo {title} {Structures and dynamics
  of small scales in decaying magnetohydrodynamic turbulence},\ }\href@noop {}
  {\bibfield  {journal} {\bibinfo  {journal} {Physics of Fluids}\ }\textbf
  {\bibinfo {volume} {25}},\ \bibinfo {pages} {105106} (\bibinfo {year}
  {2013})}\BibitemShut {NoStop}%
\bibitem [{\citenamefont {Turner}(1986)}]{turner1986conserved}%
  \BibitemOpen
  \bibfield  {author} {\bibinfo {author} {\bibfnamefont {L.}~\bibnamefont
  {Turner}},\ }\bibfield  {title} {\bibinfo {title} {Hall effects on magnetic
  relaxation},\ }\href@noop {} {\bibfield  {journal} {\bibinfo  {journal} {IEEE
  Transactions on Plasma Science}\ }\textbf {\bibinfo {volume} {14}},\ \bibinfo
  {pages} {849} (\bibinfo {year} {1986})}\BibitemShut {NoStop}%
\bibitem [{\citenamefont {Pandit}\ \emph {et~al.}(2009)\citenamefont {Pandit},
  \citenamefont {Perlekar},\ and\ \citenamefont {Ray}}]{pandit2009statistical}%
  \BibitemOpen
  \bibfield  {author} {\bibinfo {author} {\bibfnamefont {R.}~\bibnamefont
  {Pandit}}, \bibinfo {author} {\bibfnamefont {P.}~\bibnamefont {Perlekar}},\
  and\ \bibinfo {author} {\bibfnamefont {S.~S.}\ \bibnamefont {Ray}},\
  }\bibfield  {title} {\bibinfo {title} {Statistical properties of turbulence:
  an overview},\ }\href@noop {} {\bibfield  {journal} {\bibinfo  {journal}
  {Pramana}\ }\textbf {\bibinfo {volume} {73}},\ \bibinfo {pages} {157}
  (\bibinfo {year} {2009})}\BibitemShut {NoStop}%
\bibitem [{\citenamefont {Benzi}\ \emph {et~al.}(1993)\citenamefont {Benzi},
  \citenamefont {Ciliberto}, \citenamefont {Tripiccione}, \citenamefont
  {Baudet}, \citenamefont {Massaioli},\ and\ \citenamefont
  {Succi}}]{benzi1993extended}%
  \BibitemOpen
  \bibfield  {author} {\bibinfo {author} {\bibfnamefont {R.}~\bibnamefont
  {Benzi}}, \bibinfo {author} {\bibfnamefont {S.}~\bibnamefont {Ciliberto}},
  \bibinfo {author} {\bibfnamefont {R.}~\bibnamefont {Tripiccione}}, \bibinfo
  {author} {\bibfnamefont {C.}~\bibnamefont {Baudet}}, \bibinfo {author}
  {\bibfnamefont {F.}~\bibnamefont {Massaioli}},\ and\ \bibinfo {author}
  {\bibfnamefont {S.}~\bibnamefont {Succi}},\ }\bibfield  {title} {\bibinfo
  {title} {Extended self-similarity in turbulent flows},\ }\href@noop {}
  {\bibfield  {journal} {\bibinfo  {journal} {Physical review E}\ }\textbf
  {\bibinfo {volume} {48}},\ \bibinfo {pages} {R29} (\bibinfo {year}
  {1993})}\BibitemShut {NoStop}%
\bibitem [{\citenamefont {Chakraborty}\ \emph {et~al.}(2010)\citenamefont
  {Chakraborty}, \citenamefont {Frisch},\ and\ \citenamefont
  {Ray}}]{chakraborty2010extended}%
  \BibitemOpen
  \bibfield  {author} {\bibinfo {author} {\bibfnamefont {S.}~\bibnamefont
  {Chakraborty}}, \bibinfo {author} {\bibfnamefont {U.}~\bibnamefont
  {Frisch}},\ and\ \bibinfo {author} {\bibfnamefont {S.~S.}\ \bibnamefont
  {Ray}},\ }\bibfield  {title} {\bibinfo {title} {Extended self-similarity
  works for the burgers equation and why},\ }\href@noop {} {\bibfield
  {journal} {\bibinfo  {journal} {Journal of fluid mechanics}\ }\textbf
  {\bibinfo {volume} {649}},\ \bibinfo {pages} {275} (\bibinfo {year}
  {2010})}\BibitemShut {NoStop}%
\end{thebibliography}%
\end{document}